\documentclass{acmtog}

\acmVolume{VV}
\acmNumber{N}
\acmYear{YYYY}
\acmMonth{Month}
\acmArticleNum{XXX}  
\acmdoi{10.1145/XXXXXXX.YYYYYYY}

\usepackage{multirow}

\newcommand{\diff}{\mathrm{d}}
\newcommand{\I}{\mathbf{g}}
\newcommand{\II}{\mathbf{h}}

\newcommand{\diffU}[1]{{E^{#1}_{}}}

\newcommand\numberthis{\addtocounter{equation}{1}\tag{\theequation}}

\usepackage{pdfpages}
\usepackage{mystyle}
\usepackage{placeins}
\usepackage{subfig}

\usepackage{soul}

\begin{document}

\markboth{E. Corman and M. Ovsjanikov}{Functional Characterization of Deformation Fields}

\title{Functional Characterization of Deformation Fields} % title

\author{ETIENNE CORMAN {\upshape and} MAKS OVSJANIKOV
\affil{LIX, \'{E}cole Polytechnique, CNRS}}




\acmformat{}

\maketitle

\begin{bottomstuff} 
\end{bottomstuff}

\begin{abstract} 
  In this paper we present a novel representation for deformation fields of 3D
  shapes, by considering the induced changes in the underlying metric. In
  particular, our approach allows to represent a deformation field in a
  coordinate-free way as a linear operator acting on real-valued functions
  defined on the shape. Such a representation both provides a way to relate
  deformation fields to other classical functional operators and enables
  analysis and processing of deformation fields using standard linear-algebraic
  tools. {This opens the door to a wide variety of applications such as
    explicitly adding extrinsic information into the computation of functional
    maps, intrinsic shape symmetrization, joint deformation design through
    precise control of metric distortion, and coordinate-free deformation
    transfer without requiring pointwise correspondences.
% intrinsic
%   symmetrization and even improved shape matching by considering the composition of the
%   deformation with other functional operators. 
  Our method is applicable to both surface and volumetric shape representations
  and we guarantee the equivalence between the operator-based and standard
  deformation field representation under mild genericity conditions in the
  discrete setting.  We demonstrate the utility of our approach by comparing it
  with existing techniques and show how our representation provides a powerful toolbox for
  a wide variety of challenging problems. }
\end{abstract}

% !TEX root = fundeform_tog2.tex

\section{Introduction}
Designing and analyzing shape deformations is a central problem in computer graphics and geometry
processing, with applications in scenarios such as shape manipulation \cite{yu2004mesh,sorkine07}, animation and
deformation transfer \cite{sumner2004deformation}, shape interpolation \cite{kilian07,vontycowicz15},
and even anisotropic meshing \cite{panozzo14} among myriad others. Traditionally, shape deformation
has been motivated by interactive applications in which the main goal is to design a 
deformation that satisfies some user-prescribed handle constraints while preserving the main
structural properties of the shape. In other applications, such as shape interpolation and
deformation transfer, that lack handle constraints, the goal is to design a global deformation
field that would satisfy some structural properties as well as possible.

In both types of applications, most approaches are based on specifying a deformation energy and
providing a method to optimize it. On the other hand, several works have demonstrated that by
choosing an appropriate \emph{representation} for shape deformations, many tasks can become
significantly easier, and in particular can help to enforce certain properties of the deformation
field, which are otherwise very difficult to access and optimize for. In addition to the classical
per-vertex displacement vectors, such representations have included gradient-based deformations
\cite{yu2004mesh,zayer05}, Laplacian-based approaches \cite{lipman2004differential,sorkine04} and
M\"{o}bius transformations in the context of conformal deformations \cite{crane11,vaxman15} among
others.

\begin{figure}
\begin{tabular}{c}
	\includegraphics[width=\linewidth]{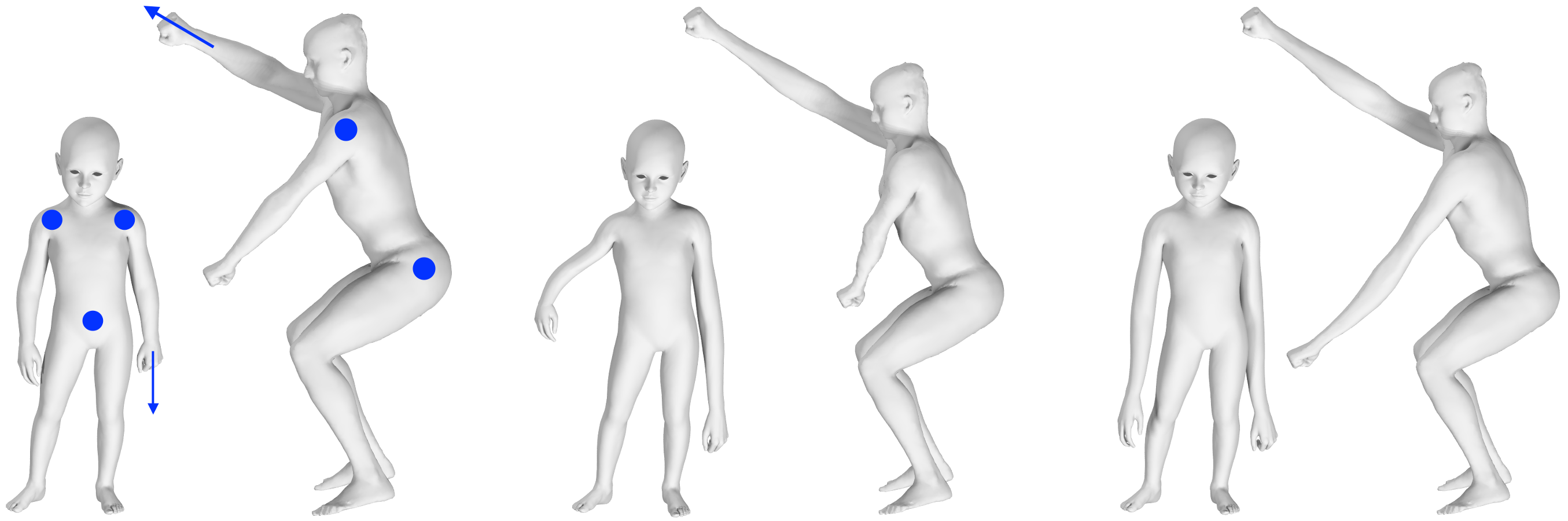}
%	\includegraphics[height=.25\textwidth]{Images/Teaser/Init2}&
%	\hspace{2em}&
%	\includegraphics[height=.25\textwidth]{Images/Teaser/Deformation_joint1}&
%	\includegraphics[height=.25\textwidth]{Images/Teaser/Deformation_joint2}
\end{tabular}
\caption{{An example of \emph{joint deformation design} using our framework where all
    objectives are easily expressed as linear constraints. Left pair: A set of local constraints for
    the deformation fields on two different shapes. Middle pair: Joint deformation design by a
    direct transfer of the coordinate functions of the deformation field. This method yields
    undesirable effects as it is rotation dependent. Right pair: Joint deformation design with a
    soft (functional) map and no pointwise correspondences between the shapes.}
% \maks{Can we add the
%     result of using transferring deformations as $x,y,z$ functions to this example? It would be nice to have it
%     early on.}\etienne{This example might not be the best to do this. I suspect it will give similar results.}
  \label{fig:teaser}
\vspace{-2mm}}
\end{figure}

At the same time, a number of recent works have shown that many basic operations
in geometry processing can be viewed as linear operators acting on real-valued
functions defined on the shapes. This includes the functional representation of
mappings or correspondences acting through composition
\cite{ovsjanikov2012functional,pokrass2013sparse}, representations of vector
fields as derivations \cite{pavlov2011structure,azencot2013operator} and
formulation of shape distortion via shape difference operators
\cite{rustamov-2013}. One advantage of these representations is that linear
operators can be naturally \emph{composed,} which makes it easy to define, for
example, the push-forward of a vector field with respect to a mapping, if both
are represented as linear operators, or to solve for Killing vector fields, by
composition between a derivation and the Laplacian operator. {Moreover,
  by using a consistent functional representation these techniques often
  alleviate the need for point-wise correspondences, which can be difficult to
  obtain, as shown very recently for example in a work on joint cross-field
  design \cite{azencot17}.}

While tangent vector fields are classically understood as operators
(derivations) in differential geometry, extrinsic vector fields do not enjoy a
similar property. Our main goal is to provide a coordinate-free representation
of extrinsic vector fields (that we also call deformation fields) as functional
operators, which will prove useful for analysis and design of shape
deformations. {As we demonstrate below our representation greatly
  simplifies certain tasks such as intrinsic symmetrization, the computation of
  mappings by composition with other operators, and joint deformation design
  without requiring point-wise mappings. Moreover, it provides an explicit link
  between deformation fields and the changes in intrinsic metric quantities,
  which can be useful in a variety of analysis and deformation processing
  tasks.}

For example, consider two shapes shown in Fig. \ref{fig:teaser} (left). By using
our framework, it is possible to combine local deformation constraints
with intrinsic objectives such as constructing a deformation field that is
as-isometric-as-possible. {Moreover, our representation allows to relate
  deformations on multiple shapes in a coordinate-free way, enabling
deformation transfer and joint design using only soft, functional
correspondences as shown in Fig. \ref{fig:teaser} (right).}

\section{Related Work}
\label{sec:related}
Shape deformation is one of the oldest and best-researched topics in computer graphics and geometry processing. We
therefore only mention works most directly related to ours and refer the interested reader to surveys including
\cite{nealen2006physically,botsch2008linear} and \cite{botsch2010polygon} (Chapter 9).

A multitude of methods exists for surface deformation starting with the seminal work of
\cite{terzopoulos1987elastically}, its early follow-ups including \cite{celniker1991deformable,welch1992variational} and
the multi-scale variants, such as \cite{zorin1997,kobbelt98,guskov99} among many others. Similarly to our approach,
many of these techniques are based on optimizing the so-called elastic thin shell energy that measures stretching and bending,
and which is often linearized for efficiency. In the majority of cases, deformations are represented explicitly as
extrinsic vector fields defined on a surface, making deformation transfer difficult in the absence of precise pointwise correspondences.

A number of methods have proposed alternative representations for deformation fields, which greatly simplify
certain tasks in design and analysis. This includes gradient-based techniques \cite{yu2004mesh,zayer05} which
consider the deformation field by aligning its gradient with a set of local per-triangle transformations. By working in
gradient space, constraints can be posed independently on the triangles and then optimized globally by solving the
Poisson equation. Similarly, Laplacian-based techniques
\cite{sorkine04,lipman2004differential,SketchBasedEditing:2005} are based on defining shape deformations
by manipulating per-vertex differential coordinates (Laplacians) in order to match some target Laplacian
coordinates. Such differential coordinates enable direct editing of local shape properties, which can be especially
beneficial for preserving and manipulating the high-frequency details of the surface. However, these
coordinates are typically not rotationally invariant and additional steps are necessary to introduce invariance
\cite{sorkine04,lipman2004differential,paries2007simple}.

More recently, a number of methods have introduced representations for mesh deformations specifically geared towards
particular shape manipulations, such as computing conformal transformations by designing special maps into the space of
quaternions \cite{crane11} or by using face-based compatible M\"obius transformations \cite{vaxman15}. These techniques
are rotationally invariant and coordinate-free, while being restricted to special types of manipulations. Another
technique, closely related to ours, designs shape deformations by constructing a continuous divergence-free vector
field \cite{von2006vector}, and applying path line integration to obtain a deformed shape. We also consider the effect
of the deformation on the metric, but both analyze the distortion of arbitrary extrinsic vector fields and show how they
can be represented in coordinate-free way as linear functional operators.

{ Our use of spectral techniques and functional maps for representing deformation fields is
  also related to previous works in spectral shape processing, including the early approaches of L\'evy and colleagues and their
  extensions \cite{levy06,levy2008,dey2012eigen} and more recent techniques such those based
  on coupled quasi-harmonic bases and functional maps 
\cite{kovnatsky2013coupled,yin2015spectral}. In these and related methods deformation fields
  are represented as triplets of functions, which encode displacement in each spatial
  coordinate. Although this representation is simple and naturally fits with the functional map
  framework, it suffers from several drawbacks. First, it is not rotationally
  invariant and induces artefacts if the shapes are not pre-aligned or are in
  different poses (see e.g., Figure \ref{fig:teaser}). Perhaps more
  fundamentally, such a
  representation is not ``shape-aware'' since it does not reflect the change in the (e.g., metric)
  structure of the shapes induced by the deformation, which reduces its utility in
  deformation analysis and design. We demonstrate through extensive experiments, that by using our
  coordinate-free representation we can avoid these limitations and open the door to entirely novel
  design and analysis applications, such as intrinsic symmetrization (Section \ref{exp:intSym}),
  which cannot be achieved using previous methods.}

Our approach of considering the deformation via its induced metric distortion is also related to the work of
\cite{eigensatz2009positional} and \cite{sela2015computational} who manipulate shapes by explicitly editing their
curvature properties. Moreover, our use of the strain tensor in characterizing metric distortion is closely related to
the applications in various physically based deformation scenarios including \cite{thomaszewski09,muller2014} among many
others (see also the surveys on physically based elastic deformable models
\cite{nealen2006physically,rumpf2014geometry}). Our approach is also related to the works that aim to design
as-isometric-as-possible shape deformations \cite{zhang2015fast,solomon2011killing,MartinezEsturo2013a}. Similarly to
the latter work, our framework is general and allows an arbitrary prescribed distortion, although our method
works directly on surface representations and moreover enables applications such as joint deformation design.

{Finally, our framework for joint design is related to the deformation transfer and
  interpolation techniques such as \protected\cite{sumner2004deformation,baran2009semantic} and
  \cite{kilian07} to name a few.} Our approach is different in that we place special emphasis on
relating deformations between shapes with only soft (or functional) correspondences, which are often
much easier to obtain than detailed point matches. Moroever, rather than transporting Jacobian
matrices associated with the deformation, which requires both a pre-alignment and an approximate
triangle-to-triangle map (as done in \cite{sumner2004deformation}) we study and transport the change
in the intrinsic metric structure directly. As we show below, this results in better joint deformation
design especially given approximate functional maps, and shapes in arbitrary poses.

Thus, in contrast to the majority of existing techniques our goal is to devise a
coordinate-free representation of extrinsic deformations as linear functional
operators, by making an explicit connection between the extrinsic deformations
and the change in intrinsic metric quantities. As such, our representation fits
within the recent line of work that represents many operations in geometry
processing as functional operators, including mappings or correspondences
\cite{ovsjanikov2012functional,pokrass2013sparse}, representations of vector
fields as derivations \cite{pavlov2011structure,azencot2013operator} and the
formulation of shape distortion via shape difference operators
\cite{rustamov-2013}. Therefore, although we build on classical constructions
such as the infinitesimal strain tensor, we show how they can be exploited to
create a functional representation of shape deformation, which can be used in
conjunction with other operators. {As we demonstrate below, our
  representation is particularly useful for analysing and manipulating the
  effect of the deformation on the shape structure and for relating deformations
  across shapes, with only soft correspondences between them. In particular, it
  enables applications such as intrinsic symmetrization, joint deformation
  design and allows to introduce extrinsic information in the computation of
  functional maps. Remarkably, we prove that together with the classical
  Laplace-Beltrami operator, our approach leads to a \emph{complete} (up to
  rigid motion) \emph{ coordinate-free functional shape representation}, which
  opens the door to new shape processing applications.}

\section{Overview}

{The rest of the paper is organized as follows: first, we define the functional deformation
  field representation using the classical notions of the Levi-Civita connection and the strain
  tensor, and list the main properties of this representation (Section \ref{sec:representation}). We
  then provide a link between this definition and the previously proposed shape difference
  operators, by considering their infinitesimal extensions, introducing a new unified operator, and
  proving the equivalence between the two definitions (Section \ref{sec:relation_shapediffs}). In
  Sections \ref{sec:discrete} and \ref{sec:discrete_inf_shapediffs} we provide a discretization of
  all of these notions, and show that they preserve the main properties of the continous
  counterparts.  Finally, we illustrate the utility of our representation by % first describing how
  % deformation fields can be recovered from their functional representation by solving a simple
  % convex optimization problem (Section \ref{sec:basis}) and then
  describing several novel application scenarios, which range from functional map inference, to
  intrinsic symmetrization and deformation field design that all exploit the properties of our representation and its relation to other previously proposed linear
  operators (Section \ref{sec:results}). Note that Sections \ref{sec:relation_shapediffs} and
  \ref{sec:discrete_inf_shapediffs} can be skipped by readers that are not interested in the
  connection to shape difference operators.}

To summarize, our main contributions include:

\begin{itemize}
\item Introducing \emph{functional deformation fields} as a way to represent
  extrinsic vector fields in a coordinate-free way as operators acting on
  functions, represented as matrices in the discrete setting.
\item Providing a link between functional deformation fields and the previously proposed shape
  difference oprators, which leads to both a new unified shape difference and alternative functional
  deformation fields, which can be made sensitive to specific (e.g., non-conformal) classes of
  distortions.
\item Showing how functional deformation can be used to naturally add extrinsic
  information (second fundamental form) into the computation and analysis of
  functional maps. We also prove that together with the Laplace-Beltrami
  operator, they provide a \emph{complete} coordinate-free shape
  characterization up to rigid motions.
\item Describing how this representation enables a number of novel applications
  including intrinsic shape symmetrization, deformation design and functional
  deformation transfer without pointwise correspondences.
\end{itemize}

% !TEX root = fundeform_tog2.tex

\section{Extrinsic Vector Fields as Operators} \label{sec:representation}

{In this section we provide a coordinate-free representation of extrinsic
  vector fields by considering their action on the underlying shape
  metric. Throughout this section we assume that we are dealing with a smooth
  surface $M$ without boundary embedded in $\mathbb{R}^3.$ The appropriate
  discretization of all the concepts introduced in this section will be given in
  Section \ref{sec:discrete}.}

\paragraph{The Levi-Civita Covariant Derivative}
We first need to introduce some fundamental notions from differential
geometry. In particular, we will use the classical Levi-Cevita connection to
define derivatives on a surface. More precisely, given a \emph{tangent vector}
$u$ at some point $p \in M$, and an extrinsic vector field $V$ on $M$, consider
an arbitrary curve $\gamma(t)$ on $M$ such that $\gamma(0) = p$ and
$\gamma'(0) = u$. Then, we let
$\bar{\nabla}_u V = \left. \frac{\partial V(\gamma(t))}{\partial t}
\right|_{t=0} $.
Here $\bar{\nabla}_u V$ is the standard covariant derivative of the ambient
space. Note that at a fixed point $p\in M$, $\bar{\nabla}_u V$ is a vector in
$\mathbb{R}^3$. We can project the covariant derivative onto the tangent plane
at $p$ to obtain a vector in the tangent plane, which is denoted simply by
$\nabla_u V$ where $\nabla$ is the Levi-Cevita connection on $M$ extended
naturally to extrinsic vector fields, (\cite{docarmo2013riemannian} p. 126). We
also remark that for any vector $x$ in the tangent space,
$\langle\nabla_u V,x\rangle = \langle\bar{\nabla}_u V,x\rangle$, which we will
use in our discretization.

The fundamental object that we consider below is the infinitesimal strain
tensor, which can be understood as a bilinear form, acting on pairs of vectors
$x,y$ in the tangent plane of a point $p \in M$. Namely, given an extrinsic
vector field $V$, the infinitesimal strain tensor
$\mathcal{L}_{V}\mathbf{g} (x,y)$ is defined as:
\begin{align}
\label{eq:straindef}
\mathcal{L}_{V} \mathbf{g} (x,y) = \langle x, \nabla_y V \rangle + \langle \nabla_x V, y \rangle 
\end{align}

This quantity has the advantage of being linear in the vector field $V$, which
makes it easy to handle for deformation and vector field design and therefore
has been used in a wide variety of works in computer graphics
\cite{nealen2006physically}.

{Physically, this tensor represents the infinitesimal stretch that the
  object undergoes at each point.  Thus, the eigenvector associated to the
  largest eigenvalue of $\mathcal{L}_{V} \I$ (which can be thought of simply as
  a symmetric 2x2 matrix) at a point $p$, corresponds to the tangent vector $x$
  that represents the local direction of maximal stretch.}

{With these definitions in hand we propose to consider a linear functional
  operator $E^V$, which we will use to capture and manipulate a deformation field
  $V$. Both the input and the output of our operator are smooth real-valued
  functions defined on the surface. This operator is defined implicitly, in the
  same spirit as the shape difference operators introduced by Rustamov et
  al. \cite{rustamov-2013} as follows: for every \emph{pair} of
  real-valued functions $f,g$ we require:}
\begin{align}
\label{eq:opdef}
\int_M \langle \nabla g, \nabla \diffU{V}(f) \rangle \diff\mu = \int_M \mathcal{L}_{V} \I (\nabla g, \nabla f) \diff\mu.
\end{align}

{The following proposition guarantees that $E^V$ is well-defined. 

\begin{proposition}
  For any extrinsic vector field $V$ there is a unique linear functional
  operator $E^V$ that satisfies Eq.~\eqref{eq:opdef} above. Moreover, this
  operator is linear in both the vector field $V$ and function $f$.
\end{proposition}

In the rest of the paper we call the linear functional operator $E^V$, a
  \emph{functional deformation field representation of $V$}. Our main goal is to
  design, manipulate and analyze extrinsic vector fields $V$ through their
  associated linear functional operators $E^V$. This approach has already proved
  useful in the context of manipulating maps or correspondences 
\cite{ovsjanikov2012functional}, tangent vector fields
\cite{azencot2013operator} and shape distortions
\cite{rustamov-2013}. In particular, these works have helped to
  establish a general formalism of shape manipulation through the associated
  linear functional operators, which can ``communicate'' by composition. This
  allows, for example, to transfer tangent vector fields
  across shapes without assuming pointwise correspondences
\cite{azencot2013operator} or to design very efficient shape matching
  algorithms using the functional map representation
\cite{ovsjanikov2016computing}.  Therefore, inspired by these works, we
  propose to extend this framework to also include \emph{extrinsic} (or
  deformation) fields. As we show below, our representation naturally fits
  within the general functional operator formalism and enables a number of novel
  applications.}

\subsection{Key Properties of Functional Deformation Fields}
\label{sec:fundeform_props}

\paragraph{Second-fundamental form representation}
One interesting special case to consider is the interpretation of $\diffU{V}$
when the deformation field is the normal field $V = n$. By using Eq.
\eqref{eq:straindef} it is possible to see (\cite{docarmo2013riemannian} p.128) that the covariant derivative of the
normal yields the second fundamental form denoted by
$\II_p : T_p M \times T_p M \rightarrow \mathbb{R}$, more precisely
$\mathcal{L}_n \I = - 2 \II$. Therefore the operator $\diffU{n}$ captures the
action of curvature on functions, since:
\begin{equation*}
	\int_{M} \langle \nabla f, \nabla \diffU{n} (g) \rangle \diff \mu = - 2 \int_{M} \II (\nabla f, \nabla g) \diff \mu .
\end{equation*}

From a theoretical point of view the knowledge of the Laplace-Beltrami operator gives access to the
first fundamental form and $\diffU{n}$ yields information about the second. Thus these two operators jointly
provide a coordinate-free representation of the embedding.

\begin{figure}[t!]
	\centering
	\begin{tabular}{cc}
		\includegraphics[width=.4\columnwidth]{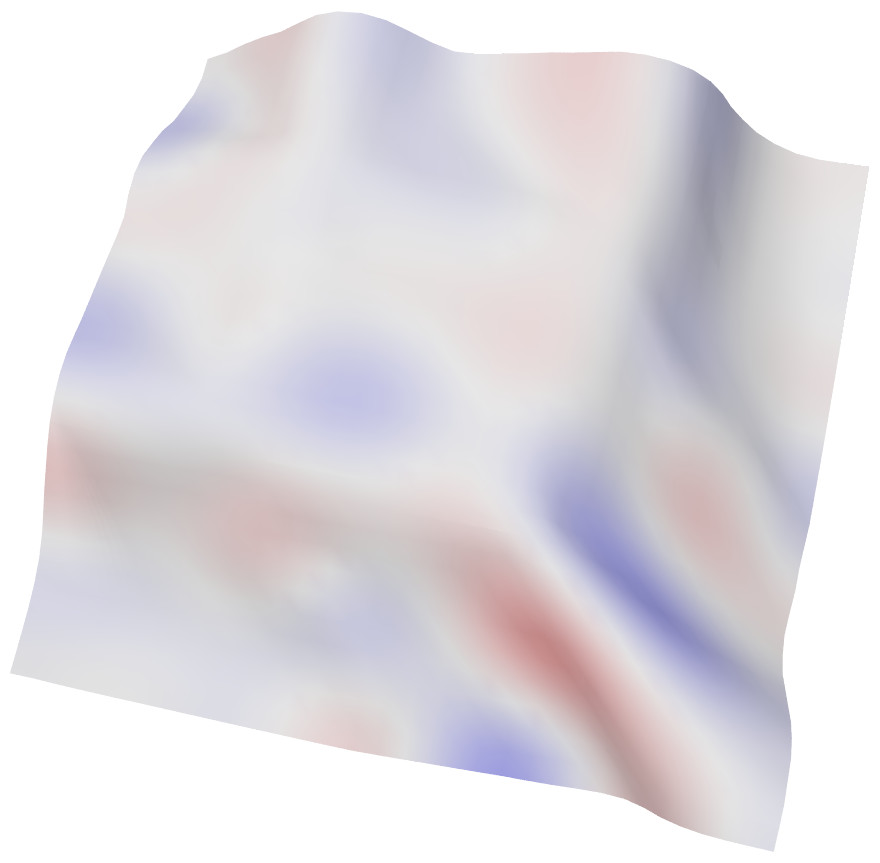}&
		\includegraphics[width=.4\columnwidth]{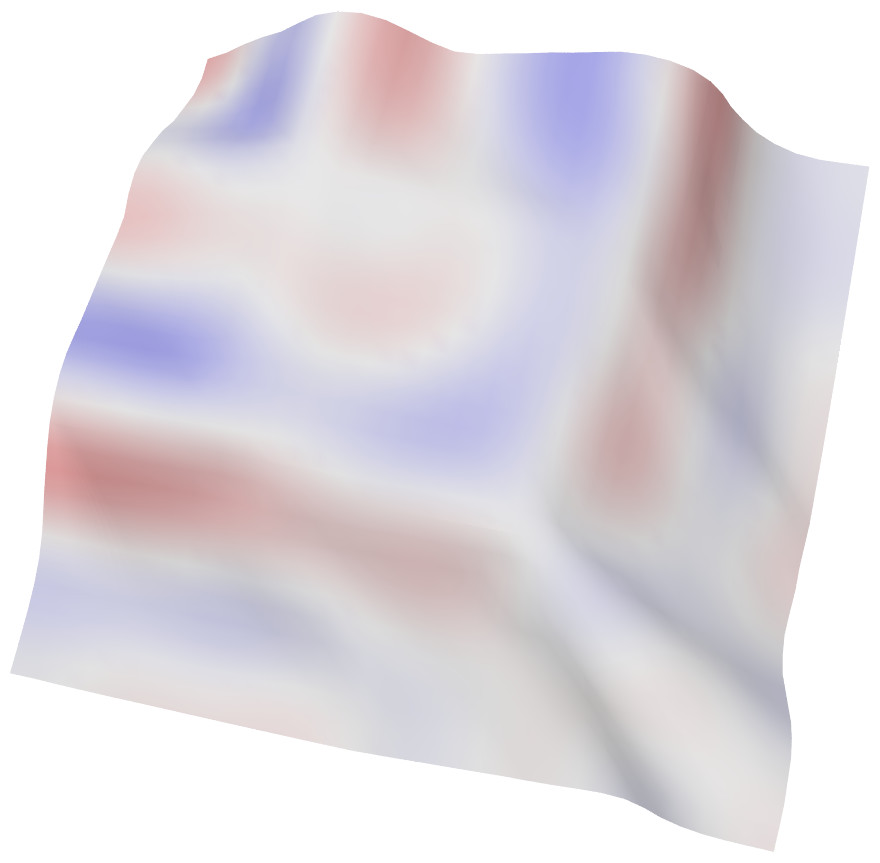}\\
		\includegraphics[width=.4\columnwidth]{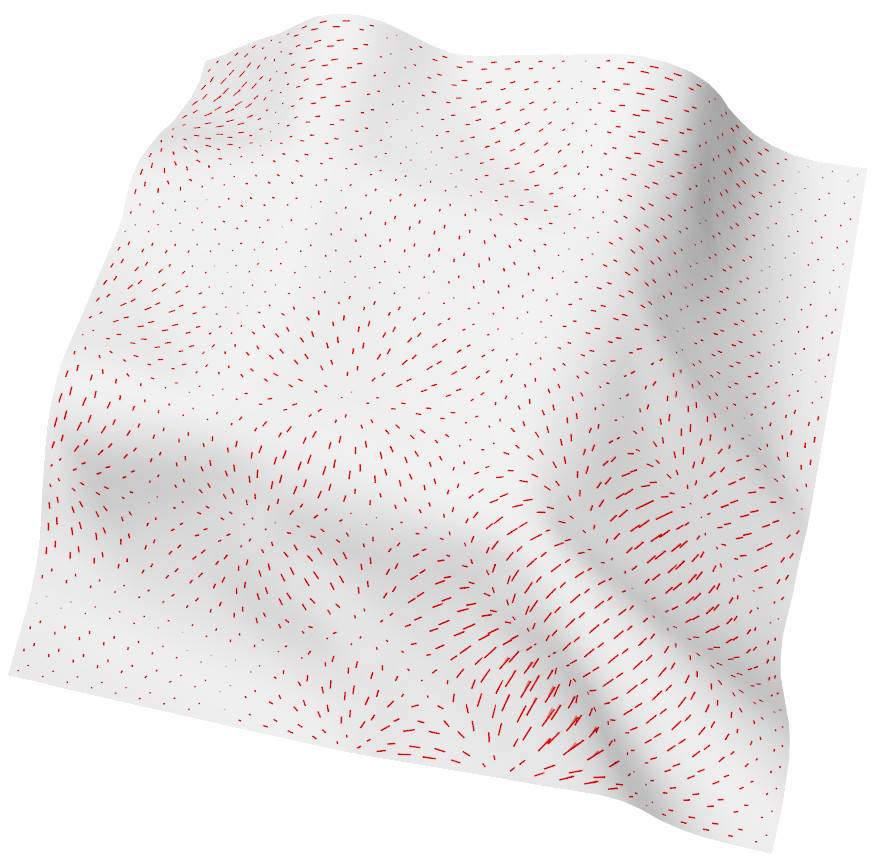}&
		\includegraphics[width=.4\columnwidth]{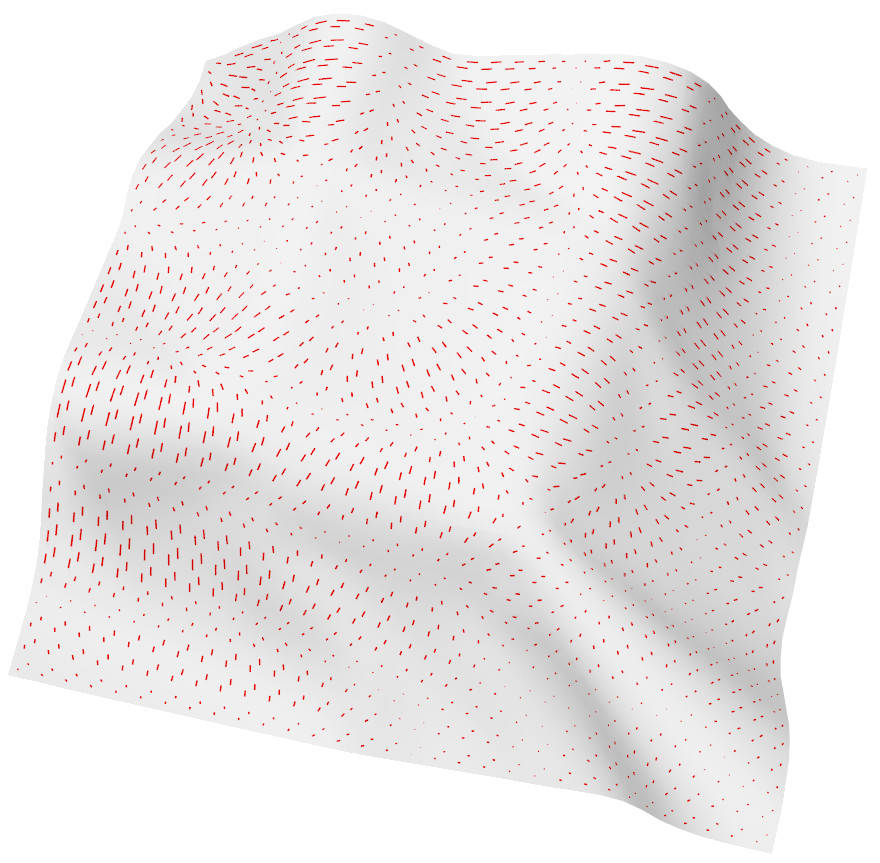}\\
		1st eigenfunction & 2nd eigenfunction 
	\end{tabular}
\vspace{-1mm}
	\caption{Two eigenfunctions associated with the largest eigenvalues of the operator infinitesimal shape
          difference $\diffU{n}$ for the normal field $n$. The gradients of these functions represent the direction of
          maximal curvature (Bottom row). \vspace{-1mm}}
	\label{fig:curvature}
\end{figure}

The operator $\diffU{n}$ can be used to obtain a multi-scale representation of curvature information on the triangle
mesh, as shown in Figure \ref{fig:curvature}. In particular, the eigenfunctions corresponding to the largest eigenvalues
of $\diffU{n}$, are those that align the best with the maximal principal curvature direction, and can be obtained even
if $\diffU{n}$ is represented in a reduced functional basis, making the computation less sensitive to noise in the
triangulation. Moreover, as we demonstrate in Section \ref{exp:fmap}, the operator $\diffU{n}$ can be used to inject
extrinsic information into the computation of functional maps.

\paragraph{Composition with mappings}
In many applications we are interested in the relation between deformations on
multiple surfaces related by a mapping. In particular given a deformation field
$U_M$ of shape $M$ and a diffeomorphism $\varphi : M \rightarrow N$ with the
associated functional map (pullback) $C_\varphi$ of functions from $N$ to $M$,
one can define a deformation field $V_N$ of shape $N$ that produces the same
metric distortion. Instead of looking directly at the deformation of the metric,
which might require a mapping between individual triangles
\cite{sumner2004deformation}, we account for the action of the metric on
functions:
\begin{equation*}
	\diffU{U_M} C_\varphi (f) = C_\varphi \diffU{V_N} (f) ~~\forall f \in C^{\infty}(N)
\end{equation*}
In other words, $V_N$ can be obtained by considering an extrinsic vector field,
whose operator representation has the same effect on functions when composed
with the functional map $C_\varphi$ as $\diffU{U_M}$. This property allows us to
relate deformation fields without requiring point-to-point correspondences
between shapes, by simply considering the commutativity of the operators
$C_\varphi$ and $\diffU{}$. We illustrate this in Figure
\ref{fig:mapComposition} and use it in Section \ref{exp:defTrans} for
deformation transfer and deformation symmetrization on meshes with different
connectivities with only a functional map known between them. Furthermore, this
approach is applicable to design deformations jointly on two shapes, such that
they are consistent with the functional map $C_\varphi$ and even as a
regularizer in map computation.

\paragraph{Vector field representation}
In general the operator $\diffU{V}$ does not uniquely define an extrinsic vector field. From
Def.~\ref{eq:opdef} it can be shown that %Prop. \ref{Thm:isometry} 
the kernel of $V \mapsto \diffU{V}$ coincides with the vector fields
satisfying $\mathcal{L}_V \I = 0$. In case of a volumetric manifold (i.e.
$M \subset \mathbb{R}^{3}$) the kernel of our operator is restricted to infinitesimal rigid motions
%the infinitesimal change in the metric $\partial_t \I$ defines the
%vector field $V$ up to rigid motion 
(see Theorem~1.7-3 in \cite{ciarlet2000theory}) and thus provides a complete representation of extrinsic
vector fields.  In the case of a surface embedded in $\mathbb{R}^{3}$ the kernel
of $\diffU{V}$ includes infinitesimal isometries such as Killing vector fields
but also local normal fields in planar areas. No rigidity result seems to be
known for smooth surfaces.  However, as we demonstrate below, in the discrete
case of shapes represented as triangle meshes, it can be shown that for almost
all surfaces the kernel of $V \mapsto \diffU{V}$ consists only of rigid
deformations (Prop. \ref{Thm:rigidity}). Note that we place \emph{no restriction
  on the magnitude of the deformation fields}. Thus, although our construction
is based on the infinitesimal strain tensor, the extrinsic vector fields
themselves are not limited to infinitesimal (or local) deformations. Finally, as
we show below, our constructions can be easily extended to the case of
tetrahedral meshes, resulting in a \emph{complete} operator-based representation
for deformation fields of \emph{volumes}, not sensitive to the exceptional
cases, present in the case of surfaces.
\section{Relation to Shape Difference Operators}
\label{sec:relation_shapediffs}

{ The functional deformation field representation introduced above is
  closely related to the previously proposed shape difference operators. In this
  section we describe this relation in detail, and highlight the following two
  key insights: 1) How our analysis leads to a novel \emph{unified} shape
  difference operator, and 2) How alternative functional deformation field
  representations can be constructed, to be sensitive to only a particular class
  of metric distortions. } {Our analysis also sheds light on the
  discretization of functional deformation fields. Nevertheless, the discussion
  in this section is not required for the understanding of either the
  implementation or the results of our approach, apart from the intrinsic
  symmetrization application (Sec. \ref{exp:intSym}), in which we use this relation. As such, this section
  can be skipped by readers not interested in these relations.}

\subsection{Shape Difference Operators}
Introduced by \cite{rustamov-2013}, the shape difference operators describe a
shape deformation by considering the change of inner products between
functions. Namely, given a pair of shapes $M,N$ and a diffeomorphism
$\varphi : N \rightarrow M$, with the associated linear functional map
(pullback) defined by $C_{\varphi}(f) = f \circ \varphi$, the authors introduce
the area-based and conformal shape difference operators $D_A$ and $D_C$
respectively, as linear operators acting on (and producing) real-valued
functions on $M$ implicitly via the following equations:
\begin{align}
	\langle f, D_A(g) \rangle_{L^2(M)} &:= \langle C_{\varphi}(f), C_{\varphi}(g)
        \rangle_{L^2(N)} &\forall f,g\\
	\langle f, D_C(g) \rangle_{H^1_0(M)} &:= \langle C_{\varphi}(f), C_{\varphi}(g)
        \rangle_{H^1_0(N)} & \forall f,g
	\label{Def:shapeDiffArea}
\end{align}
where the inner products are defined as $\langle f, g \rangle_{L^2 (M)} := \int_M f g \diff \mu$ and 
$\langle f, g \rangle_{H^1_0 (M)} := \int_M \langle \nabla f, \nabla g \rangle \diff \mu.$

The existence and the linearity of the operators $D_A$ and $D_C$ is guaranteed by the Riesz
representation theorem. As shown in \cite{rustamov-2013}, for smooth surfaces, the map $\varphi$ is
area-preserving (resp. conformal) if and only if $D_A$ (resp. $D_C$) is the identity map between
functions. From this it follows that $\varphi$ is an isometry if and only if $D_A$ and $D_C$ are
\emph{both} identity.

{Note that in the discrete setting the shape difference operators are obtained simply by
  considering transposes and inverses of the functional map and Laplacian matrices, as highlighted
  in \cite{rustamov-2013}. This makes properties such as existence and linearity trivial to
  see. Below we adopt the continuous (surface) formulation proposed in the original article as it
  helps to highlight both the generality of these concepts and also the relation to our
  representation of extrinsic vector fields.}

\paragraph{Infinitesimal Shape Difference Operators}
Our main goal in this section is to consider a one-parameter family of shapes $M_t$, given by
displacing the points of a base shape along some fixed deformation field.  Specifically, given a
surface $M$ embedded in $\mathbb{R}^{3}$ we consider a family $M_t$,
parameterized by a scalar $t$ and given by $p_t = p_0 + t V(p_0),$ where $p_0$ is a fixed point in
$\mathbb{R}^{3}$, and $V(p)$ is a vector in $\mathbb{R}^{3}$ that represents the displacement of the
point.

Now consider the family of maps $\varphi_t : M \rightarrow M_t$, given trivially via $\varphi_t(p) =
p_t$, and the associated functional maps $C_{\varphi_t^{-1}}$ mapping functions from $M_0$ to $M_t$. This gives rise
to a one-parameter family of shape difference operators $D^{V}_t$ (which can be taken either to be the area or
conformal-based operators). We then introduce \emph{the infinitesimal
shape difference operator} as follows:
\begin{definition}
  The \emph{infinitesimal area-based shape difference operator} associated with an extrinsic vector
  field $V$ on a surface $M$ is defined as:

\vspace{-5mm}
\begin{align}
\label{eq:define-e}
E_{A}^V := \left. \frac{\partial {D_A^{V}}_t}{\partial t} \right|_{t=0}, 
\end{align}
\end{definition}
{We define the \emph{infinitesimal conformal shape difference operator} $E_C^V$ similarly by
replacing $D_A^V$ by $D_C^V$ on the right side of Eq. (\ref{eq:define-e}).}

{Remark that since both $E_A^V$ and $E_C^V$ are defined as derivatives of
  a one-parameter family of linear operators acting on real-valued functions on
  a surface, both the range and the domain of these operators are also
  real-valued functions on $M$. Moreover, as $D_A^V$ and $D_C^V$ reflect (or,
  equivalently, are sensitive to) changes in the area and conformal metric
  structure, this implies that $E_A^V$ and $E_C^V$ will only reflect extrinsic
  vector fields up to infinitesimally area-preserving or conformal
  deformations. This naturally raises the question of whether there exists
  another ``unified'' shape difference operator $D_I$, which would be sensitive
  to general (non-isometric) metric changes. If so, would such $D_I$ lead to an
  infinitesimal shape difference $E_I$ that would agree with the definition of
  $\diffU{V}$ given in Eq. (\ref{eq:opdef})? Below, we provide precisely such a
  definition which both extends the applicability of shape difference operators
  and helps to establish a deeper link with our functional deformation fields.}

\paragraph{Unified shape difference}
The main reason for which $D_C$ is only sensitive to conformal changes is that both the inner
product and the integration are taken on the target shape. To define a unified shape difference
taking into account all intrinsic changes one should compare the pullback metric to the metric on
$M$ while keeping the integrating measure fixed. We thus propose a unified shape difference operator
$D_I$ that fully characterizes isometric distortion.

\begin{definition}
	Assuming that $\varphi : N \rightarrow M$ is a diffeomorphism, the unified shape difference $D_I : \mathcal{C}^\infty (M) \rightarrow \mathcal{C}^\infty (M)$ is defined implicitly by:
	\begin{equation*}
		\langle f, D_I(g) \rangle_{H^1_0(M)} := \int_{M} C_{\varphi^{-1}} \left( \langle \nabla  C_{\varphi}(f), \nabla  C_{\varphi}(g) \rangle \right) \diff \mu^M .
	\end{equation*}
	\label{Def:OpIso}
\vspace{-3mm}
\end{definition}
The existence of $D_I$ is once again guaranteed by the Riesz
representation theorem. Moreover, as we claimed above, the following proposition (proved in the
supplemental material) shows that the unified shape difference fully characterizes isometric deformation.

\begin{proposition} $D_I (f) = f$ for all $f \in \mathcal{C}^\infty (M) $ if and only if $\varphi$
  is an isometry. \label{Thm:isometry} 
\end{proposition}

\begin{figure}[t!]
\begin{centering}
\begin{tabular}{cc}
\includegraphics[width=0.45\columnwidth]{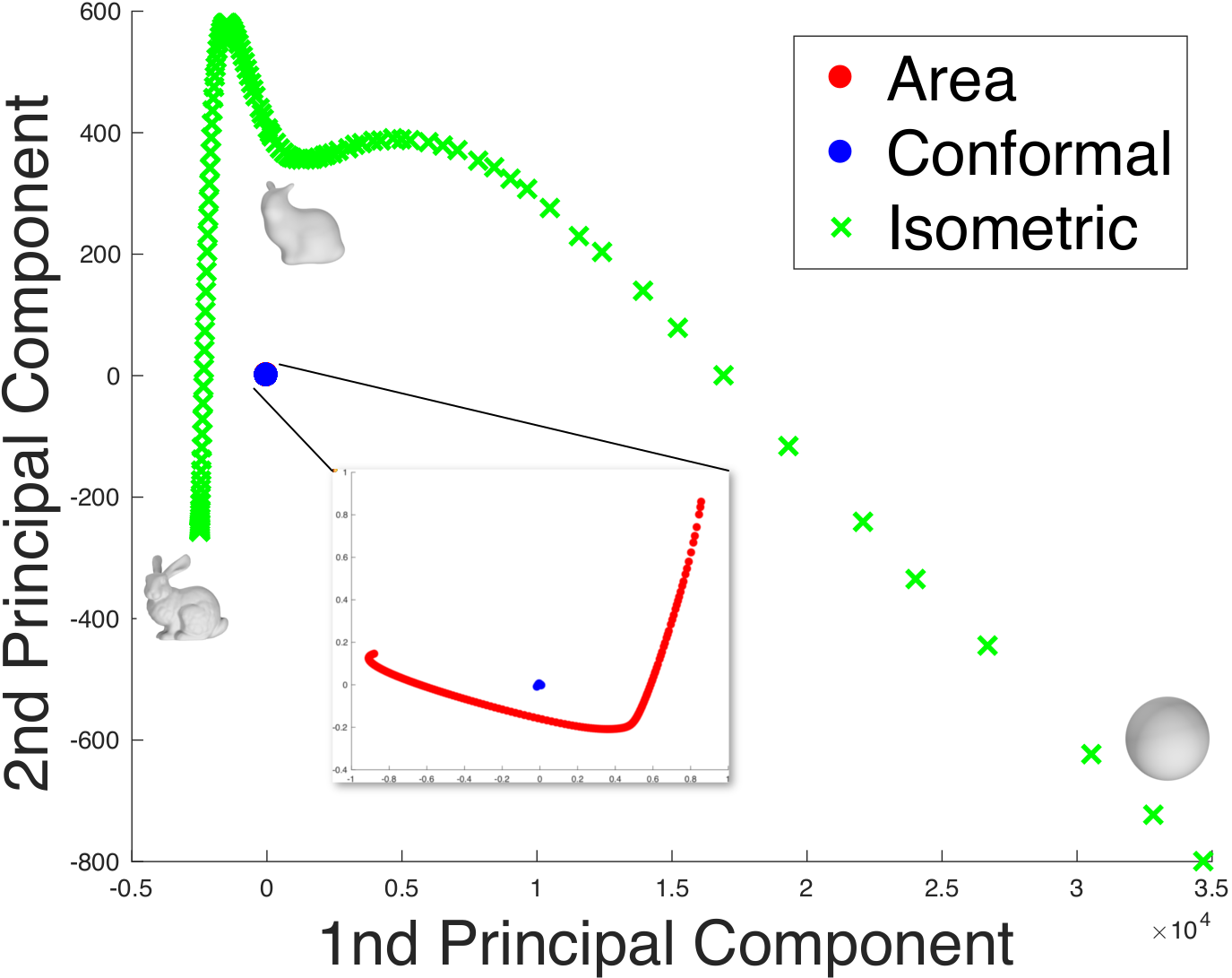}&
\includegraphics[width=0.45\columnwidth]{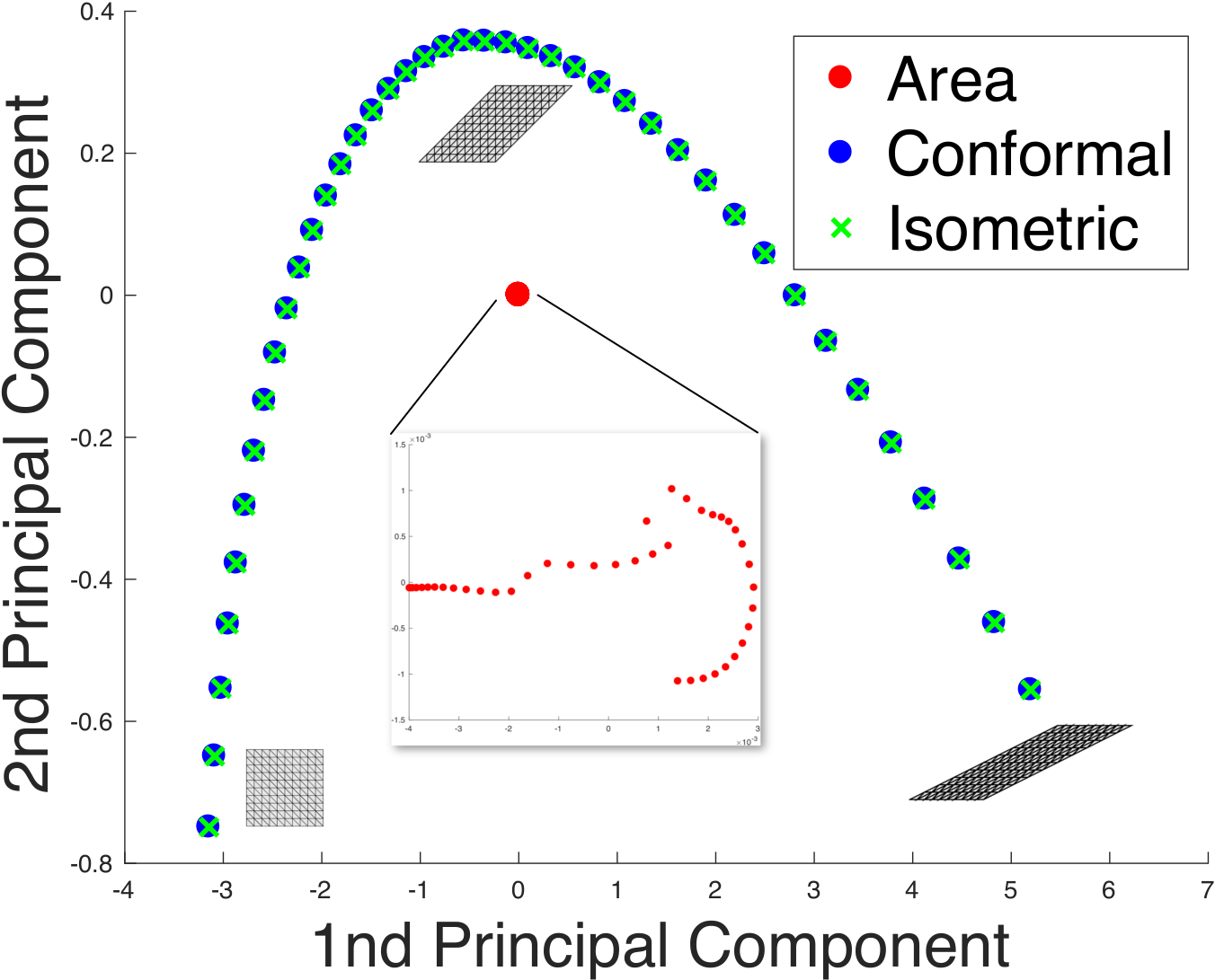}\\
Conformal deformation&
Area preserving deformation
\end{tabular}
\vspace{-1.5mm}
\par\end{centering}
\caption{\label{fig:all-shapediffs}{Left: Approximately conformal deformation of a bunny into a sphere. 
  The PCA applied to shape differences confirms the presence of large area and isometric distortion in contrast
  to small conformal distortion. Right: Area preserving deformation of a plane. The area-based shape difference
  is almost constant while conformal and isometric differences agree.}\vspace{-1.5mm}
}
\end{figure}

To illustrate the properties of the three shape differences we use a simple
low-dimensional description of a shape collection in Figure
\ref{fig:all-shapediffs}. Here we choose a fixed base shape and compute the
shape difference matrices with respect to the remaining shapes in a
collection. Then, we represent each shape by its shape difference matrix and
plot them as points in PCA space. Figure \ref{fig:all-shapediffs} represents the
conformal deformation of a bunny into a sphere as viewed by the three shape
differences. As expected $D_C$ is almost identity while the area and isometric
shape differences both capture the distortion. In the second experiment, shown
in Figure \ref{fig:all-shapediffs}, we explore another collection obtained by
the shearing of a plane patch. As this deformation is area preserving, the
area-based shape difference provides no information, unlike the other two
operators.

% \paragraph{Infinitesimal Shape Differences}
With the definition of the unified shape difference $D_I$ in hand, we introduce its infinitesimal
counterpart $E_I$ by following the same construction as done in \eqref{eq:define-e} above.
% \ref{subsec:shapediffs}, shape differences operators can be expressed using the pullback metric and pushforward
% measure. Using Proposition \ref{prop:metricderiv} we have all the tools to describe the infinitesimal shape difference
% operators. 
The following proposition (proved in the supplemental material) characterizes these new operators.
\begin{proposition}
  Let $V$ be a smooth deformation field on $M$, the derivatives of $D_A$, $D_C$ and $D_I$ at time
  zero satisfy for all smooth functions $f,g$:
	\begin{align*}
		& \langle f, E^V_A(g) \rangle_{L^2(M)} = \int_{M} \mathrm{div} (V) f g \diff \mu , \\
		& \langle f, E^V_C(g) \rangle_{H^1_0(M)} = \int_{M} \mathrm{div} (V) \langle \nabla f, \nabla g \rangle - \mathcal{L}_V \I (\nabla f, \nabla g) \diff \mu , \\
		& \langle f, E^V_I(g) \rangle_{H^1_0(M)} = - \int_M \mathcal{L}_V \I (\nabla f, \nabla g) \diff \mu .
	\end{align*}
	\label{Thm:derivShapeDif}
\end{proposition}
As can be seen, the infinitesimal shape differences inherit the properties of the original
operators. Namely, $E^V_A$ vanishes if and only if $\mathrm{div} (V)$ is equal to zero, i.e.,
whenever $V$ infinitesimally preserves the volume form. On the conformal side, finding an extrinsic
vector field $V$ such that $E^V_C = 0$ is equivalent to solving the \textit{conformal Killing
  equation}: $\mathcal{L}_V \I = \mathrm{div} (V) \I$ characteristic of a conformal vector
field. Both properties combined lead to an isometric deformation induced by the vector field $V$
captured by $E_I$.

Moreover Prop. \ref{Thm:derivShapeDif} reveals a clear link between shape differences:
\begin{equation}
	\langle f, E^V_I(g) \rangle_{H^1_0} = \langle f, E^V_C(g) \rangle_{H^1_0} - \langle 1, E^V_A(\langle \nabla f, \nabla g \rangle) \rangle_{L^2} .
	\label{operatorDecomp}
\end{equation}
Thus, intuitively, the operator $E_I$, representing isometric distortion, can be
decomposed into an area and a conformal part. We note that linear dependence between shape operators shown in Eq. \eqref{operatorDecomp} can be
understood as the decomposition of the matrix $\mathcal{L}_V \I$ into a trace free part, linked to
the conformal Killing equation, and a divergence part, related to the change in area.

{Finally, this proposition shows that the functional deformation field representation
  introduced in Section \ref{sec:representation} is exactly the same as the infinitesimal shape
  difference operator $E^V_I$ arising from the unified shape difference. Remarkably, this relation
  also holds exactly in the discrete setting as we show in Section
  \ref{sec:discrete_inf_shapediffs}.}

%
% \paragraph{Lie derivative representation}
% From Proposition \ref{Thm:isometry} we deduce that $\diffU{V}(f) = 0$ for all $f$ if and only if
% $\mathcal{L}_V \I = 0$, i.e., if the extrinsic vector field $V$ preserves the intrinsic shape metric
% to first-order.  The operator $\diffU{V}$ clearly quantifies how the Lie derivative of the metric
% affects gradients of functions. So the

\paragraph{Summary}{
To summarize, in this section we first showed that an alternative way for constructing a linear
functional operator representation of extrinsic vector fields consists in considering a family of
deformations of the shape, constructing the associated shape difference operators, and
taking their derivative at zero, which leads to infinitesimal shape differences. This also suggests
alternative functional deformation field operators, sensitive only to specific kinds of
deformations (e.g., non area-preserving or non-conformal). Finally, we showed that by modifying the
definition of shape differences, a new, unified difference operator can be constructed and that its
derivative at time zero leads precisely to the functional deformation field formulation introduced
in the previous section.}

% !TEX root = fundeform_tog2.tex

\section{Discrete Setting} \label{sec:discrete} {In this section we
  provide the discretization of functional deformation fields. For this, we
  first propose a particular discretization of the Levi-Civita connection and
  the Lie derivative of the metric on the triangle mesh, which leads to a simple
  formula for the operator $\diffU{V}$. In the following section,
  Sec. \ref{sec:discrete_inf_shapediffs}, which can be skipped similarly to 
  Sec. \ref{sec:relation_shapediffs}, we demonstrate that the
  deep connection between functional deformation fields and infinitesimal shape
  difference operators also holds in the discrete setting.}

{Throughout this section, we assume that we are given a manifold triangle
  mesh. We denote by $( \mathcal{X}, \mathcal{E}, \mathcal{F})$ respectively the
  set of vertices, edges and faces. We will consider the deformation field $V$,
  which we also call an extrinsic vector field, to be given as a
  three-dimensional vector per vertex. }

\subsubsection{Discrete connection} \label{par:connection} To build the discrete operator $\diffU{V}$ we need a consistent
discretization of the Levi-Civita connection. While several discrete
connections have been proposed (e.g. 
\cite{azencot2015discrete,liu2016discrete}), because of the special nature of our problem, we choose to build our own. This is because, 
applications such as parallel transport require that the vectors $u$, $v$ and $\nabla_u v$ are expressed in
the same space (at vertex or face or edge) so often an averaging step has to be introduced to transfer, for example, a
face-based representation of a vector to an edge based representation. In our setting such a requirement is not needed
and it is easier to distinguish tangent vector fields that will be expressed by one vector per face and extrinsic vector
fields expressed at vertices. Thus, our goal is to obtain a connection of the ambient space $\bar{\nabla}_u V$ where $u$
is a tangent vector and $V$ is an extrinsic vector field :
\begin{equation*}
	\begin{array}{llll}
		\bar{\nabla} : & \mathbb{R}^{3 |\mathcal{F}|} \times \mathbb{R}^{3 |\mathcal{V}|} & \rightarrow & \mathbb{R}^{3 |\mathcal{F}|} \\
		& (u, V) & \mapsto & \bar{\nabla}_u V
	\end{array}
\end{equation*}

We build the connection $\bar{\nabla}$ using finite differences as follows. Since extrinsic vector fields are defined
at vertices the differences are taken along the edges.
\begin{definition}
	In a given triangle $T \in \mathcal{F}$ the ambient covariant derivative along the edge $e_{ij}$ is defined by
	\begin{equation*}
		\left(\bar{\nabla}_{\frac{e_{ij}}{\|e_{ij}\|}} V\right)_T = \frac{V_i - V_j}{\|e_{ij}\|} .
	\end{equation*}
	
	Thus the ambient connection in the directions $E = (e_{ij}, e_{jk})$ can be stored in a matrix
	\begin{align}
\label{eq:ambientconnection}
		(\bar{\nabla}_E V)_T = \left(\begin{matrix} V_i - V_j & V_j - V_k \end{matrix}\right) .
	\end{align}
	Then, given any tangent vector $x = E \alpha$, the covariant derivative in its direction can be computed as
        $\bar{\nabla}_x V = (\bar{\nabla}_E V) \alpha$.
	\label{def:connection}
\end{definition}

Given the expression above, the discrete Lie derivative of the metric at
triangle $T$ follows immediately, using Eq. \eqref{eq:straindef}. Namely for any
pair of tangent vectors $x = E \alpha,y = E \beta$ in the triangle $T$, we have:
\begin{align}
	\mathcal{L}_V \I (x, y)_T &= \langle x, (\bar{\nabla}_E V) \beta \rangle + \langle (\bar{\nabla}_E V) \alpha, y \rangle .
	\label{disc:infOpIso}
\end{align}

% \maks{This description seems incomplete. Did we introduce $W_H$?} \etienne{There use to be the shape differences before this part...}
If $W_M$ denotes the cotangent-weight Laplacian, which classically represents the inner products of
$H_0^1$ (and is also called \emph{stiffness matrix}),
%After integration 
we obtain the discrete functional deformation field operator from its definition~ \eqref{eq:opdef}:
\begin{equation*}
	f^\top W_M \diffU{V} g = -\sum_{T \in \mathcal{F}} \mathcal{L}_V \I (\nabla f, \nabla g)_T \mu(T) .
\end{equation*}
Then we obtain $\diffU{V}(u) = W_M^{-1} H$, where $H$ is a Laplacian matrix whose weights depend on the extrinsic vector field:
\begin{align*}
	& (H)_{ij} = \frac{1}{2} \sum_{j \sim i} ( c(T_{\alpha_{ij}}) + c(T_{\beta_{ij}}) ),\\
	& c(T) = (\langle e_{jk}, V_j - V_i \rangle + \langle e_{ij}, V_j - V_k \rangle)\frac{1}{4 \mu(T)} \\ & - \mathrm{Tr} \left( (E^\top E)^{-1} E^\top (\nabla_E V) \right) \frac{\langle e_{jk}, e_{ki} \rangle}{\mu(T)} .
\end{align*}

The computations can be found in the supplemental material.

\subsection{Properties} \label{subsec:discrete_prop}

Interestingly, many of the properties of the continuous operators are satisfied exactly by their discrete counterparts.

\paragraph{Linearity}
The discretization $\diffU{V}(f)$ naturally preserves the linearity with respect to both $V$ and $f$ which is very convenient for practical purposes.

{
In practice, it is often convenient to use a functional basis, so
that any function can be represented as a linear combination of some basis functions
$\phi_i$. Given such a basis, the operator $\diffU{V}$ can be seen as the (possibly
infinite) matrix: $\diffU{V}_{ij} = \langle \phi_i, \diffU{V}(\phi_j) \rangle_{L^2(M)}$. 
%For a function $f = \sum_i \alpha_i \phi_i$ the action of the operator reduced to:
% $\diffU{V}(f) = \sum_i \alpha_i \diffU{V}(\phi_i)$.
The choice of basis depends on the application. Since
we are interested in smooth deformations of a surface, we take a subset of the smoothest
functions given by the first $k$ eigenfunctions of the Laplace-Beltrami operator. In that
case, $\diffU{V}$ will be represented simply as a $k \times k$ matrix.  As shown in
Figure~\ref{fig:basisSize} the size of the basis $k$ affects the deformation field that we
can represent and recover. Increasing $k$ allows a more faithful representation of high
frequency deformation fields.

The linearity with respect to $V$ allows the same operation for vector fields. Therefore, if the deformation field is given in some
basis $V = \sum_i \alpha_i X_i$ then the operator reads $\diffU{V} = \sum_i \alpha_i
\diffU{X_i}$. This means that when designing a deformation field $V$ we can consider an objective as a function
  of the coefficients $\alpha$.%, and moreover if the objective is quadratic in $\alpha$ then
%recovering $V$ can be done by solving a single linear system of equations. Note that this allows
%us to combine constraints both on the vector field itself (e.g., handle constraints) and on its
%operator representation, such as enforcing commutativity with other operators, while remaining
%easy to optimize for.
}

\paragraph{Vector Fields representation}
In the continuous setting the kernel of $V \mapsto \diffU{V}$ is the set of
infinitesimal isometries. However, to the best of our knowledge, there is no
characterization of how often this set is reduced to rigid motion. In the
particular setting of our discretization some standard results can be applied,
however.

\begin{proposition}
    For almost all triangle meshes $M$ without boundary, the operator $\diffU{V}$ uniquely defines the
        extrinsic vector field $V$ up to rigid motion.
	\label{Thm:rigidity}
\end{proposition}

{Thanks to this proposition, we can guarantee that $\diffU{V}$ is almost always a complete
  coordinate-free representation of extrinsic vector fields $V$. Triangle meshes containing
  perfectly flat neighborhoods fall in the category of shapes on which the map
  $V \mapsto \diffU{V}$ is not injective.  Namely, since by definition of the strain tensor
  (Eq. \ref{eq:opdef}), whenever $\nabla_x V$ and is normal to the surface for all $x$, (as is the
  case when e.g. $V$ is a normal field on a flat part and zero elsewhere), the tensor
  $\mathcal{L}_V$ will lead to the zero operator. Although we have found that for organic and
  natural shapes, such vector fields are rare or non-existent, they can nevertheless be important
  for coarse meshes or man-made objects with flat areas.}

\subsection{Construction for Tetrahedral Meshes}
To remedy this problem, we extend our discretization to tetrahedral meshes thus
avoiding ill-defined vector fields as the kernel of $V \mapsto \diffU{V}$ is of
dimension $6$ (translations and infinitesimal rotations). For this we follow the
construction provided in Section \ref{par:connection}, by adapting it to tet
meshes. Namely, we extend the ambient covariant derivative matrix $E$ in
Eq. \eqref{eq:ambientconnection} to three dimensions, by considering the
covariant derivative along three directions of a tet mesh, and thus storing a
3x3 matrix $\bar{\nabla}_E V$ per simplex. We then use Eq. \eqref{disc:infOpIso}
without any modifications to obtain a discretization of the functional
deformation fields on tet meshes. The final resulting formula for
the matrix $E_I^V$ is provided in the supplementary material.

\begin{figure}[t!]
	\centering
	\includegraphics[width=1\columnwidth]{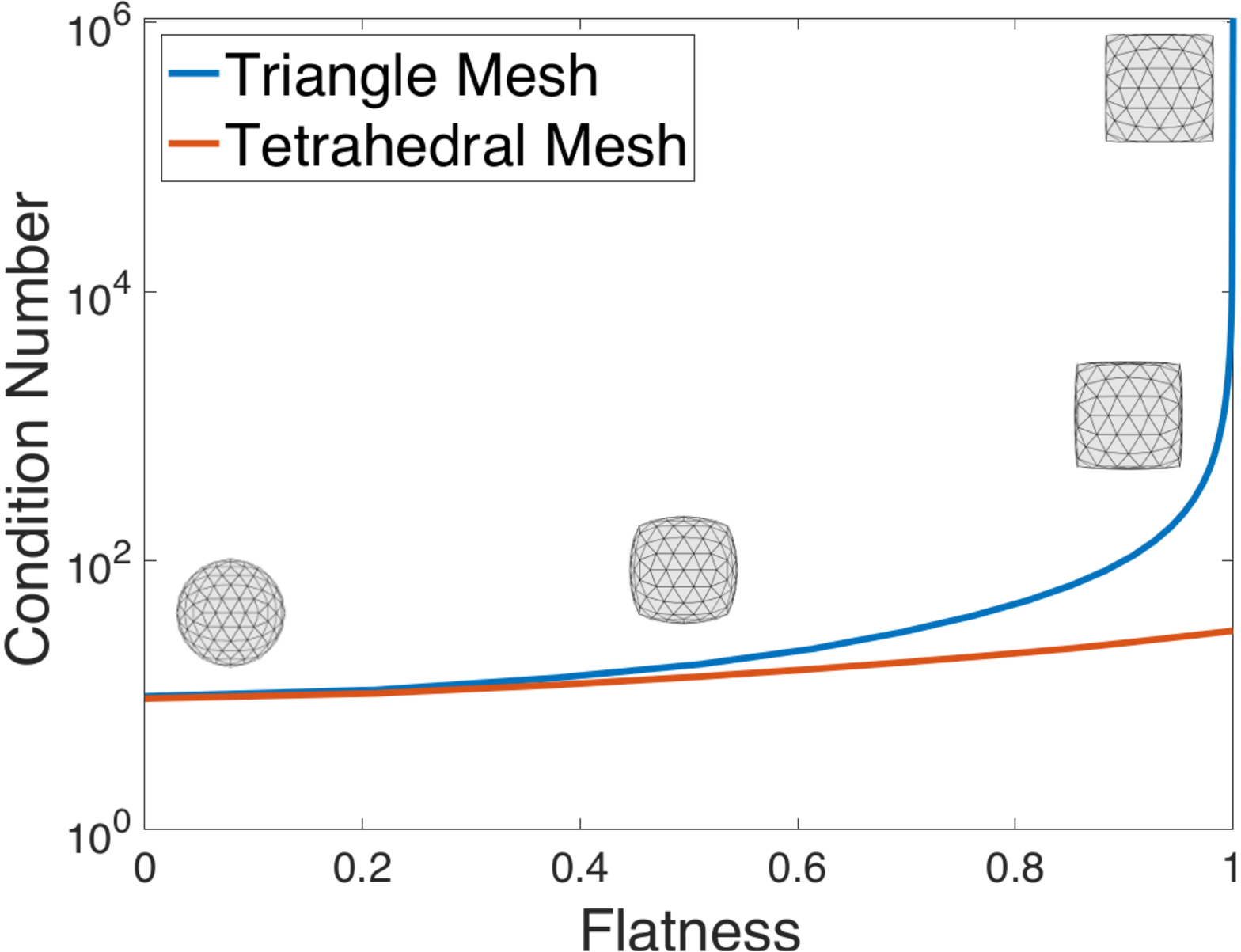}
	\caption{Comparison of the condition number of the linear map $V \mapsto \diffU{V}$ when computed on triangle and tetrahedral meshes with respectively $162$ and $163$ vertices. The condition number is computed as the ratio of the largest and the 7th lowest singular value to avoid infinitesimal rotations and translations naturally mapped to zero. We consider the collection of meshes formed by a sphere morphing into a cube. For a triangle mesh the condition number goes to infinity as the shape becomes increasingly flatter. When considering a tetrahedral mesh the condition number remains bounded. 
\vspace{-2mm}}
	\label{fig:condNum}
\end{figure}

\begin{figure}[t!]
	\centering
	\begin{tabular}{cc}
		\includegraphics[width=.35\columnwidth]{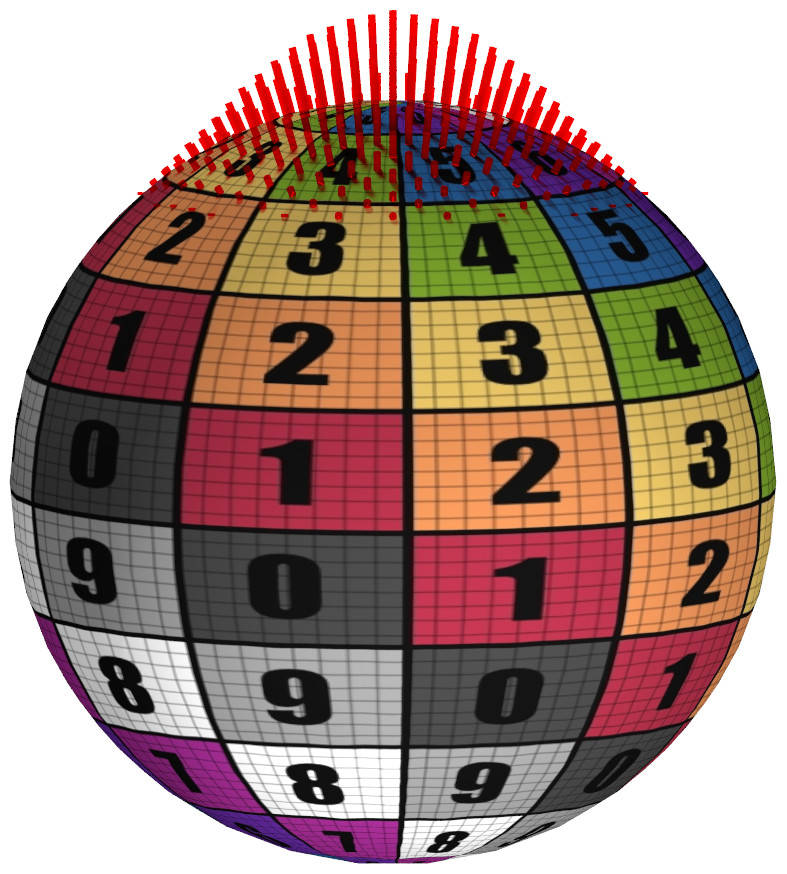}&
		\includegraphics[width=.35\columnwidth]{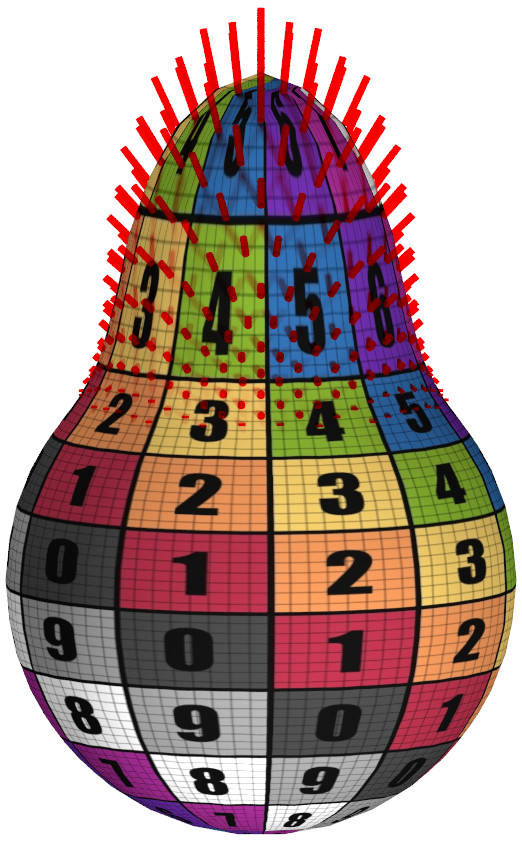}\\
	\end{tabular}
\vspace{-1mm}
	\caption{Example of deformation fields that commute with the diffeomorphism represented by texture transfer. Note that both the
          direction and the magnitude of the vector field have to adapt to the underlying geometry to produce the same
          metric distortion. 
\vspace{-1mm}}
	\label{fig:mapComposition}
\end{figure}

We compare the stability of our representation between tetrahedral and triangle meshes in
Figure~\ref{fig:condNum}, by plotting the condition number of the linear system for recovering the
vector field $V$ from its operator representation $E^V_I$ in the case of surface (triangle) and tet
mesh reprentations of a cube. We note that although the condition number becomes unbounded for the
triangle mesh representation as the shape approaches a flat cube, it nevertheless remains remarkably
stable: even at $0.9$ where the sphere is almost a cube the condition number is about $100$. In
contrast the condition number for tet meshes remains bounded even for a perfectly flat shape.

An important consequence of Proposition~\ref{Thm:rigidity} is that
deformation fields are fully encoded by the operator $\diffU{V}$ up to
\emph{infinitesimal} rigid motions. Therefore any deformation can be recovered
regardless of its scale and nature. For instance Figure~\ref{fig:basisSize}
shows that non-infinitesimal \emph{global rotations} are correctly encoded and
recovered from our operator representations.

\section{Discrete Infinitesimal Shape Differences}
\label{sec:discrete_inf_shapediffs}
{Similarly to the link between established in Section
  \ref{sec:relation_shapediffs} between our initial definition for functional
  deformation fields and infinitesimal shape differences, we can consider an
  alternative discretization to the one above by considering a family of
  deformed meshes and taking the derivative of shape difference operators. In
  this section we show that this approach leads to exactly the same result,
  which means that remarkably Proposition \ref{Thm:derivShapeDif} is satisfied
  \emph{exactly} in the discrete setting. To demonstrate this result we first
  provide a discretization of the unified shape difference operator and then
  highlight the link between the infinitesimal shape difference operators and
  functional deformation fields. Similarly to Section
  \ref{sec:relation_shapediffs}, this section is primarily of conceptual
  interest and can be skipped by readers who wish to proceed to the practical
  results.}

To compute the shape differences we start from the discretization of the inner
product $\langle . , . \rangle_{H^1_0}$ using standard first order finite
elements. We will denote by $L$ the classical cotangent Laplacian matrix, $W$
the inner product of $H^1_0$ and $A$ the lumped mass matrix such that
$L = A^{-1} W$.  As before $\mu$ is a measure and $\mu(T)$ denotes the area of
triangle $T$.

\subsection{Discrete unified shape difference}

The discretization of the unified shape difference is straightforward when $N$ and $M$ are triangle meshes and share the
same connectivity. In Definition \ref{Def:OpIso} given above, the gradients and the
point-wise scalar products are taken on $N$ while the measure $\diff \mu^M$ comes from $M$. Therefore the right hand side
can be discretized by a modified cotangent weight formula:
%
%\begin{equation}
%	f^\top W_M D_I g = -\sum_{T \in \mathcal{F}} \frac{\mu^M(T)}{\mu^N(T)} \frac{1}{4 \mu^N(T)} \left(\begin{matrix} f_i - f_j \\ f_j - f_k \end{matrix}\right)^\top \I_T^N \left(\begin{matrix} g_i - g_j \\ g_j - g_k \end{matrix}\right) .
%	\label{Disc:OpIso}
%\end{equation*}
\begin{align*}
	W_M D_I &= W^M_{N}, \textrm{ where }  \\
(W^M_N)_{i,j} &= 
\frac{1}{2} \left( \frac{\mu^M(T_\alpha)}{\mu^N(T_\alpha)} \cot \alpha^N_{ij} + \frac{\mu^M(T_\beta)}{\mu^N(T_\beta)}
  \cot \beta^N_{ij} \right) . \numberthis
	\label{Disc:OpIso}
\end{align*}
Here $T_\alpha,T_\beta$ are the two triangles adjacent to edge $i,j$, which is opposite to angles $\alpha$ and $\beta$,
while $\mu^M$ and $\mu^N$ are the triangle areas on shapes $M$ and $N$ respectively. Note that $W^M_N$ differs from the standard
cotangent weight matrix $W_N$ only by the ratio of weights per triangle. Moreover,notice that if the
transformation is area preserving for all triangles then $D_I$ reduces to the conformal shape difference defined in
\cite{rustamov-2013} (Option 1 in Section 5).

From the expression above it follows that $D_I = W_M^{-1} W^M_N$.

\paragraph{Expression in a basis} Similarly to the construction given in \cite{rustamov-2013} we can also express the
unified shape difference when the basis $\Phi_M$ on the source shape $M$ is given by the eigenfunctions of the Laplace-Beltrami
operator. In that case, using a diagonal matrix $\Lambda_M$ of eigenvalues, $D_I$ becomes:
\begin{align*}
D_I = \Lambda_M^{-1} \Phi_M^T W^M_N \Phi_M.
\end{align*}
This expression has the advantage of avoiding the inverse of a large sparse
matrix, and can be used to analyze deformation of a shape with fixed
connectivity in multi-scale basis, which can make the computations resilient to
local perturbations (see Option 3 in Section 5 of \cite{rustamov-2013} ).

\paragraph{Approximation with a functional map}
Note that both expressions above assume that the source and target meshes share the same connectivity. When the meshes
have different connectivity this discretization requires a map between triangles making it challenging to use in
practice. To overcome this problem we approximate this discrete formulation by transferring the weights on triangles to
lumped weights on vertices. The approximation then reduces to the usual discrete quantities:
%\begin{equation*}
%	f^\top W_M D_I g \approx - f^\top A_M A_N^{-1} W_N g .
%\end{equation*}
\begin{equation*}
	(\tilde{W^M_N})_{ij} \approx \frac{\sum_{t \sim i} \mu^M(T_t)}{\sum_{t \sim i} \mu^N(T_t)} \frac{1}{2} \sum_{j \sim i} ( \cot \alpha^N_{ij} + \cot \beta^N_{ij} ).
\end{equation*}

We recognize here the cotangent Laplacian $L_N$ with lumped area weights, namely $A_M L_N$. In the case of meshes with
different connectivity, this remark suggests the following approximation of the isometric shape difference, valid only in
a discrete sense, for an arbitrary linear functional map $C$ between $M$ and $N$:
\begin{equation*}
	f^\top A_M L_M D_I g \approx f^\top A_M C^{-1} L_N C g .
\end{equation*}

In the reduced basis of the Laplacian eigenvectors, the approximation of the shape difference becomes $D_I \approx
\Lambda_M^{-1} C^{-1} \Lambda_N C$, which preserves the principal property of the operator: $D_I$ is identity if and
only if the deformation is an isometry since the Laplacian on $N$ has to be equal to the Laplacian on $M$. We used this
discretization in Figure \ref{fig:all-shapediffs} and observed that the two expressions given above typically produce
similar results.

\subsection{Shape difference derivative} \label{par:timederiv} Suppose that each
vertex $p_i$ of the mesh is displaced by the vector $V_i$ by
$p^t_i = p_i + t V_i$. This produces a family of triangle meshes
$( \mathcal{X}^t, \mathcal{E}, \mathcal{F})$ with identical connectivity. 
{It is now possible to take the derivative with respect to $t$ of Eq.~\eqref{Disc:OpIso} at time $0$. This way we obtain a discretization of the
infinitesimal shape differences. Remarkably the resulting discretization is strictly identical to
the discrete functional operator $E^V$ proposed in Section~\ref{sec:discrete} based on the discrete Levi-Civita connection.}
\begin{proposition}
	The discretization of $\diffU{}$ based on the discrete Levi-Civita connection is equivalent to the one obtained by
        differentiating the unified shape difference operator.% consistent with continuous definition of the infinitesimal shape difference.
        \label{thm:uniqueDisc}
\end{proposition}

\paragraph{Shape difference decomposition}
Since the discretization using a discrete connection and through the time derivative agree, the decomposition described by Eq.~\eqref{operatorDecomp} is also satisfied exactly. 
%With the family of meshes conformal shape difference is:
%\begin{equation*}
%	f^\top W D^t_C g := f^\top W_t g = \sum_{T \in \mathcal{F}} \langle \nabla^t f, \nabla^t g \rangle^t_T \mu^t(T) ,
%\end{equation*}
%%
%where $W_t$ is the cotangent weight matrix associated to the deformed mesh at time $t$. As noted
%above the derivative of $\langle \nabla^t f, \nabla^t g \rangle^t_T$ will lead to $\diffU{V}$. The
%derivative of the area will produce a term close to
%$\int_{M} \mathrm{div} (u) \langle \nabla f, \nabla g \rangle \diff \mu$ thanks to Proposition
%\ref{thm:discDeriv}. 
Namely, the matrix $\diffU{V}_C$ representing the discrete infinitesimal conformal shape difference splits into the discrete functional deformation field $\diffU{V}$ and an appropriately defined discrete divergence:
\begin{align*}
	f^\top W E^V_C g = f^\top W \diffU{V} g + \sum_{T \in \mathcal{F}} \mathrm{div} (V)_T \langle \nabla f, \nabla g \rangle_T  \mu(T).
\end{align*}
Thus, the decomposition of $\diffU{V}$, representing isometric distortion, into area and conformal parts given in
Eq. \eqref{operatorDecomp} in the continuous case holds exactly in the discrete case as well.

% !TEX root = fundeform_tog2.tex

\section{Experiments}
\label{sec:results}
In this section we apply our constructions to various tasks in shape correspondence, deformation
design and analysis. As our framework relies on manipulating and inverting moderately-sized
matrices, all of the applications are very efficient, even when combining multiple objectives.

{ In some applications (Sec. \ref{exp:intSym} - \ref{exp:defTrans}), it is necessary to
  recover the deformation field from its function operator representation. For this, we use a
  reduced basis and recover the coefficients of the deformation field by solving a convex problem
  similar to basis pursuit. Namely, given a target functional deformation field operator $E^V$
  represented as a matrix, we solve for $V$ via:
\begin{align*}
\min_{\alpha} \|\sum_i \alpha_{i} E^{X_i} - E^V\|^2_F + \|\alpha\|_1,
\end{align*}
where $\alpha$ is a vector of coefficients and $E^{X_i}$ are the functional representation of the
$i^{\text{th}}$ deformation field in an overcomplete basis. Of course, the choice of basis is
application dependent. %For example, a basis for deformations, computed from the movement of a
%skeleton will be best at describing humanoid deformation. 
The simplest and most general choice would
be to consider a basis which consists of independent displacements at each vertex of the given
mesh. For a mesh with $n_V$ vertices, this would result in 3$n_V$ unknowns when solving for a
deformation field, which is feasible when $n_V$ is small (and was used in the experiment in Figure
~\ref{fig:basisSize}), but can be inefficient for larger meshes. When needed (Sec. \ref{exp:intSym}
- \ref{exp:defTrans}) we use the following deformation bases in the experiments below:}
\begin{itemize}
\item The simplest option is to take the eigenfunctions of the Laplace-Beltrami operator
  as the basis for each component of the deformation field. While simple, this basis might
  not preserve rotation invariance.
\item Alternatively we construct a basis via modal analysis of a deformation energy. In
  particular we consider an energy of the form
  $V \mapsto \int_M \| \bar{\nabla} V \|^2 \diff \mu$. This corresponds to the energy on a
  particular discretization of the Bochner Laplacian of extrinsic vector fields. To obtain
  the basis we take the eigenvectors of the Hessian of the energy, which
  correspond to smooth deformation fields.
\item Lastly, we use the handle-based deformation model described in
  \cite{adams2008meshless}. Unlike the other families, the deformation fields arising from
  this model are compactly supported and therefore better suited to reproduce local
  deformations.
\end{itemize}

\subsection{Functional map inference} \label{exp:fmap} 

\begin{figure}[t!]
	\centering
	\begin{tabular}{cc}
		\includegraphics[width=.47\columnwidth]{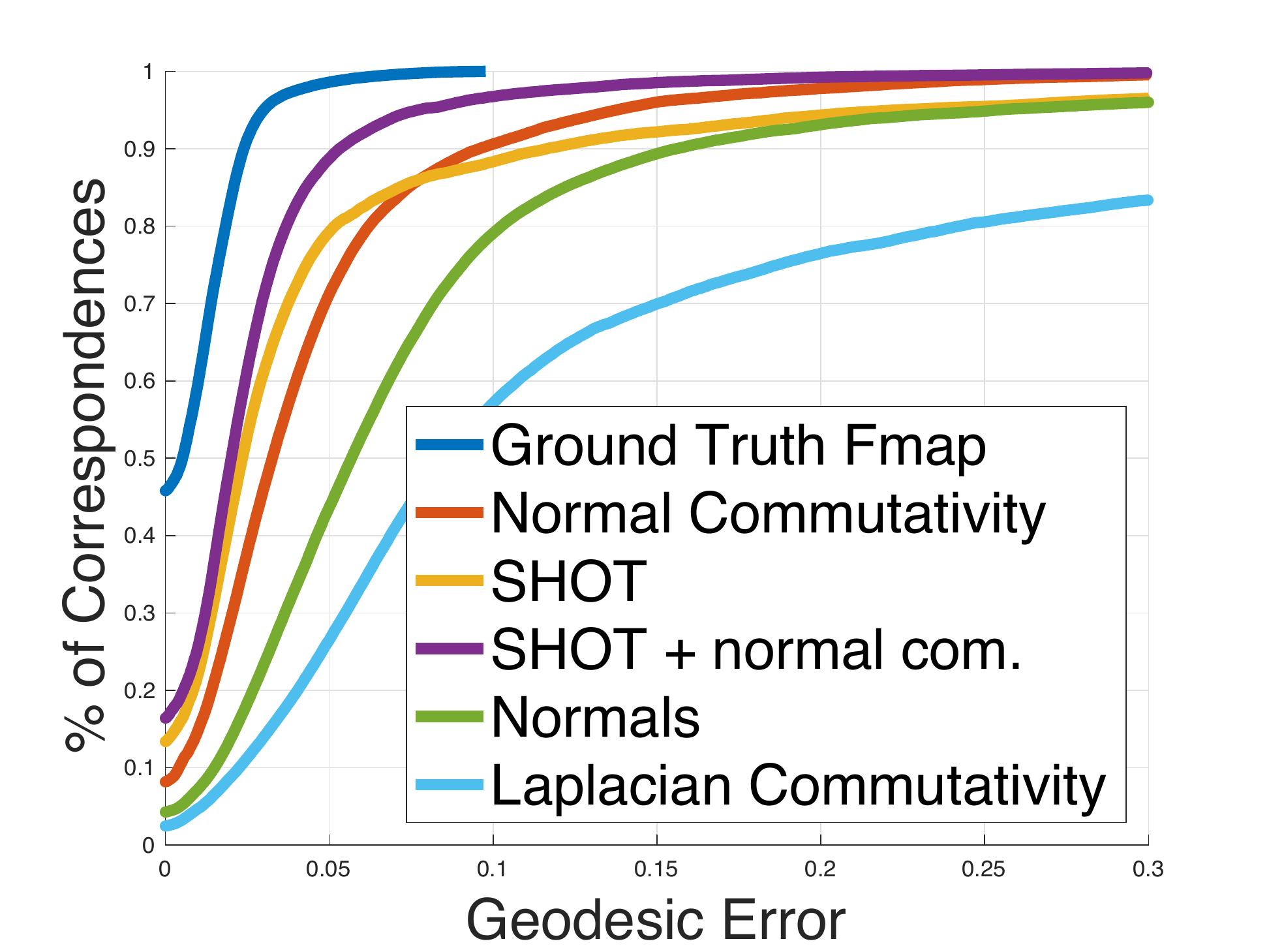}&
		\includegraphics[width=.47\columnwidth]{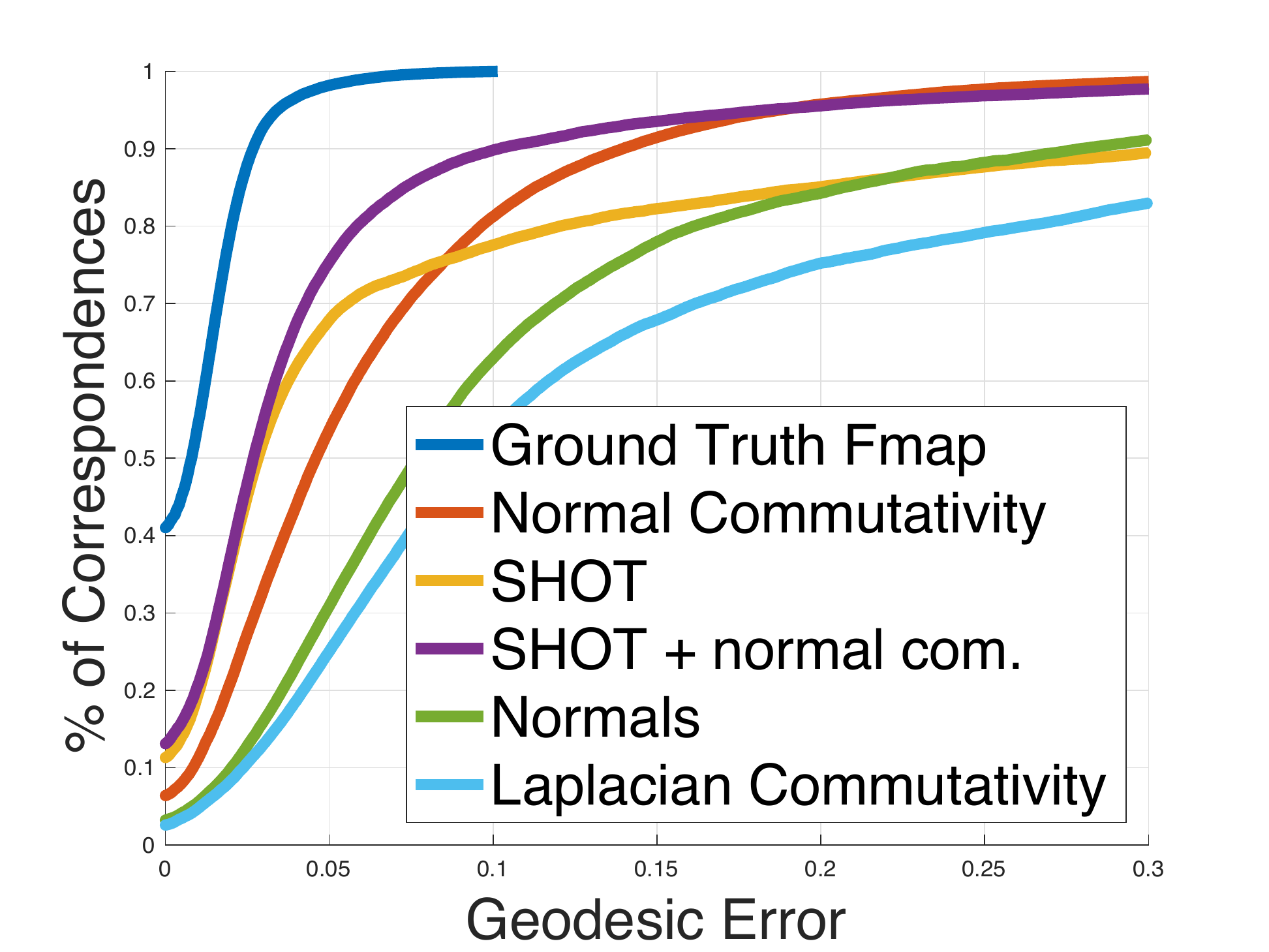}\\
		Same pose & Different Pose
	\end{tabular}
	\caption{{Percentage of correspondences within a geodesic ball averaged on a 50 pairs for a two shape matching problems: left different character taking the same pose, right different character taking different pose.}}
	\label{fig:fmapsCorresp}
\end{figure}

{In our first application, we show how
  our functional deformation field representation can be used as a regularization in shape matching
  problems. In particular, we show how this representation can be used to add extrinsic information
  to the computation of functional maps
\cite{ovsjanikov2012functional}. %All those methods mostly relies on intrinsic information such as spectral descriptors (HKS, WKS) and the nearly isometric assumption.
The vast majority of the existing methods for shape correspondence with functional maps use
  the assumption of approximate intrinsic isometries (see \cite{ovsjanikov2016computing} for
  an overview) and are either purely intrinsic or inject extrinsic or embedding-dependent
  information by adding extrinsic descriptors. On the other hand, our functional deformation field
  representation provides a natural coordinate-free way to add embedding-dependent information into
  the map estimation pipeline. In particular, our approach below is based on the following key
  observation: }
{
\begin{proposition}
  Given a pair of surfaces $M,N$ embedded in 3D, and a diffeomorphism $T: N\rightarrow M$, let $C$
  be the corresponding functional map $L^2(M) \rightarrow L^2(N).$ Then $M$ and $N$
  are related by a rigid motion in space if and only if:
$$ \| C \Delta_M - \Delta_N C \| + \| C E^n_M - E^n_N C \| = 0, $$
where $\Delta$ are the LB operators, while $E^n$ are functional deformation fields arising from the
normal fields.
\label{prop:embedding_const}
\end{proposition}
}

\begin{figure}[t!]
	\centering
\vspace{-2mm}
	\begin{tabular}{cc}
		\includegraphics[width=.47\columnwidth]{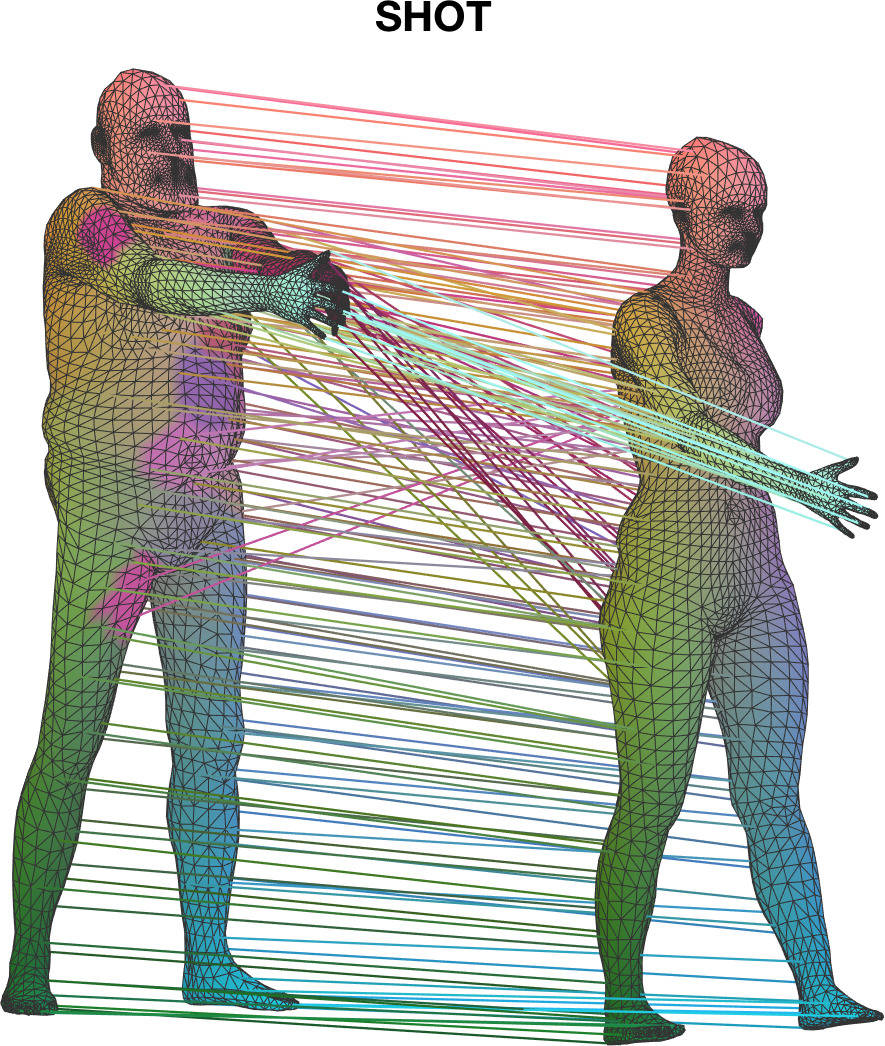}&
		\includegraphics[width=.47\columnwidth]{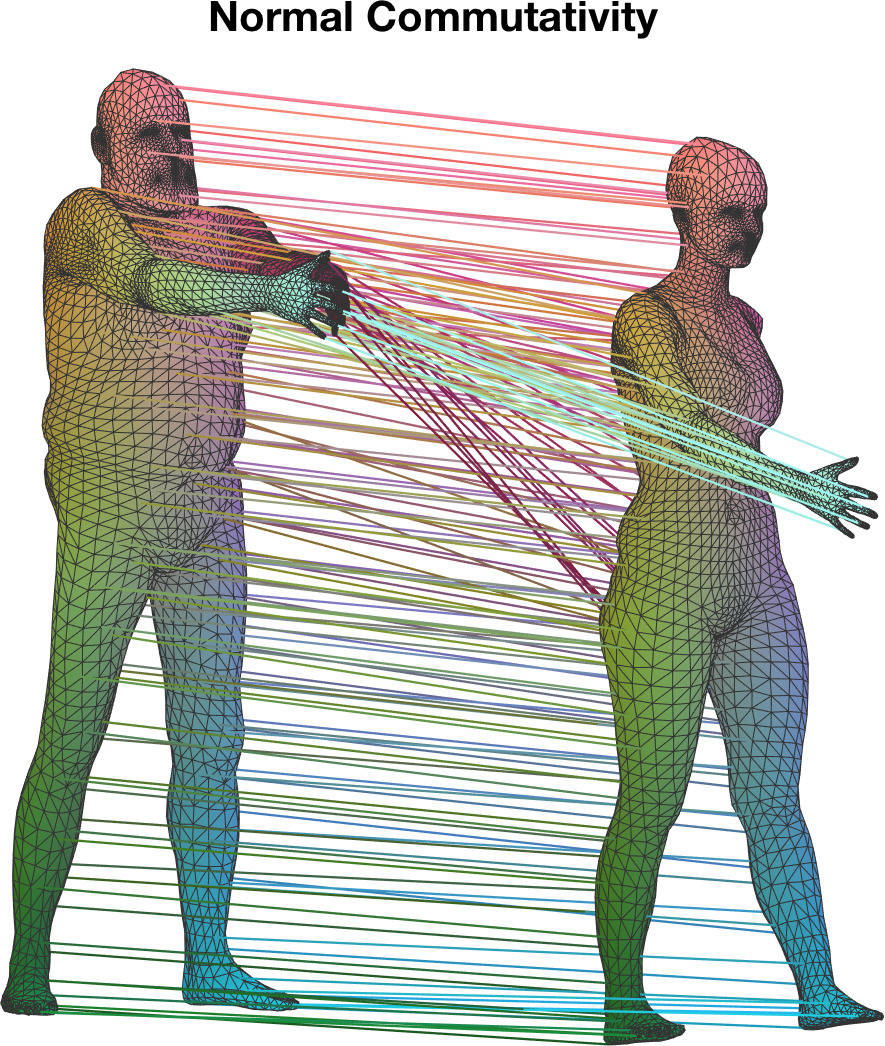}
	\end{tabular}
	\caption{{Representation of the point-to-point map evaluated in Figure
            \ref{fig:fmapsCorresp}. The RGB channel (left column) represents the $xyz$-coordinates,
            which are transferred using the recovered point-to-point map. The correspondences
            obtained with SHOT (left) are precise near sharp features (e.g., on the fingers) whereas
            our constraint (right) is informative on the entire shape.}
	\label{fig:fmapsRgbTrans}}
\end{figure}

{This proposition is simply a consequence of the fundamental theorem of surface theory and the
  relation between functional deformation fields and the second fundamental form described in
  Section \ref{sec:fundeform_props}. Note that enforcing the condition of this proposition in
  practice reduces simply to penalizing the lack of commutativity of the functional map $C$ with
  predefined operators, which can be done efficiently in practice. Therefore, we can see that
  functional deformation fields provide an effective way to capture embedding-dependent information
  in a coordinate-free way that \emph{fully characterizes the shape geometry} up to rigid motions.}

{Inspired by this observation, we propose to solve the following problem: given two shapes and
  a sparse set of correspondences recover a dense map. The shapes come from the Faust dataset}
\cite{Bogo:CVPR:2014} and we are given five corresponding landmarks at the hands, feet and head. The
baseline method following the logic of the original paper is to represent the landmark points as
delta functions $\delta_M$ and $\delta_N$ and look for the most isometric functional map
$C : L^2(M) \rightarrow L^2(N)$, by enforcing commutativity with the Laplace-Beltrami
operator. Thus, the straightforward approach would be to solve the optimization problem:
\begin{equation*}
	\min_C \, \| C \Delta_M - \Delta_N C \|_F^2 \quad \text{s.t.} \quad C \delta_M = \delta_N .
\end{equation*}

{The basic way to add extrinsic information to this problem is to constraint the map to preserve extrinsic descriptors noted $F_M,F_N$ respectively on $M,N$:
\begin{equation*}
	\min_C \, \| C \Delta_M - \Delta_N C \|_F^2 + \| C F_M - F_N \|_F^2  \quad \text{s.t.} \quad C \delta_M = \delta_N .
\end{equation*}
}
{We evaluate two commonly used descriptors:, 1) the normal vector field encoded as three independent functions and 2) the purely extrinsic descriptor SHOT \cite{tombari2010unique} successfully used for solving partial matching problems \cite{litany2017fully}.}

{We compare these descriptor-based approaches to our coordinate-free extrinsic
  constraint. According to Prop.~\ref{prop:embedding_const} if a diffeomorphism commutes with the
  Laplace-Beltrami operator and $\diffU{n}$ then the shapes admit the same embedding. Our new
  optimization problem thus reads:
\begin{align*}
	& \min_C \, \| C \Delta_M - \Delta_N C \|_F^2 + \| C \diffU{n}_M - \diffU{n}_N C \|_F^2 
	 ~\text{s.t.}~ C \delta_M = \delta_N .
\end{align*}

Once the functional map are obtained they are converted to a point-to-point map using the
  knn-algorithm as described in \cite{ovsjanikov2012functional}. The results are shown for two
  non-isometric shape matching problems: different characters taking the same pose and taking
  different pose. Figure \ref{fig:fmapsCorresp} shows the percentage of correspondences within a
  given geodesic distance. Interestingly, SHOT provides valuables information on sharp features
  allowing accurate matching near salient points but tends to fail on featureless regions. In
  comparison our constraint provides information everywhere on the shape making the results less
  subject to obvious mismatching but it is less informative on the placement of salient
  features. This intuition is confirmed by Figure \ref{fig:fmapsRgbTrans} which provides a
  visualization of the point-to-point correspondences by transferring the coordinates functions
  encoded as RGB channels. Finally, the combination of those two constraints overcomes the
  limitations of both methods taken independently.}

\subsection{Intrinsic Symmetrization} \label{exp:intSym}

\begin{figure}[t!]
	\centering
	\begin{tabular}{ccccc}
		\rotatebox{90}{\hspace{1cm}Front}&
		\includegraphics[width=.17\columnwidth]{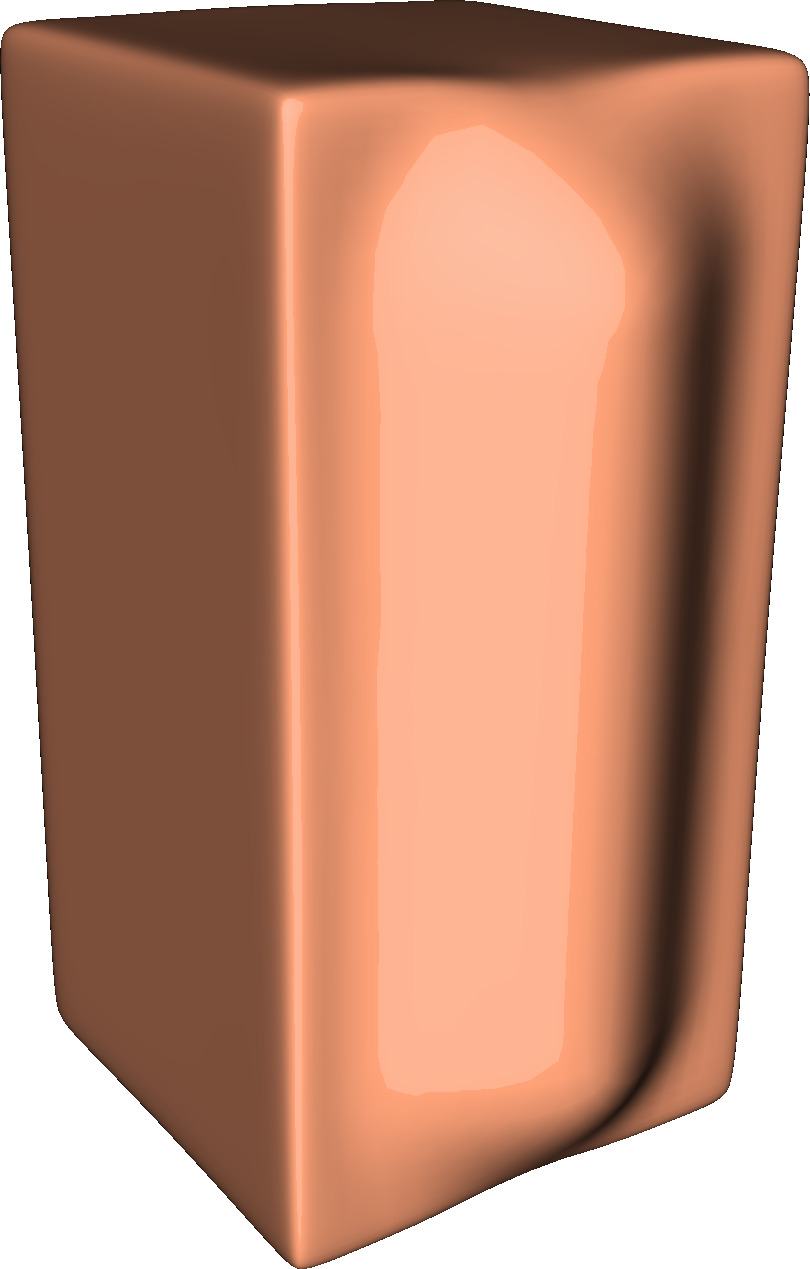}&
		\includegraphics[width=.17\columnwidth]{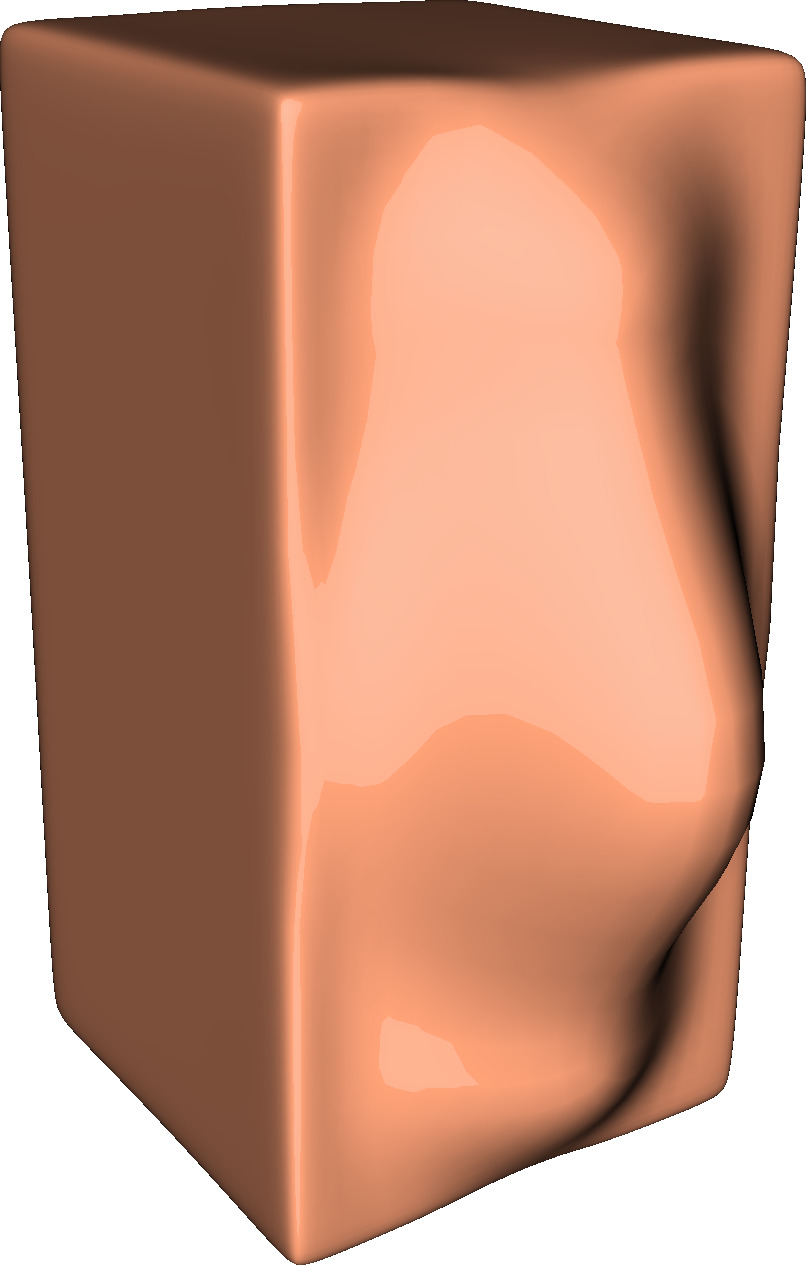}&
		\includegraphics[width=.17\columnwidth]{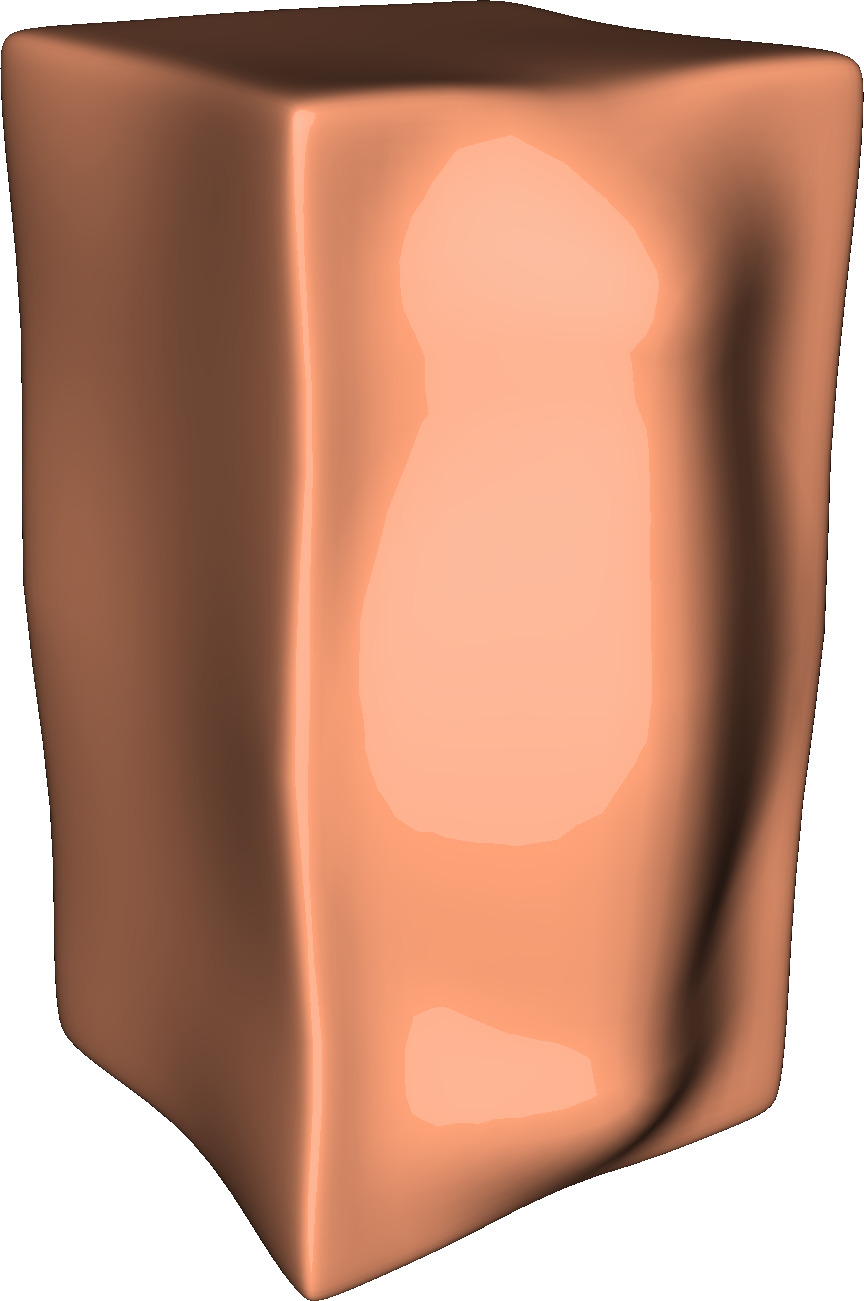}&
		\includegraphics[width=.17\columnwidth]{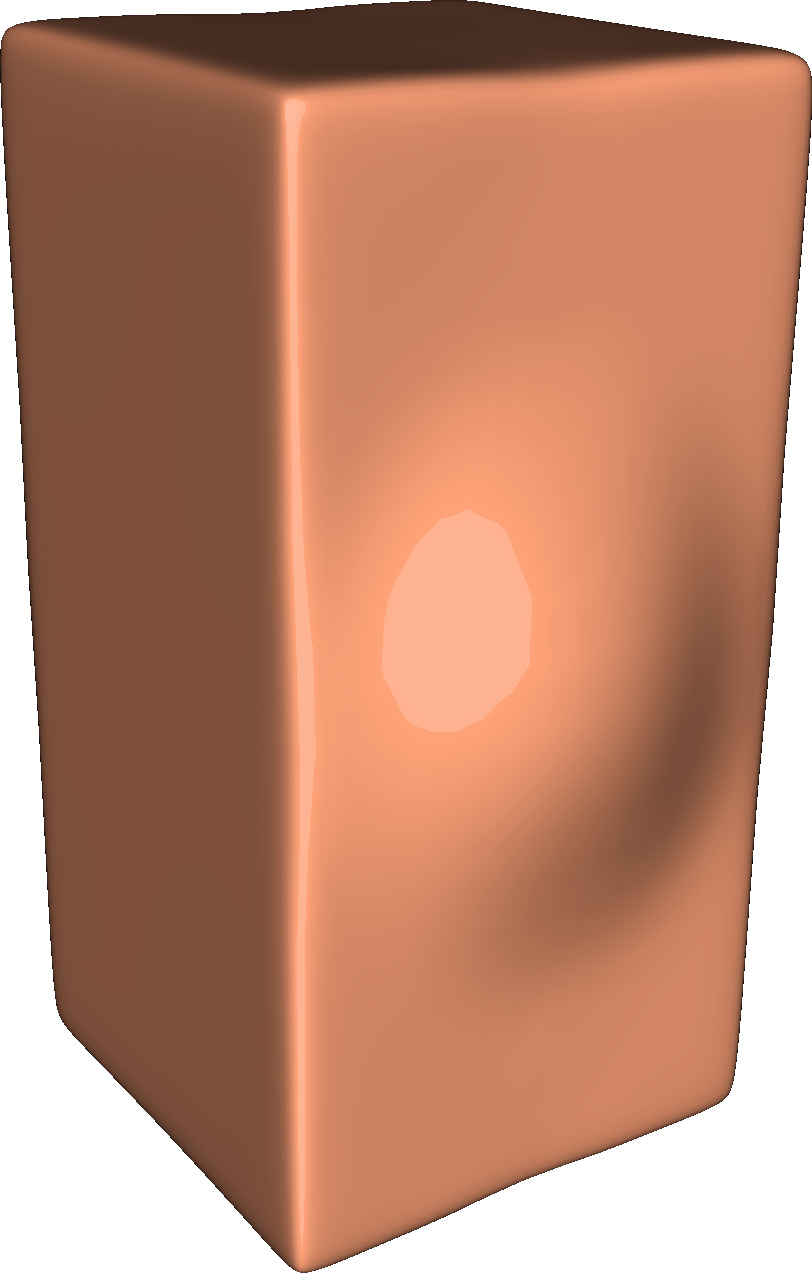}\\
		\rotatebox{90}{\hspace{1cm}Back}&
		\includegraphics[width=.17\columnwidth]{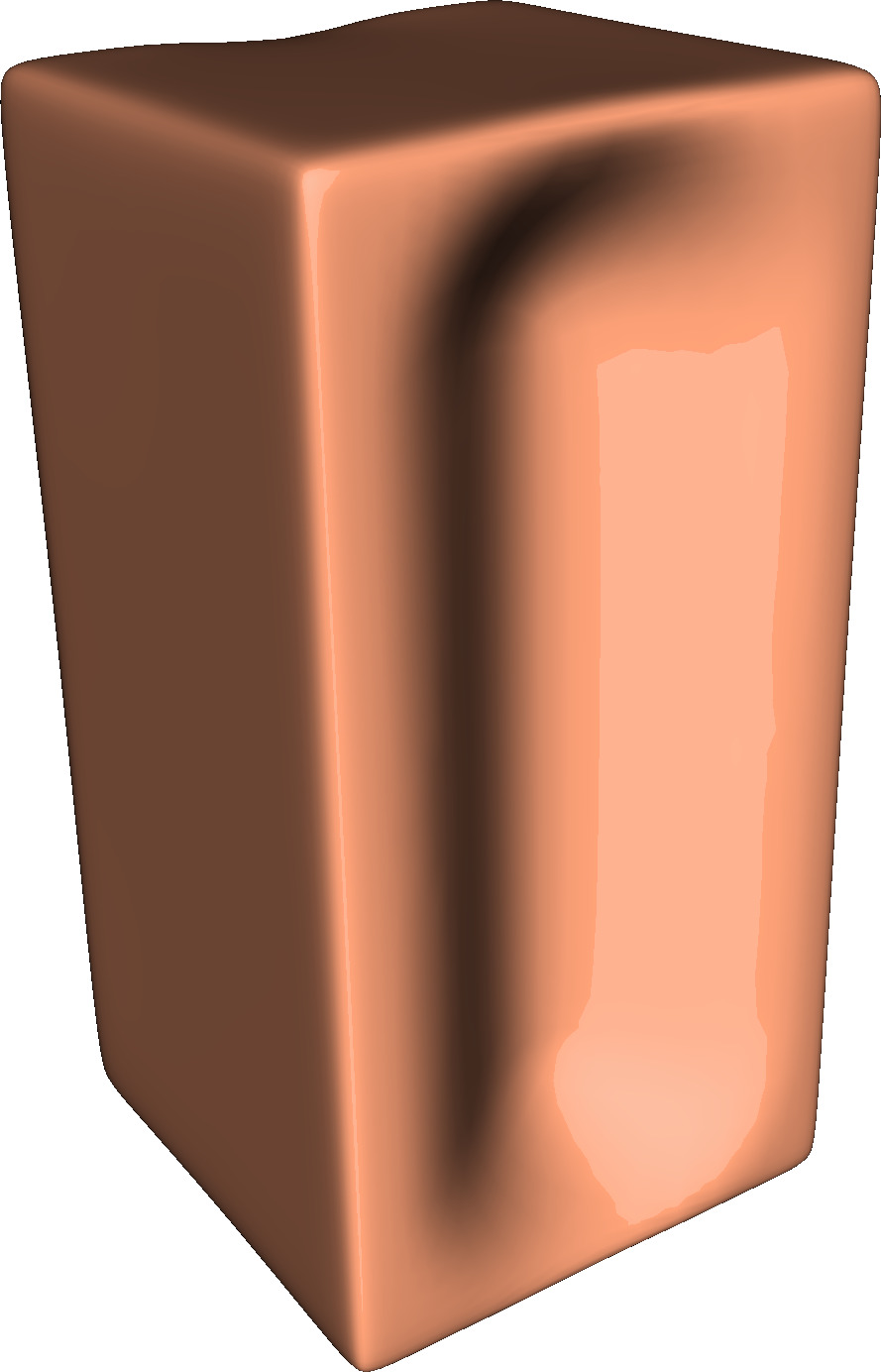}&
		\includegraphics[width=.17\columnwidth]{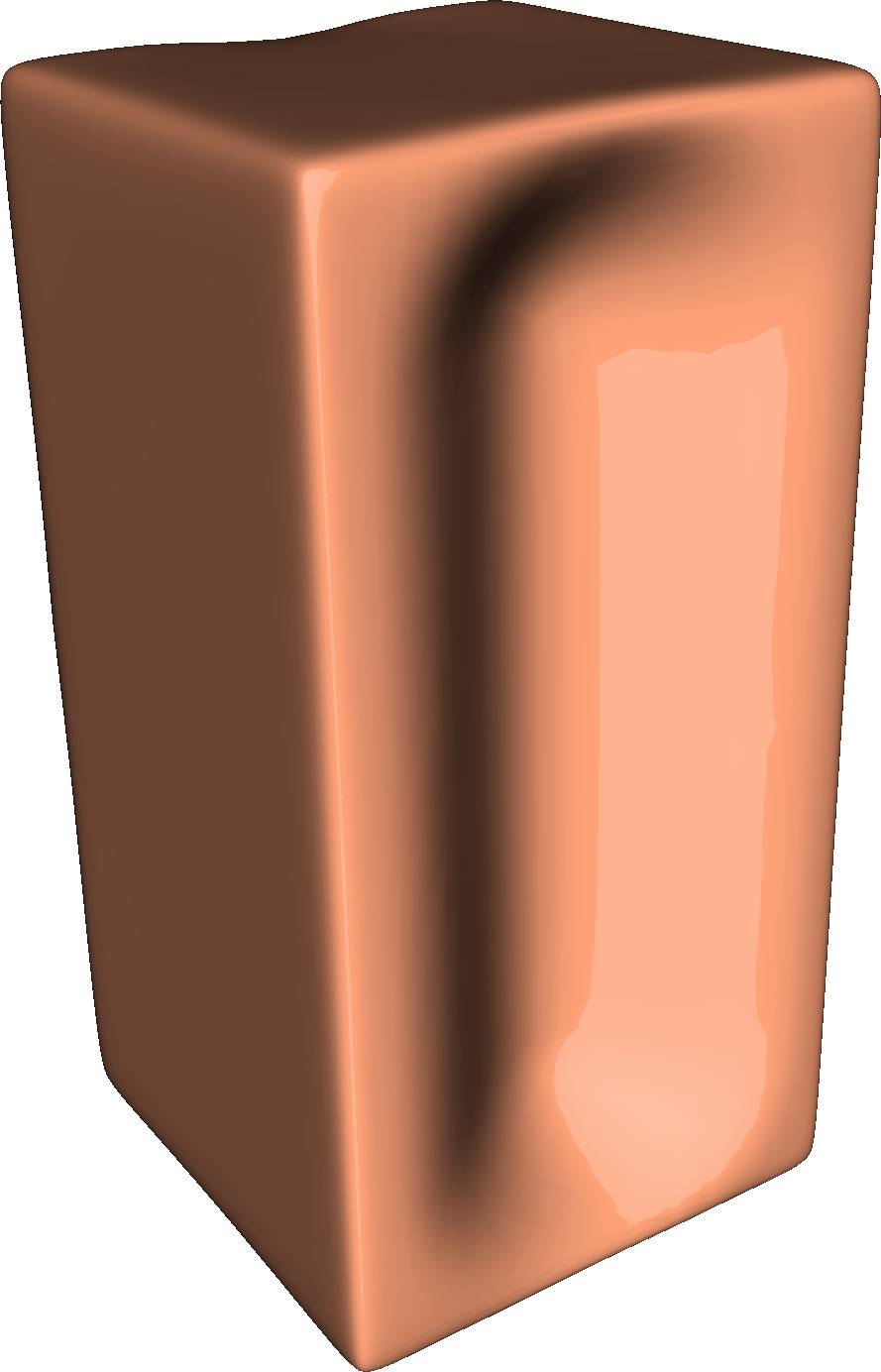}&
		\includegraphics[width=.17\columnwidth]{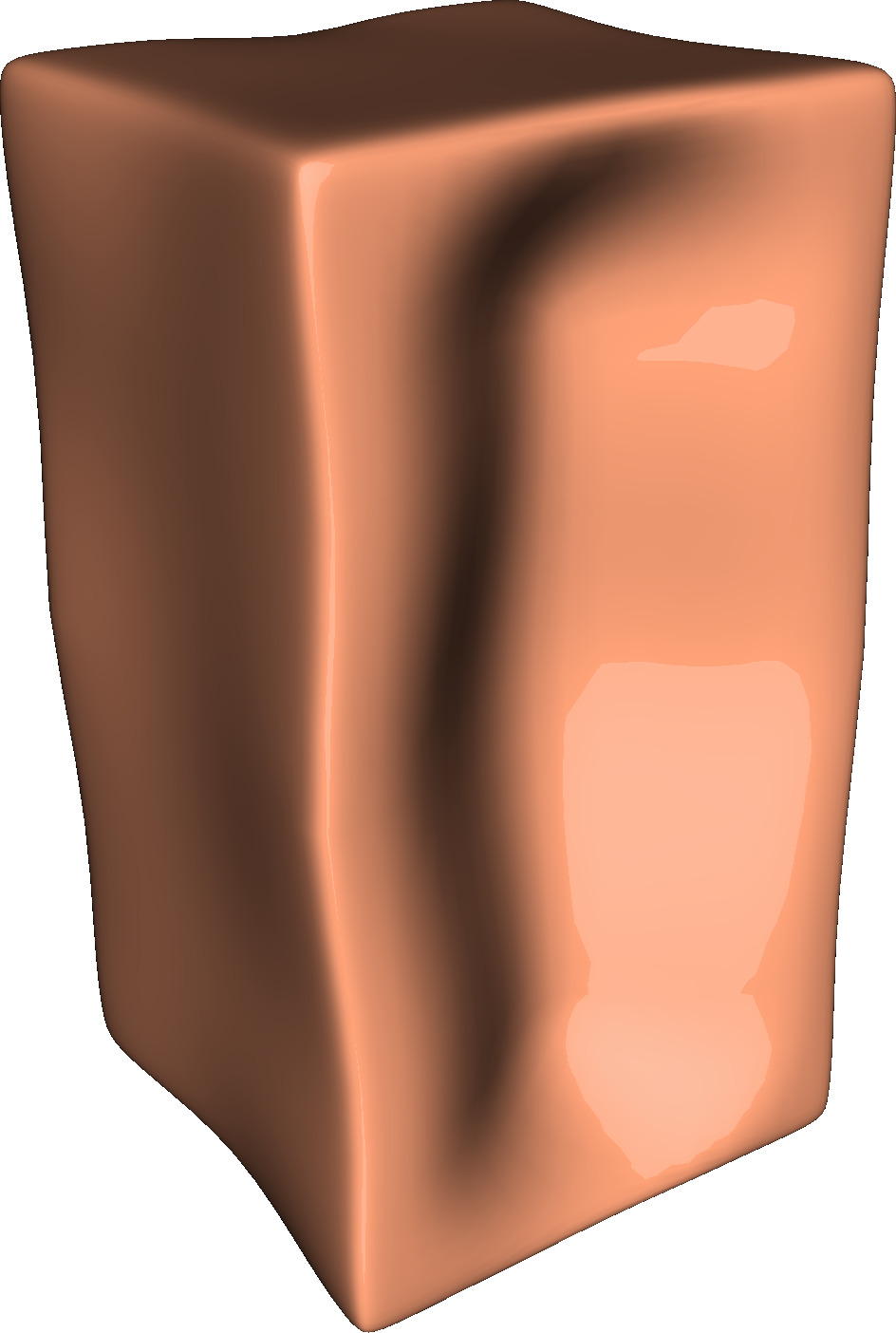}&
		\includegraphics[width=.17\columnwidth]{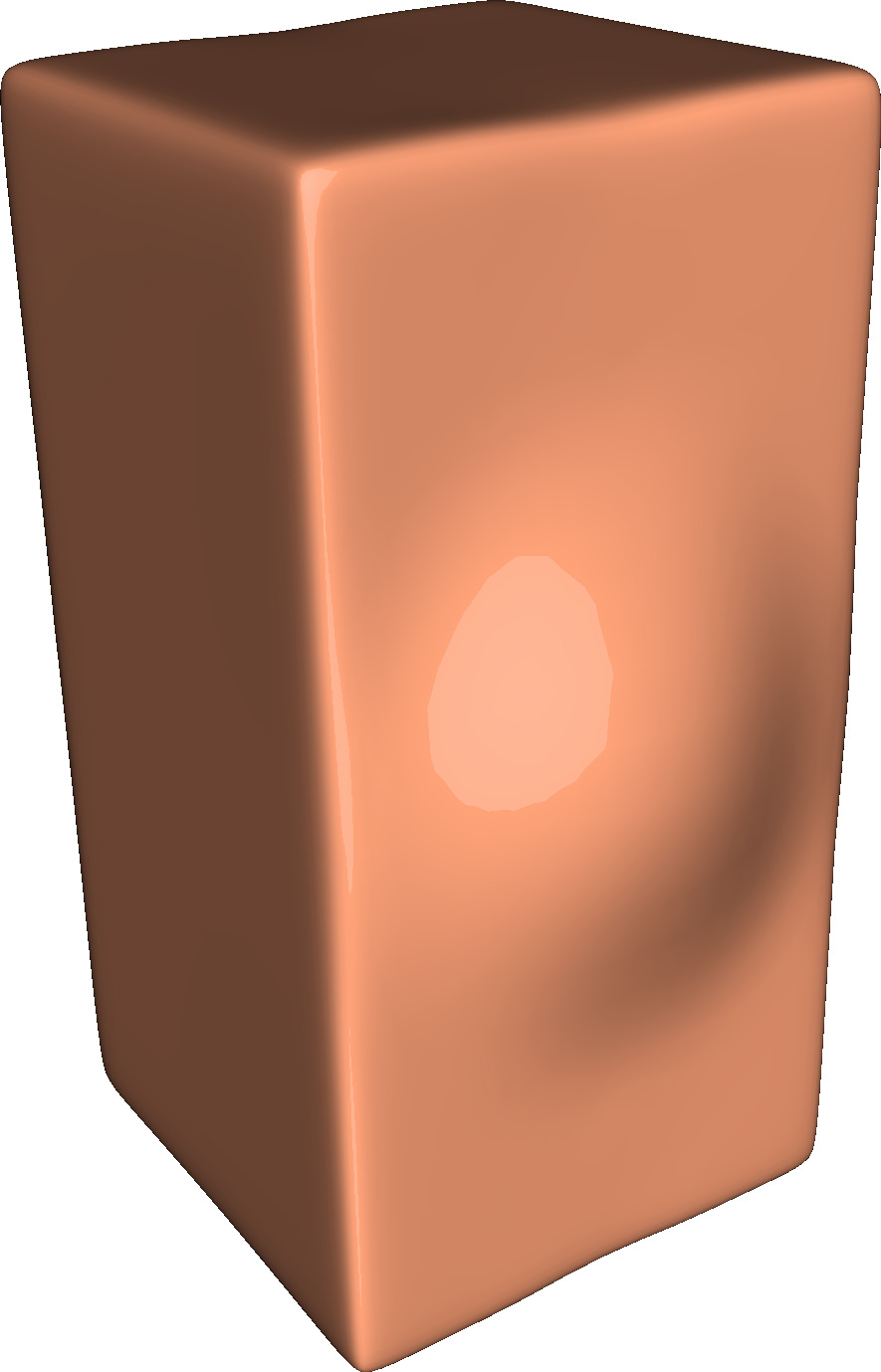}\\
		& Initial & Deformed & Intrinsic Sym. & Extrinsic Sym.
	\end{tabular}
	\caption{{An initially intrinsically symmetric bar (far left) is artificially made asymmetric (middle left). Our algorithm (middle right) is able to recover the symmetry while maintaining the intrinsic structure of the shape. In comparison an extrinsic symmetrization scheme (far right) would erase the intrinsic structure.}}
	\label{fig:cube_sym}
\end{figure}

In this section we show how our representation of deformation fields can be used to deform shapes to
make them more intrinsically symmetric, while keeping their general pose. For example,
Figure~\ref{fig:cube_sym} shows a shape with important features which would be lost by an
\textit{extrinsic} symmetrization scheme. However an \textit{intrinsic} symmetrization algorithm
would preserve those features while recovering the symmetry. This way, our goal is similar to the
one of \cite{zheng2015skeleton} although our approach, unlike theirs, avoids the computation of a
skeleton and is purely intrinsic. More precisely, given a base shape $M$ and a self-map
$\pi : M \rightarrow M$ we would like to compute the shape $M'$ such that the self-map $\psi$ on
$M'$ is an isometry. If we denote by $\varphi : M' \rightarrow M$ the map from $M$ to $M'$ then the
symmetry map on the deformed shape is given by $\psi = \varphi^{-1} \circ \pi \circ \varphi$. Using
Prop.~\ref{Thm:isometry} the isometric constraint is satisfied if and only if the unified shape
difference $D_I^{\psi}$, computed with the map $\psi$, equals identity. If $C_{T}$ is the functional
map representation of a map $T$, then after simplification this is equivalent to
$D_I^{\pi} C_{\pi}^{-1} D_I^{\varphi} C_{\pi} = D_I^{\varphi}$ (see supplementary material).
%\begin{align*}
%	D_I^{\psi} &= D_I^{\varphi^{-1}} C_\varphi D_I^{\pi \circ \varphi} C_\varphi^{-1} \\
%			&= D_I^{\varphi^{-1}} C_\varphi D_I^{\pi} C_{\pi}^{-1} D_I^{\varphi} C_{\pi} C_\varphi^{-1} \\
%			&= C_\varphi \left( D_I^{\varphi} \right)^{-1} D_I^{\pi} C_{\pi}^{-1} D_I^{\varphi} C_{\pi} C_\varphi^{-1} \\
%\end{align*}

Note, however that every intrinsically symmetric shape would be a solution of this equation. Therefore we regularize the problem by imposing that $\varphi$ should be as-isometric-as possible. The equality conditions are enforced in the least squares sense leading to the optimization problem:
\begin{align}
	\min_{\varphi} \| D_I^{\pi} C_{\pi}^{-1} D_I^{\varphi} C_{\pi} - D_I^{\varphi} \|^2_F + \tau \| D_I^{\varphi} - I \|^2_F .
	\label{eq:intSymOpt}
\end{align}

The optimization \eqref{eq:intSymOpt} is restricted to the set of diffeomorphisms so a direct
approach is challenging to use in practice. A more tractable method is to use functional deformation
fields as a first order approximation of shape differences and thus find the deformation field that
solves \eqref{eq:intSymOpt} to first order. After linearization, Eq. \eqref{eq:intSymOpt} becomes:
\begin{align}
	\min_{V} \| D_I^{\pi} C_{\pi}^{-1} \diffU{V} C_{\pi} - \diffU{V} - I + D_I^{\pi} \|^2_F + \tau \| \diffU{V} \|^2_F .
	\label{eq:intSymLin}
\end{align}

\begin{algorithm}[t!]
\caption{\scshape Intrinsic Symmetrization}\label{algo:intSym}
    \SetKwInOut{Input}{Input}
    \SetKwInOut{Output}{Output}

    \Input{Triangle mesh with vertices $p$ and self-map $\pi$}
    \Output{New vertices $p^t$}
    
    \Repeat{$\| p^{t+1} - p^{t} \| < \epsilon$}{
    	Find $V^{t+1}$ solution of \eqref{eq:intSymLin} \;
	Compute new embedding: $p^{t+1} = p^{t} + V^{t+1}$ \;
	Recompute $D_I^{\pi}, C_{\pi}$ \;
    }
\end{algorithm}
\begin{figure}[t!]
	\centering
	\begin{tabular}{c|ccc}
%		\includegraphics[height=.3\columnwidth]{Images/IntSym/mesh000_sym_init}&
%		\includegraphics[height=.3\columnwidth]{Images/IntSym/mesh000_sym_IntSym_i1}&
%		\includegraphics[height=.3\columnwidth]{Images/IntSym/mesh000_sym_IntSym_i2}&
%		\includegraphics[height=.3\columnwidth]{Images/IntSym/mesh000_sym_IntSym_i3}\\
%		Initial & Iter. 1 & Iter. 2 & Iter. 3 \\
%		2.5633 & 1.8585 & 1.4950 & 1.3515 \\
%		\hline\\
		\includegraphics[height=.3\columnwidth]{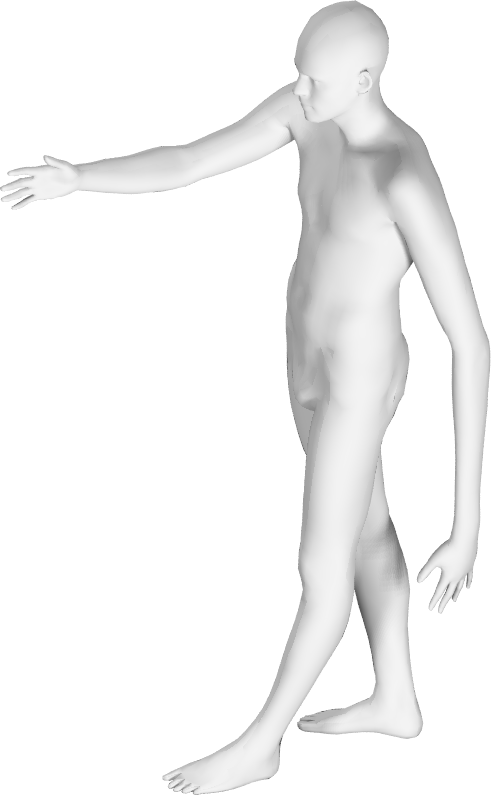}&
		\includegraphics[height=.3\columnwidth]{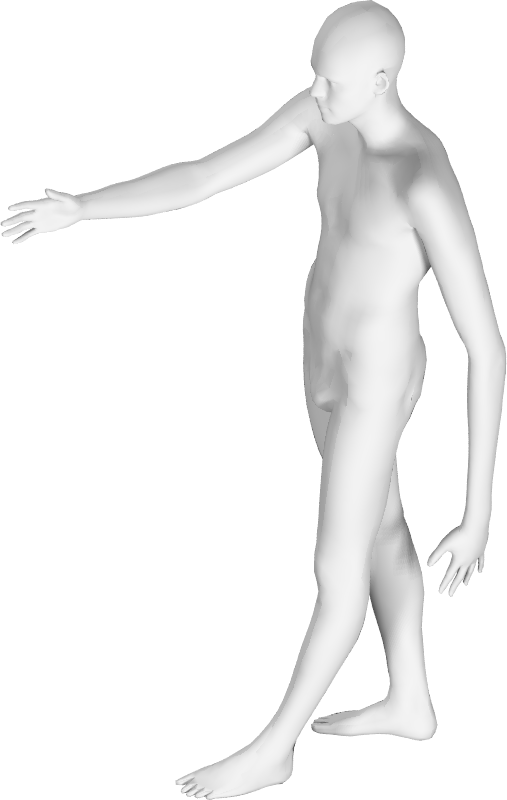}&
		\includegraphics[height=.3\columnwidth]{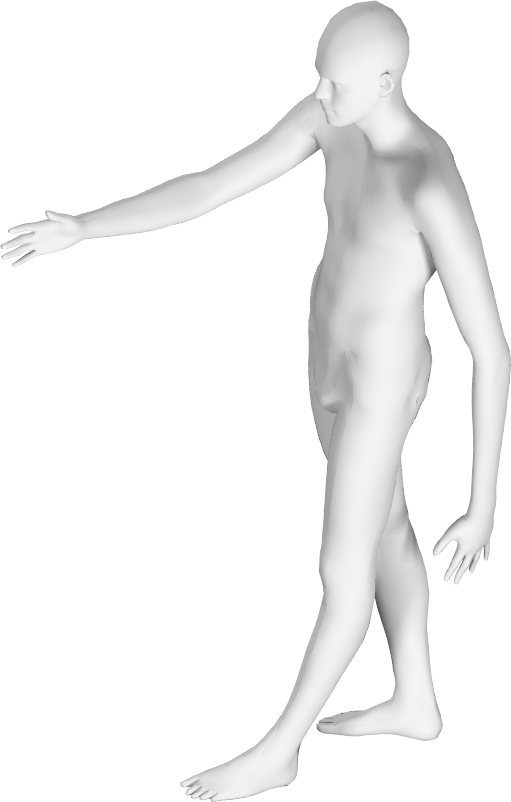}&
		\includegraphics[height=.3\columnwidth]{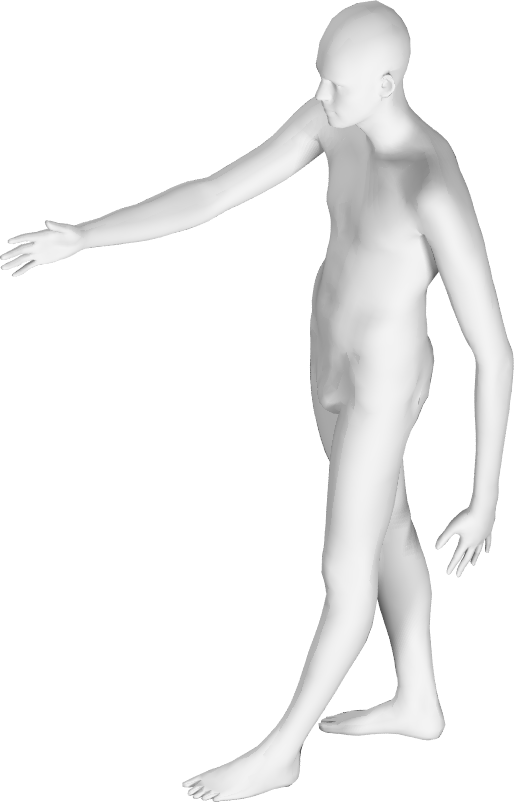}\\
		Initial & Iter. 1 & Iter. 2 & Iter. 3 \\
		3.7533 & 3.1831 & 2.8511 & 2.6788 \\
		\hline\\
		\includegraphics[height=.3\columnwidth]{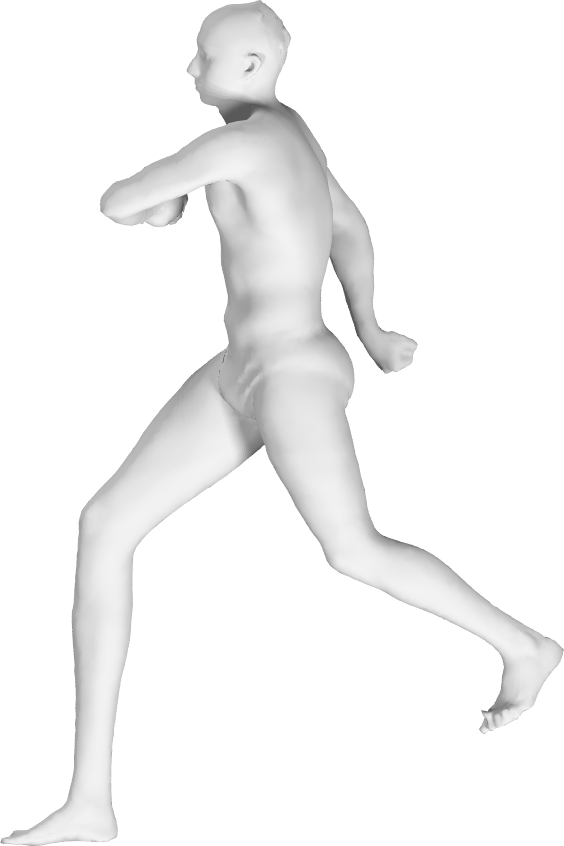}&
		\includegraphics[height=.3\columnwidth]{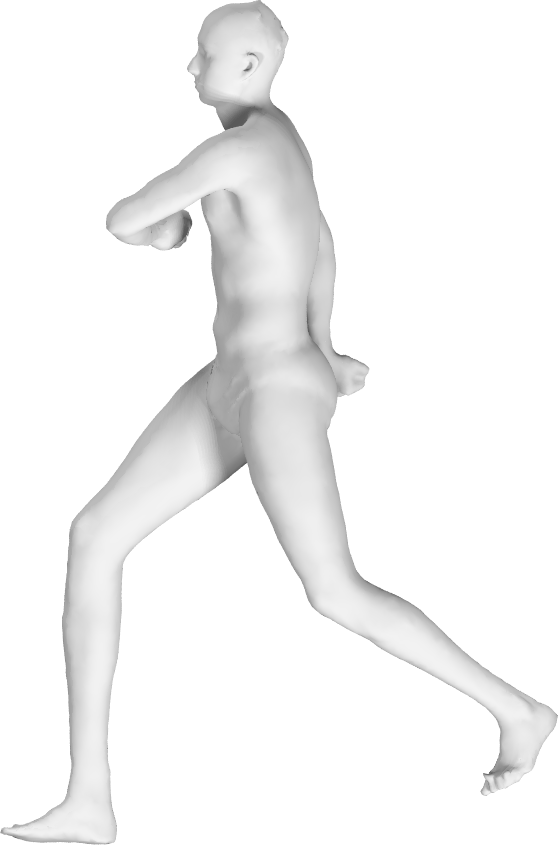}&
		\includegraphics[height=.3\columnwidth]{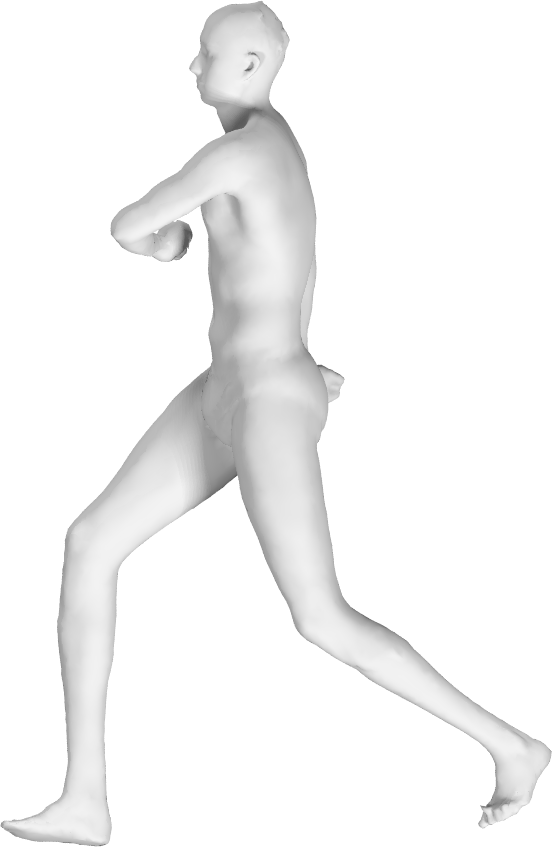}&
		\includegraphics[height=.3\columnwidth]{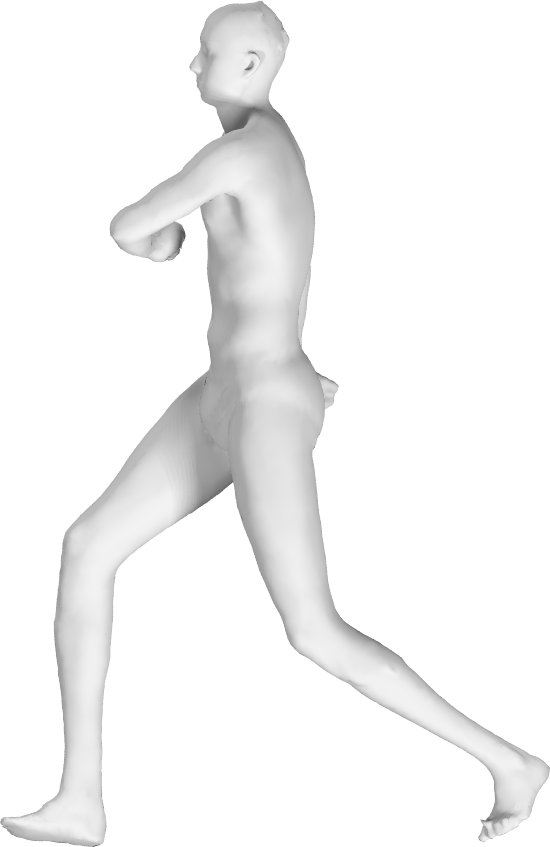}\\
		Initial & Iter. 1 & Iter. 2 & Iter. 3 \\
		4.9964 & 3.8179 & 3.2998 & 3.1256 \\
		\hline\\
		\includegraphics[width=.2\columnwidth]{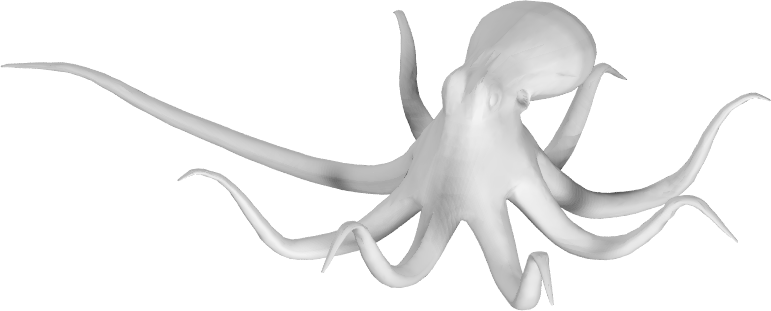}&
		\includegraphics[width=.2\columnwidth]{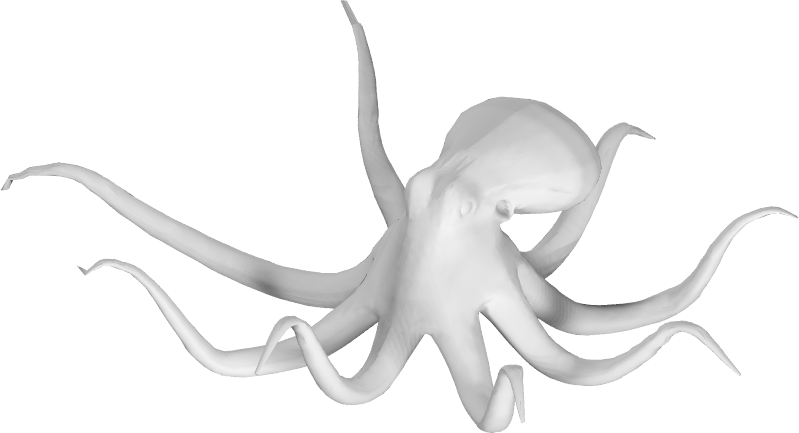}&
		\includegraphics[width=.2\columnwidth]{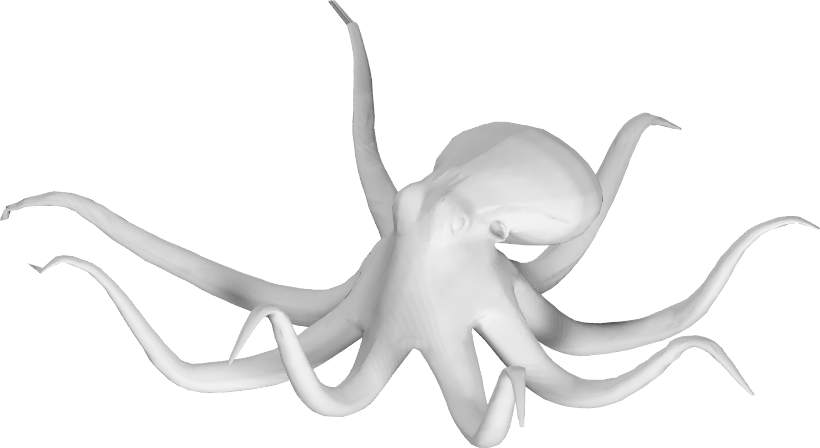}&
		\includegraphics[width=.2\columnwidth]{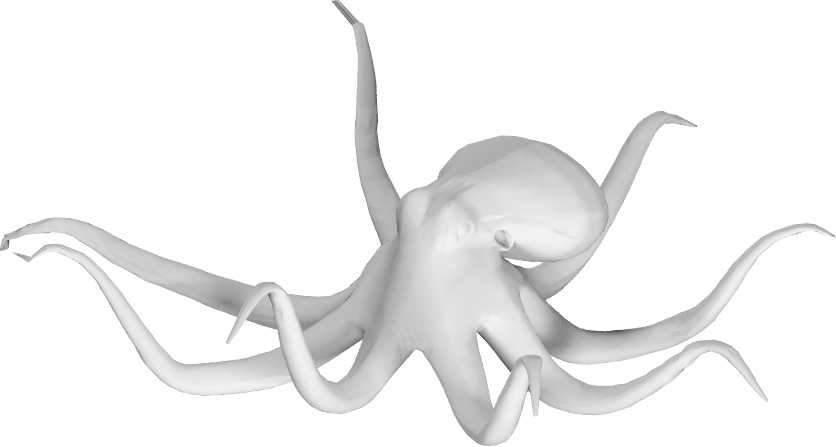}\\
		Initial & Iter. 1 & Iter. 2 & Iter. 3 \\
		6.4424 & 5.3074 & 4.8237 & 4.5554 \\
	\end{tabular}
        \vspace{-1mm}
	\caption{Three iterations of our intrinsic symmetrization method, Algorithm
          \ref{algo:intSym}, given an approximate symmetry map. At each step we measure the distance
          to the symmetry by the Frobenius norm between the intrinsic shape difference induced by
          the symmetry map and identity, namely $\| D_I - I \|_F$. Although not directly taken into
          account, this energy decreases at each iteration. Note that our algorithm works with any
          type of symmetries: see bottom row for a non-reflectional symmetry.}
	\label{fig:intSym}
\end{figure}
This linearization suggests an iterative algorithm (described in Algorithm \ref{algo:intSym}) which
alternates between solving the linearized problem \eqref{eq:intSymLin} and computing the new vertex
positions. In practice, we construct an over-complete dictionary of deformation fields, composed of
the three bases described at the beginning of Section~\ref{sec:results}, compute the optimal
deformation field by solving for the coefficients $\alpha$.

\begin{figure*}[t!]
	\centering
	\begin{tabular}{ccccc}
		\multirow{2}{*}{\includegraphics[width=.17\textwidth]{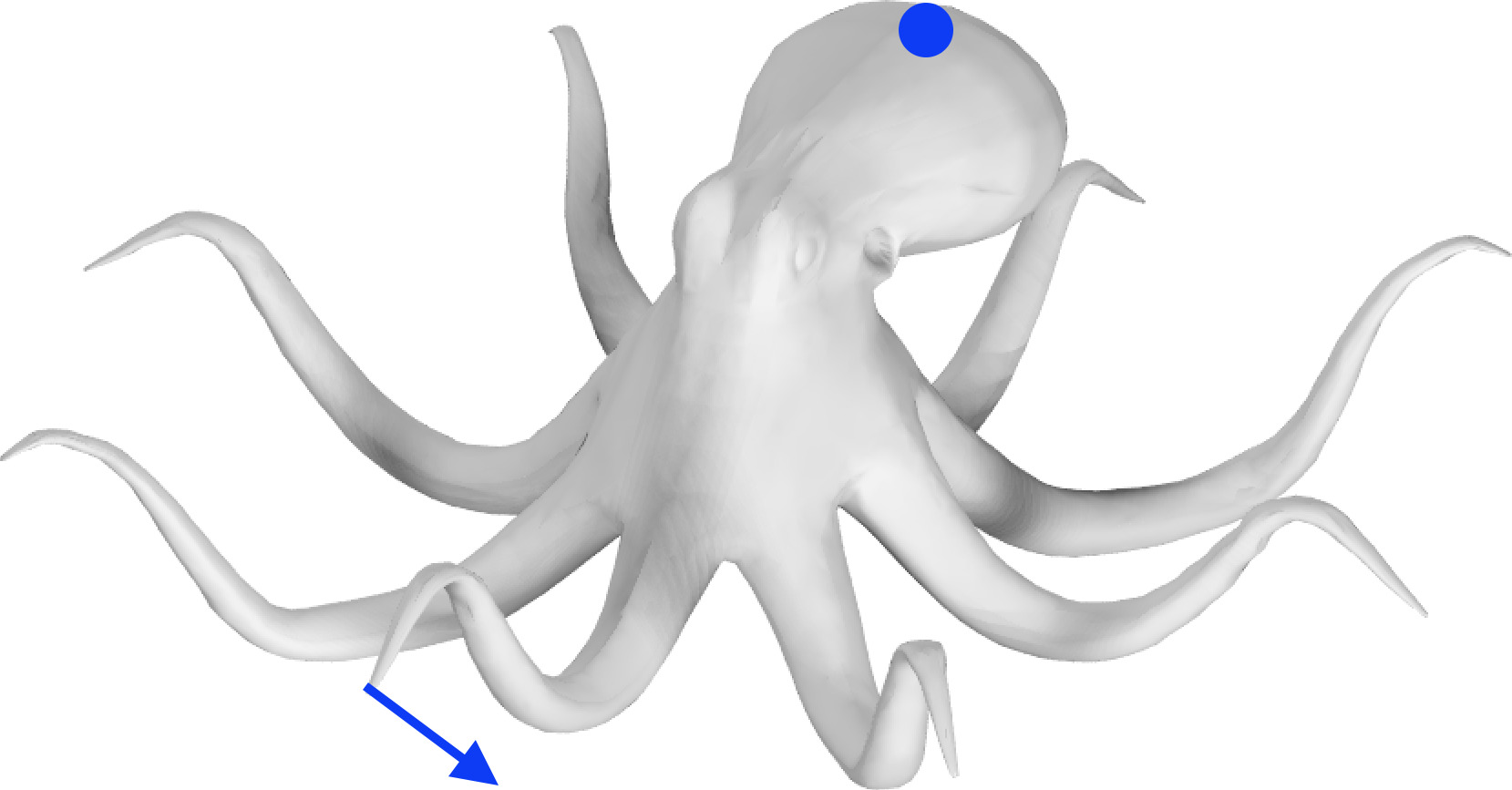}}&
		\includegraphics[width=.17\textwidth]{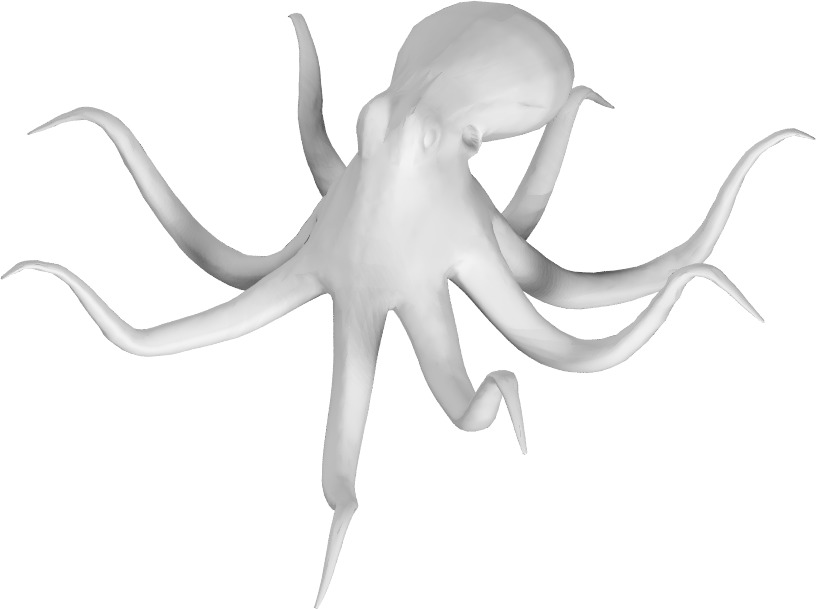}&
		\includegraphics[width=.17\textwidth]{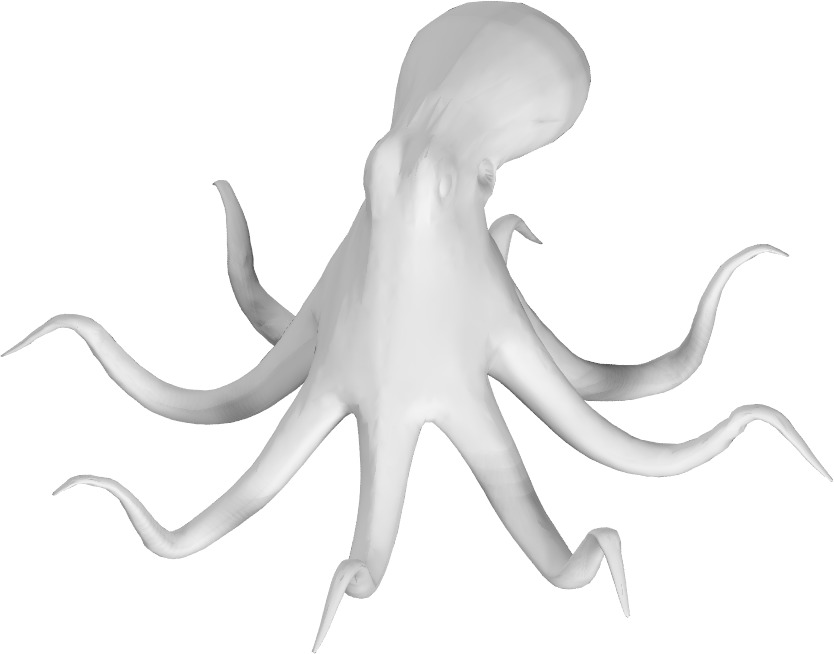}&
		\includegraphics[width=.17\textwidth]{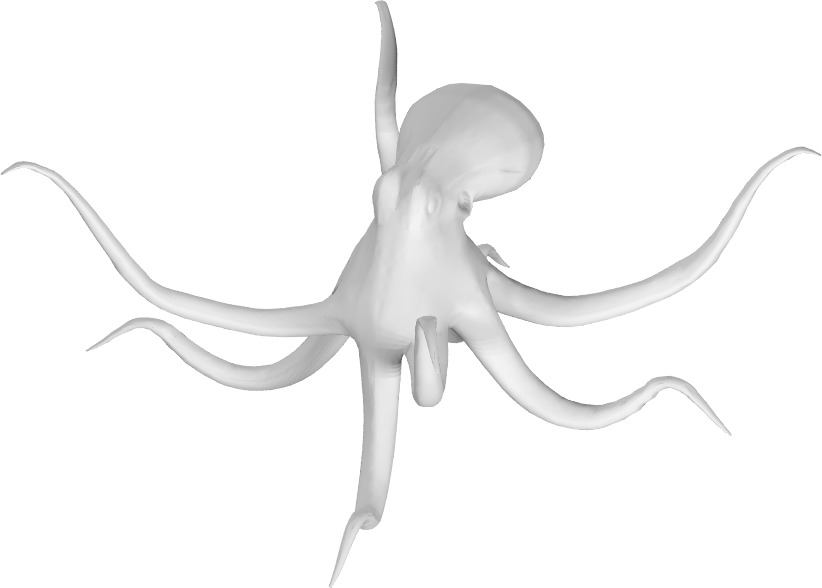}&
		\includegraphics[width=.17\textwidth]{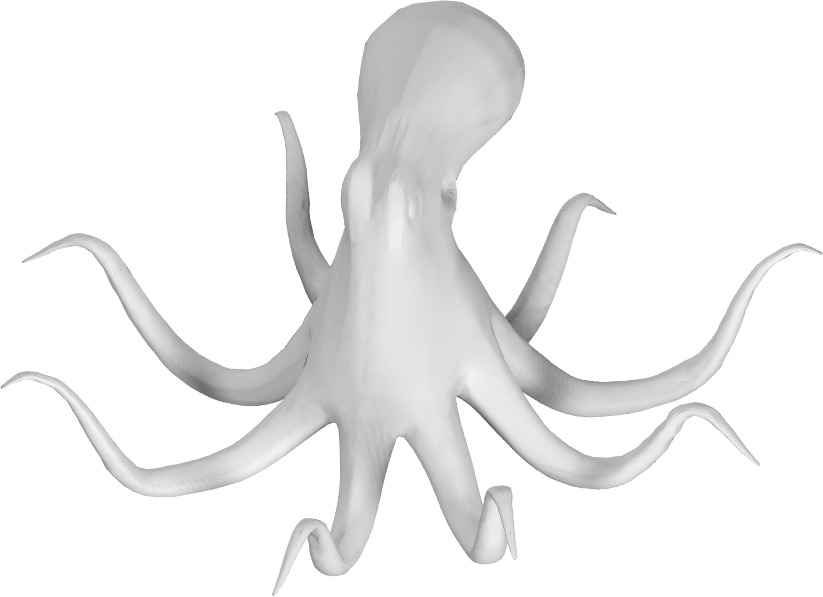}\\
		&%\includegraphics[width=.17\textwidth]{Images/VFdesign/Octopus/Octopus}&
		\includegraphics[width=.17\textwidth]{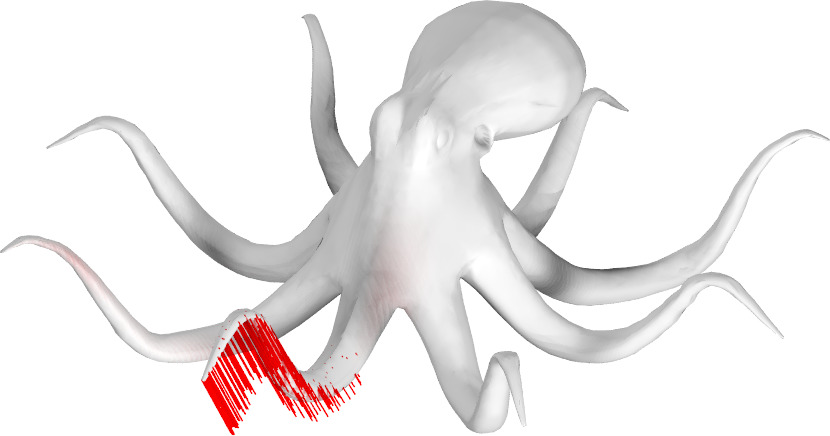}&
		\includegraphics[width=.17\textwidth]{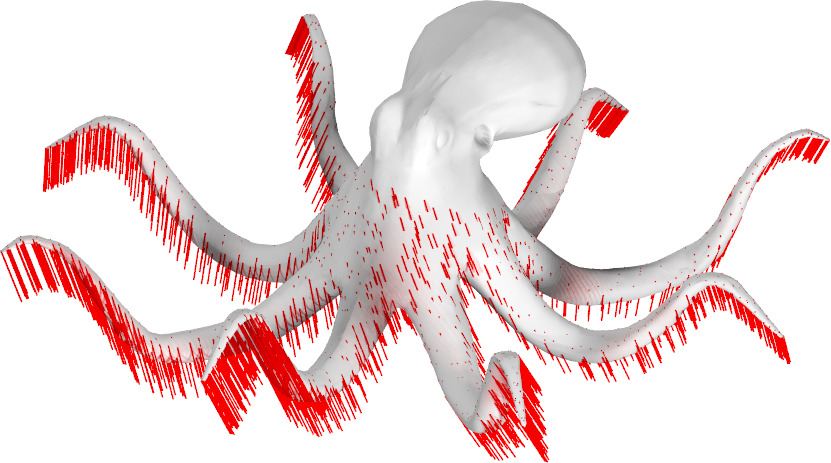}&
		\includegraphics[width=.17\textwidth]{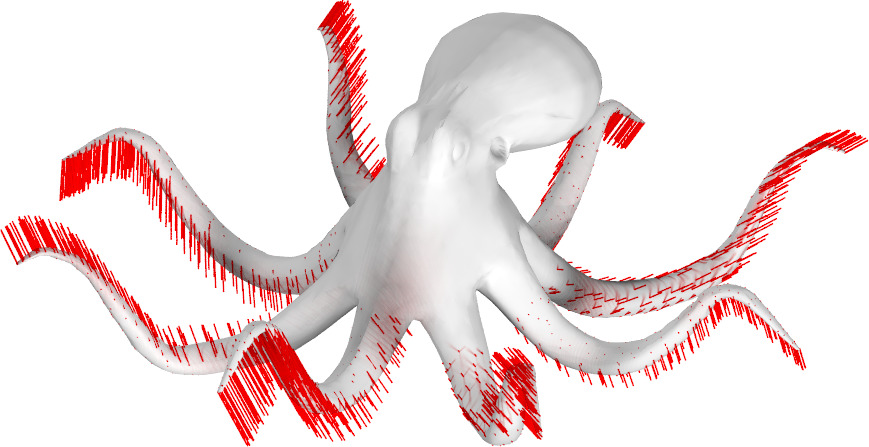}&
		\includegraphics[width=.17\textwidth]{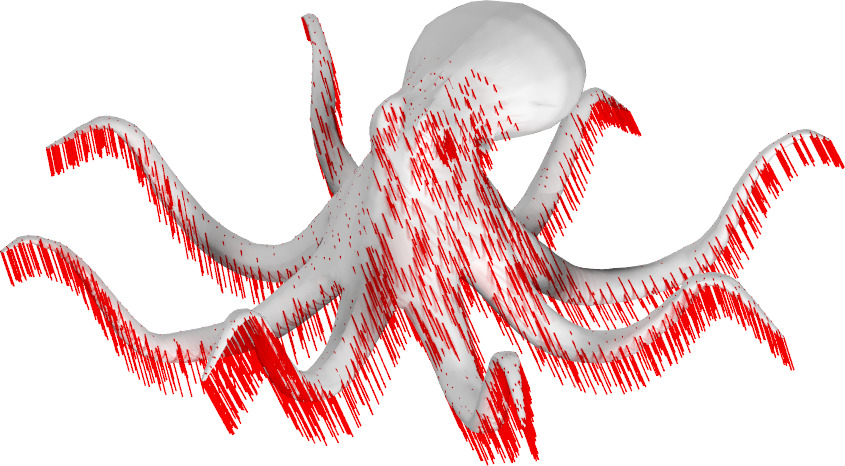}\\
		Constraints & As-isometric-as possible & Symmetry & Anti-symmetry & Laplacian Reg.
	\end{tabular}
\vspace{-2mm}
	\caption{We design deformations respecting the directional constraints shown on the far left and minimizing various criteria (from left to right): the infinitesimal shape difference leading to the most isometric vector field, the commutativity with a self-map, the anti-commutativity with the same self-map and the commutativity with the Laplace-Beltrami operator.\vspace{-4mm}}
	\label{fig:octopus}
\end{figure*}

\begin{figure}[t!]
	\centering
	\begin{tabular}{ccc}
		\includegraphics[width=.13\textwidth]{Images/VFdesign/Octopus/Octopus_deformation}&
		\includegraphics[width=.13\textwidth]{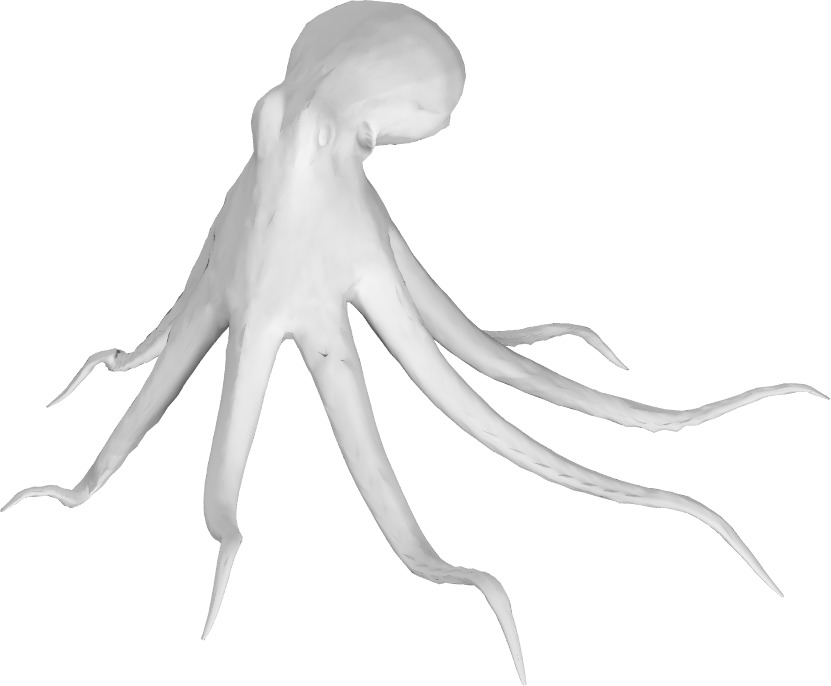}&
		\includegraphics[width=.13\textwidth]{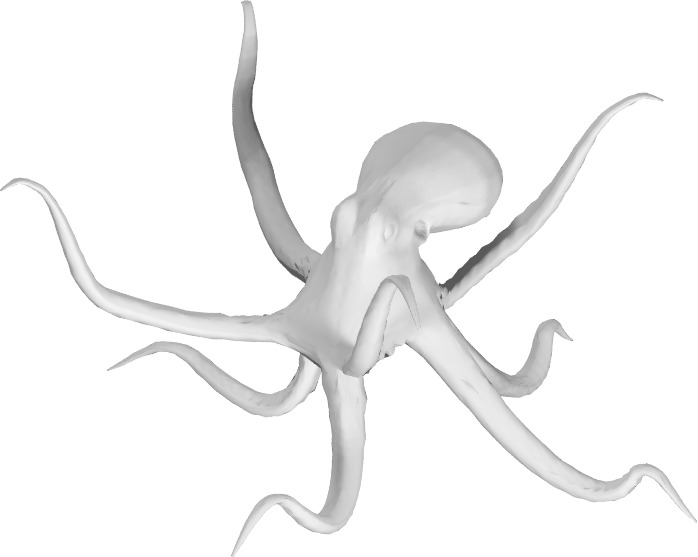}\\
		\includegraphics[width=.13\textwidth]{Images/VFdesign/Octopus/Octopus_deformation_vf}&
		\includegraphics[width=.13\textwidth]{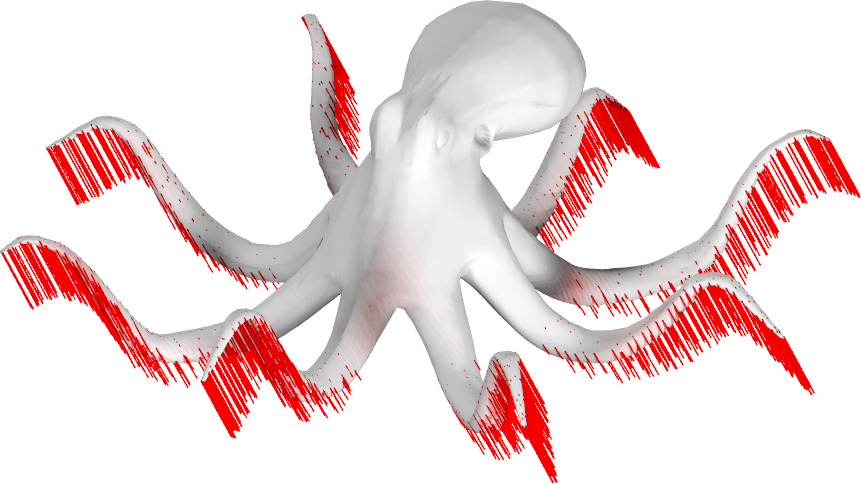}&
		\includegraphics[width=.13\textwidth]{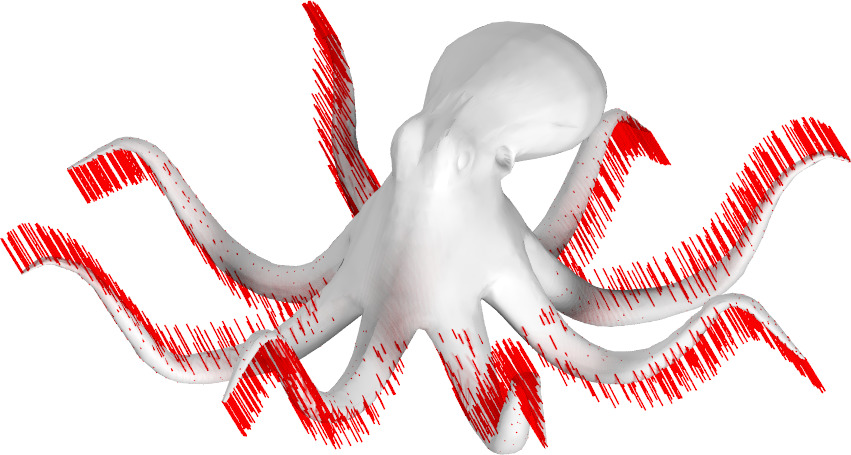}\\
		Vector field & Symmetry & Anti-symmetry
	\end{tabular}
	\caption{{We design deformations by projecting the vector field shown on the far left onto the space of symmetric and anti-symmetric functions. This direct approach does not deliver the expected results found in Figure~\ref{fig:octopus}.}}
	\label{fig:octopus_naive}
\end{figure}

{Figure~\ref{fig:intSym} shows two examples where our method successfully recovers intrinsic
  symmetry from meshes with outstretched parts. In \cite{zheng2015skeleton} the authors propose a
  method based on skeleton driven deformation to achieve intrinsic symmetry but limited to
  reflectional symmetries. Our method does not require such assumptions and works for any given
  self-mapping (e.g. bottom row in Figure~\ref{fig:intSym}). Note also that our deformation field
  representation is essential in this scenario, since for example, representing deformation fields
  through displacement functions would not provide information on the necessary (or induced) metric
  distortion.}

\subsection{Deformation design}

Since our operator is linear with respect to the deformation field one can easily combine multiple
constraints to the deformation vector field. In Figure~\ref{fig:octopus} we show how multiple
different constraints can be combined using our representation. First, we can easily require that at
a point $p$ the deformation field matches a given vector $v$, by setting $V(p) = v$, in addition to
other global constraints. Second, we can find the most \emph{isometric} (preserving the intrinsic
metric) deformation by minimizing $V \mapsto \| \diffU{V} \|_F^2$. At the same time, given a
self-map represented as a functional map $S$, we design a symmetric vector field by imposing a
constraint of the form $\diffU{V} C_S = C_S \diffU{V}$. 
%When using this constraint, we observe a similar deformation on each tentacles. 
Similarly, we can impose an \emph{anti-symmetry} constraint by requiring
$\diffU{V} C_S + C_S \diffU{V} = 0$. {In comparison, Figure~\ref{fig:octopus_naive} shows an
extrinsic deformation design method consisting in projecting each vector field component into the
space of symmetric (respectively anti-symmetric) functions. The resulting shapes look quite
distorted compared to our solution. Moreover the distance of the conformal shape difference
(resp. area-based shape difference) to identity, measuring how far the map
$S$ is from an isometry, is of $0.51$ (resp. $0.46$) for our design and $0.63$ (resp. $0.56$) for
the extrinsic design. Thus, the extrinsic deformation design tends to distort the intrinsic
structure of the shape.  Finally, we test a regularization technique for the deformation field by
imposing the commutativity with the Laplace-Beltrami operator, which tends to spread to the entire
shape. Note that despite the diversity of these constraints,  they can all be
enforced easily using our operator-based representation. In contrast, the straightforward method
shown in Figure~\ref{fig:octopus_naive}, consisting in projecting the vector field onto the space of
symmetric or anti-symmetric functions, fails in this tasks as it is fully extrinsic.}

\begin{figure}[t!]
	\centering
	\begin{tabular}{cc|cc}
		\includegraphics[height=.35\columnwidth]{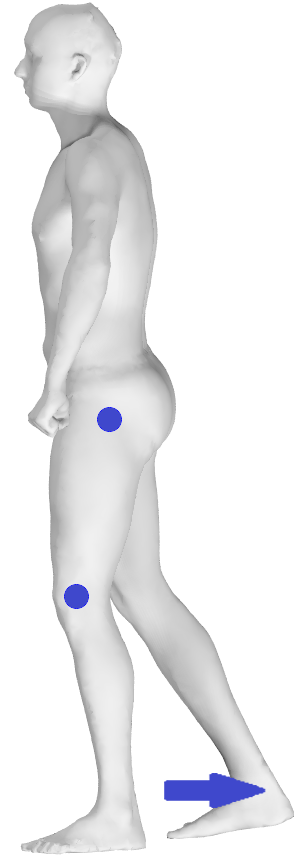}&
		\includegraphics[height=.35\columnwidth]{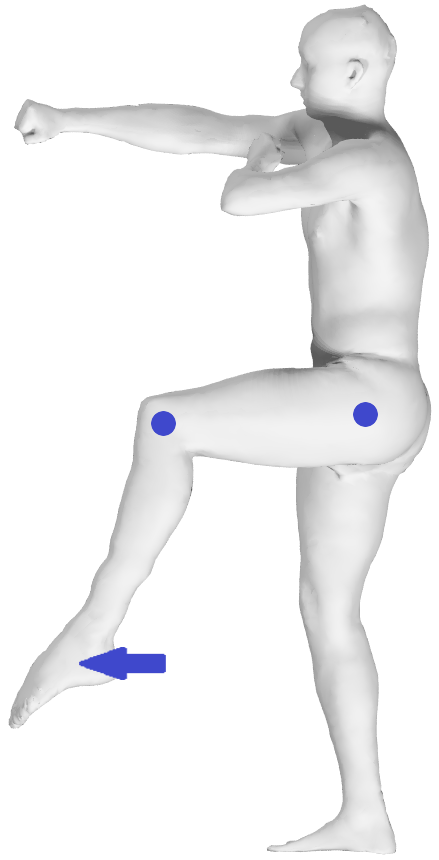}&
		\includegraphics[height=.35\columnwidth]{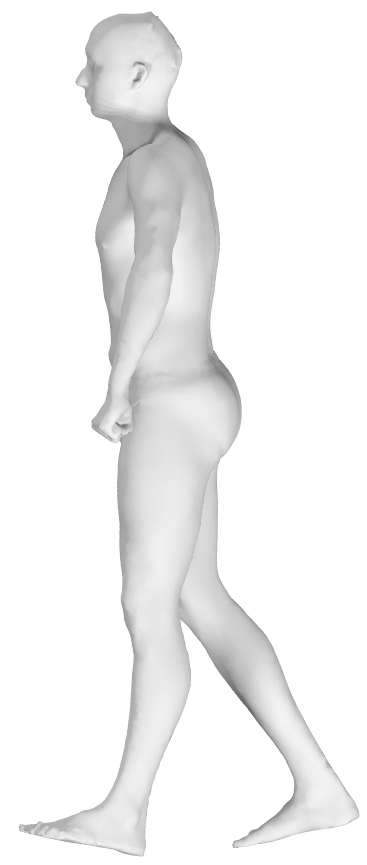}&
		\includegraphics[height=.35\columnwidth]{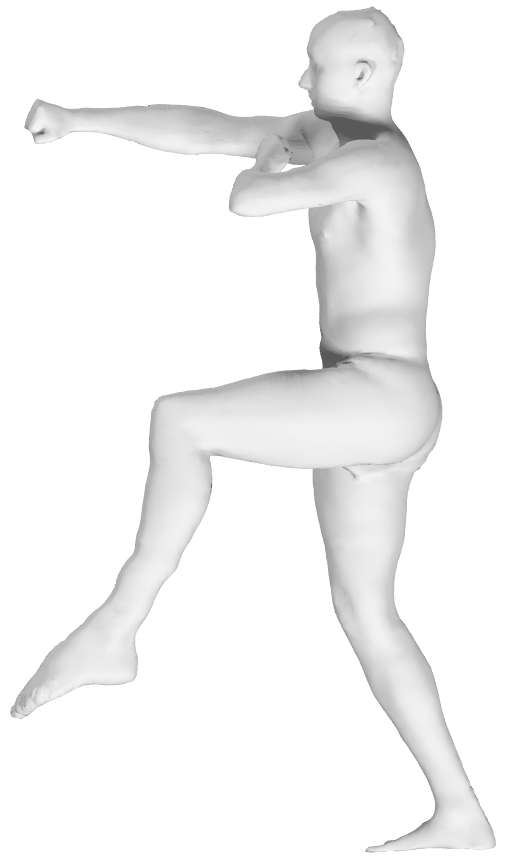}\\
		\includegraphics[height=.35\columnwidth]{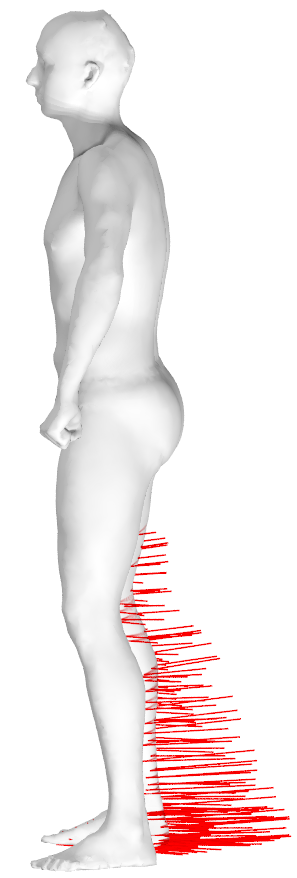}&
		\includegraphics[height=.35\columnwidth]{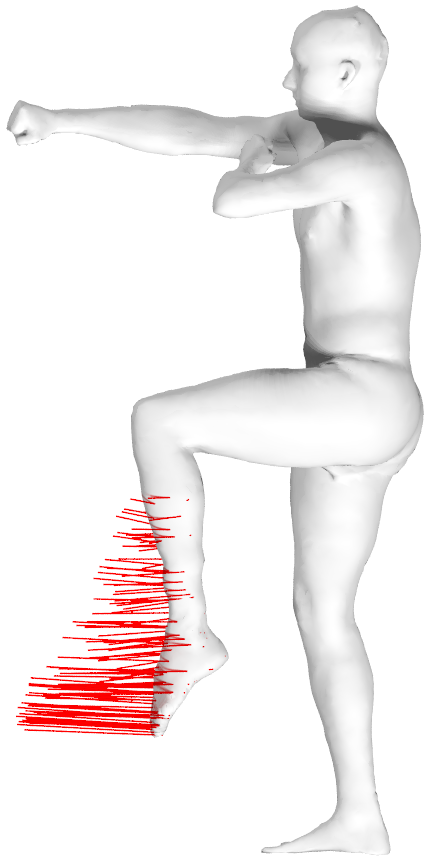}&
		\includegraphics[height=.35\columnwidth]{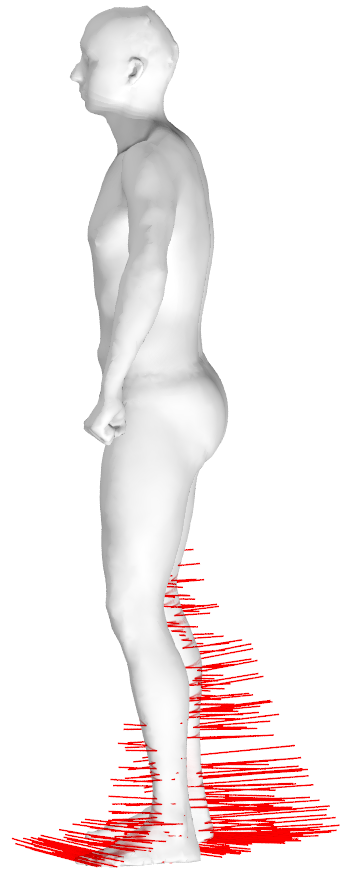}&
		\includegraphics[height=.35\columnwidth]{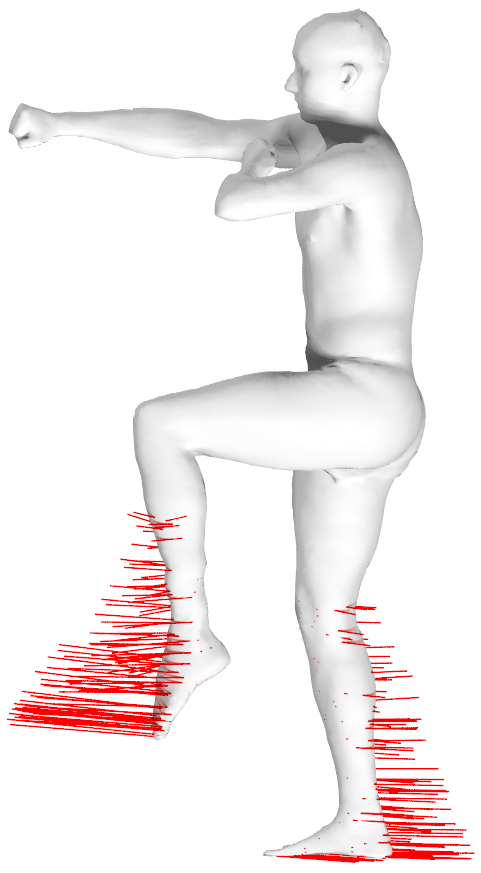}\\
		\multicolumn{2}{c}{Separate design} & \multicolumn{2}{c}{Joint design}
	\end{tabular}
        \vspace{-1mm}
	\caption{Left: The deformations are designed separately by minimizing the smoothness term
          $\langle V, \Delta V \rangle$. Right: joint deformation design by adding the commutativity
          with the mapping to the optimization. Note that the constraints on one shape tend to be
          transferred to the other.\vspace{-2mm}}
	\label{fig:jointVF}
\end{figure}

\begin{figure*}[t!]
	\centering
	\begin{tabular}{cc|cccc}
		\includegraphics[width=.11\textwidth]{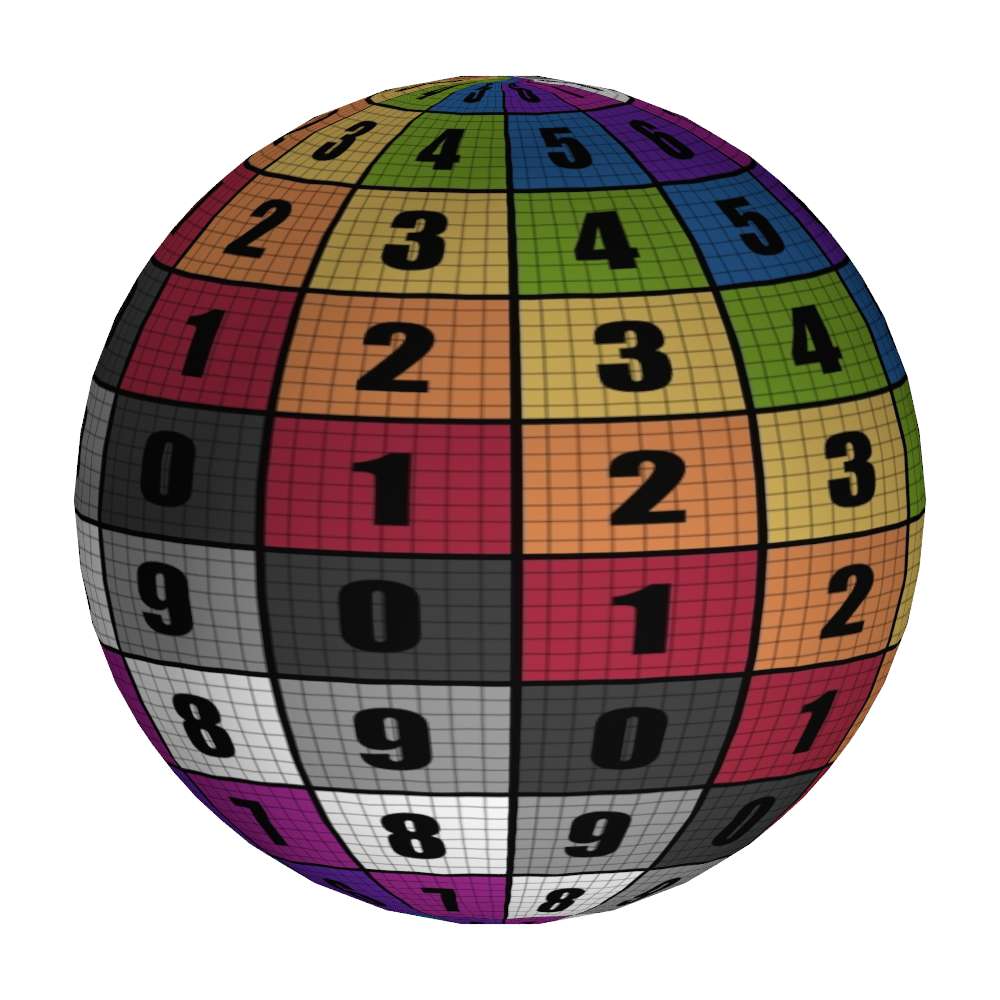}&
		\includegraphics[width=.16\textwidth]{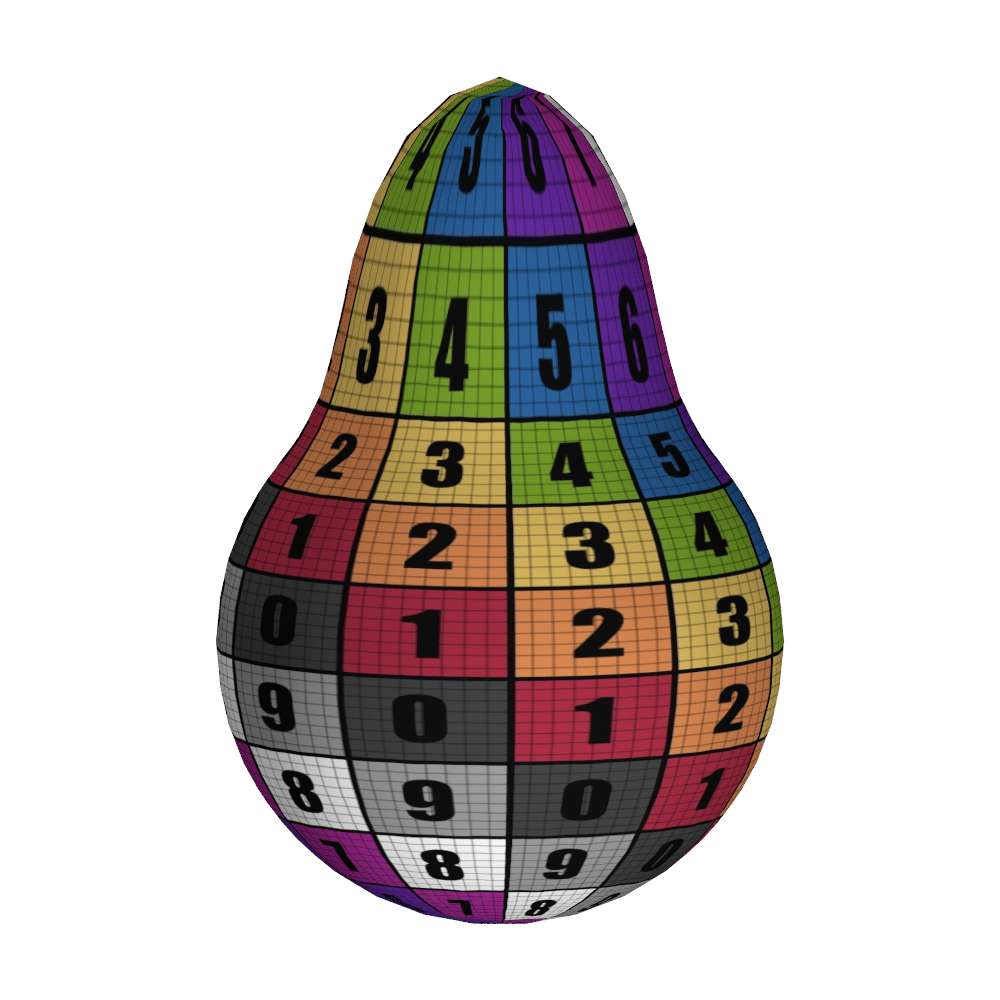}&
		\includegraphics[width=.16\textwidth]{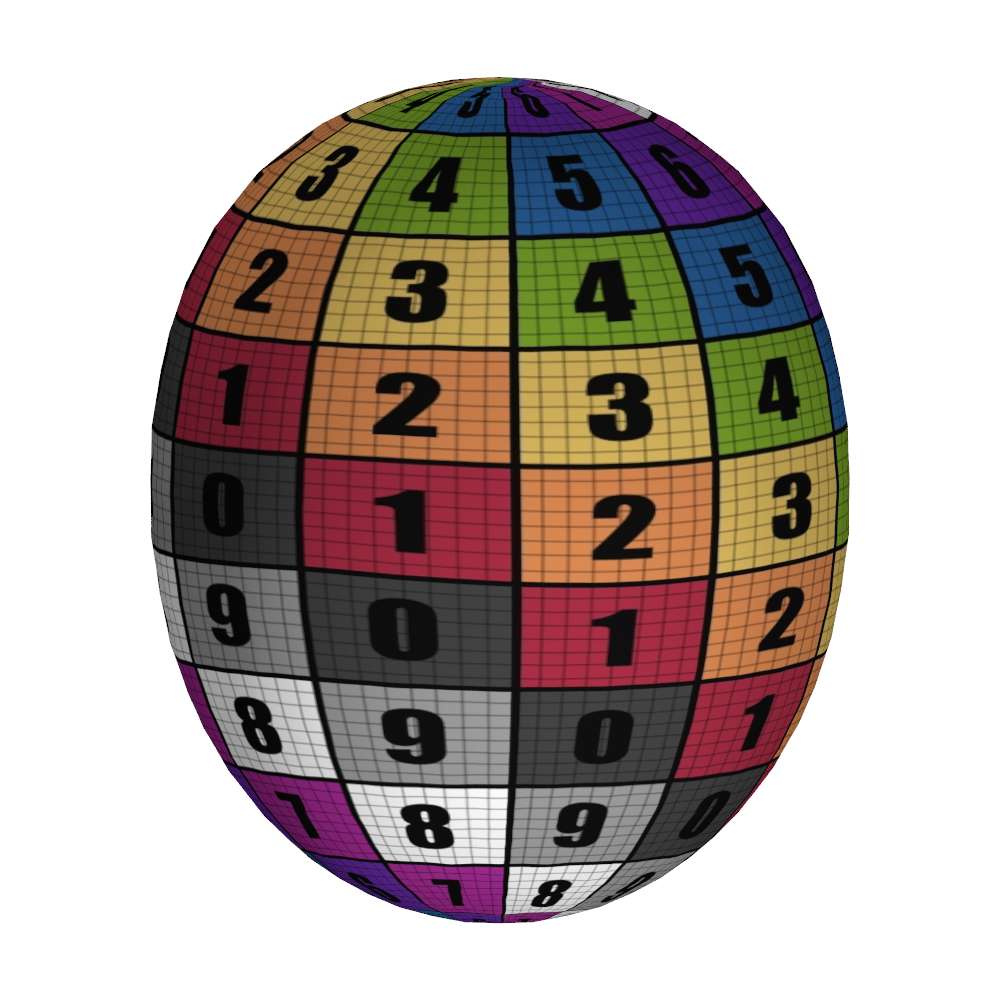}&
		\includegraphics[width=.16\textwidth]{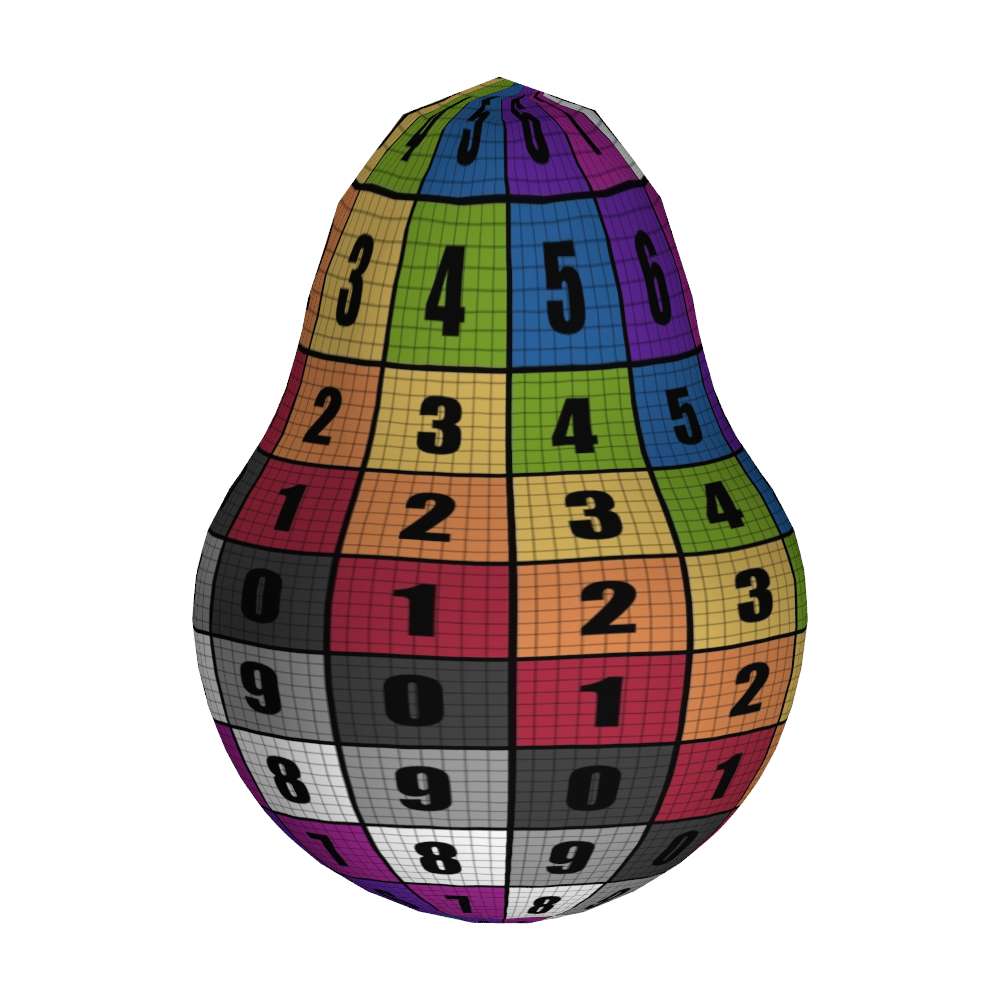}&
		\includegraphics[width=.16\textwidth]{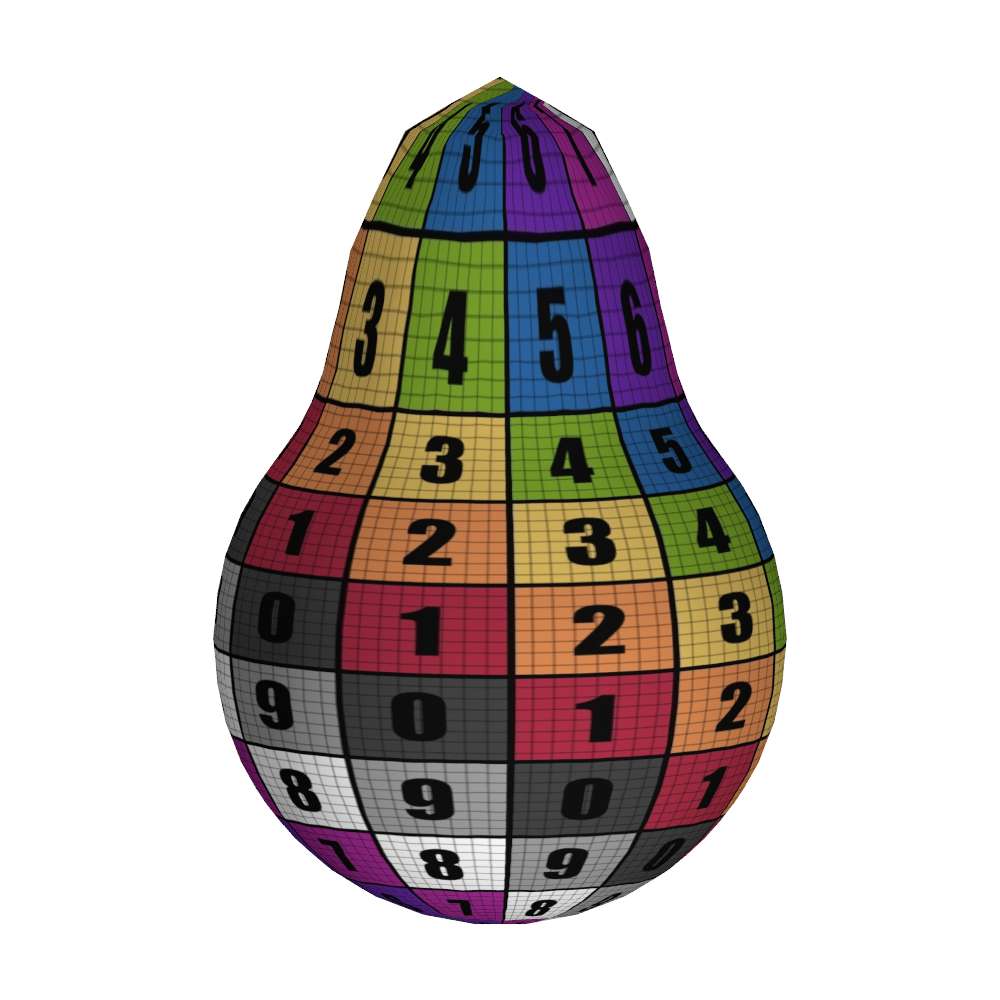}&
		\includegraphics[width=.16\textwidth]{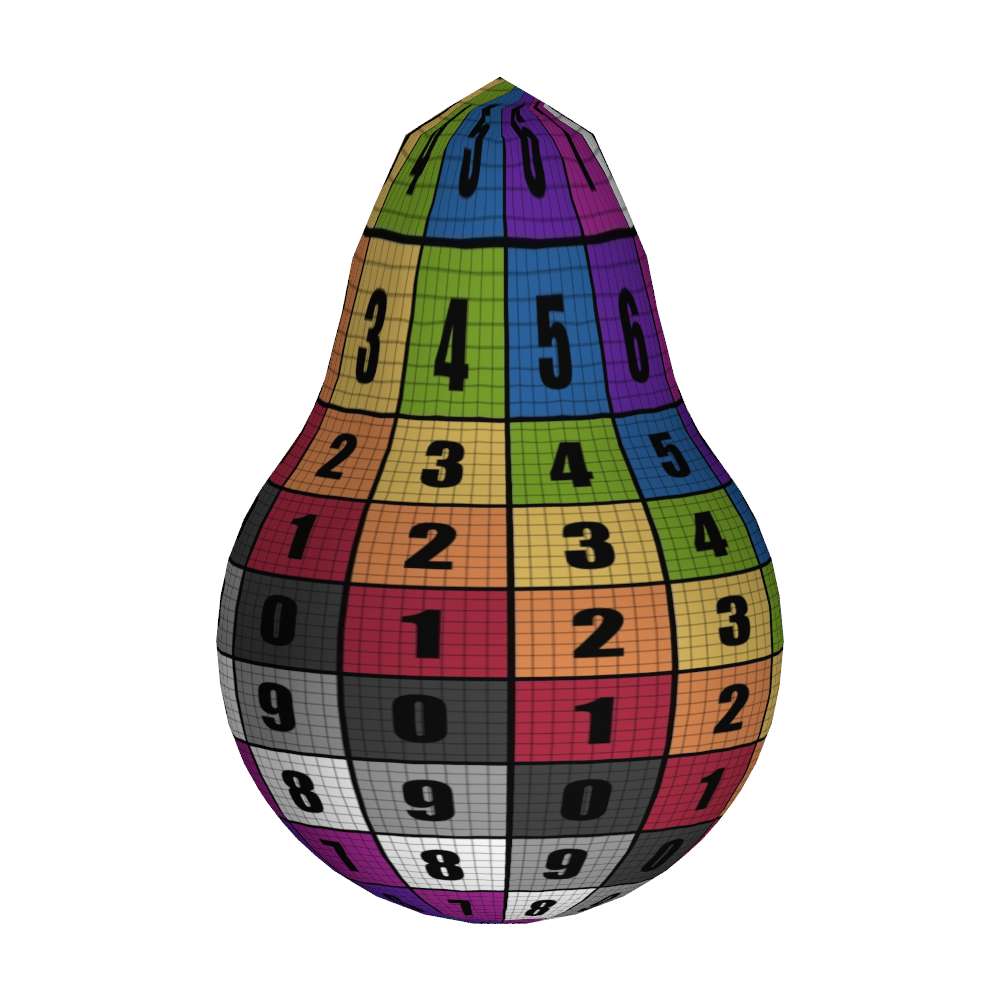}\\
          \includegraphics[width=.11\textwidth]{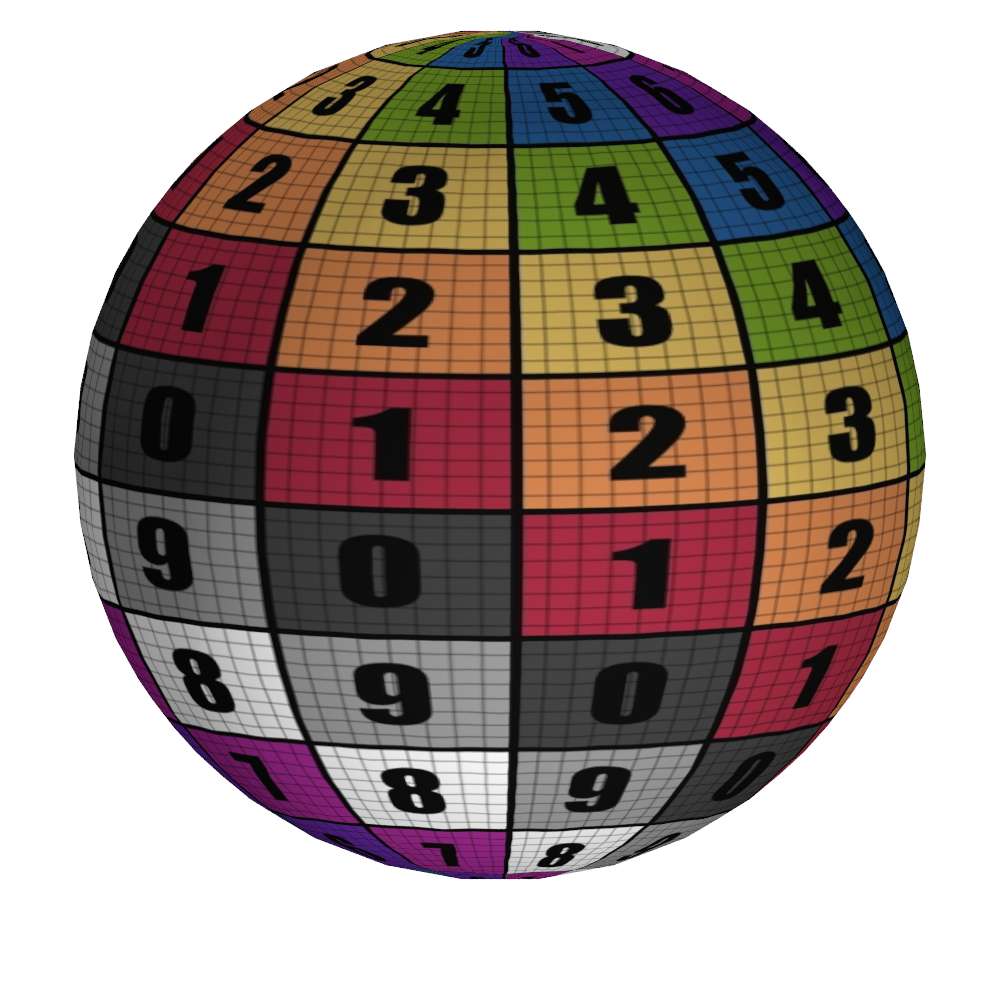}&
		\includegraphics[width=.16\textwidth]{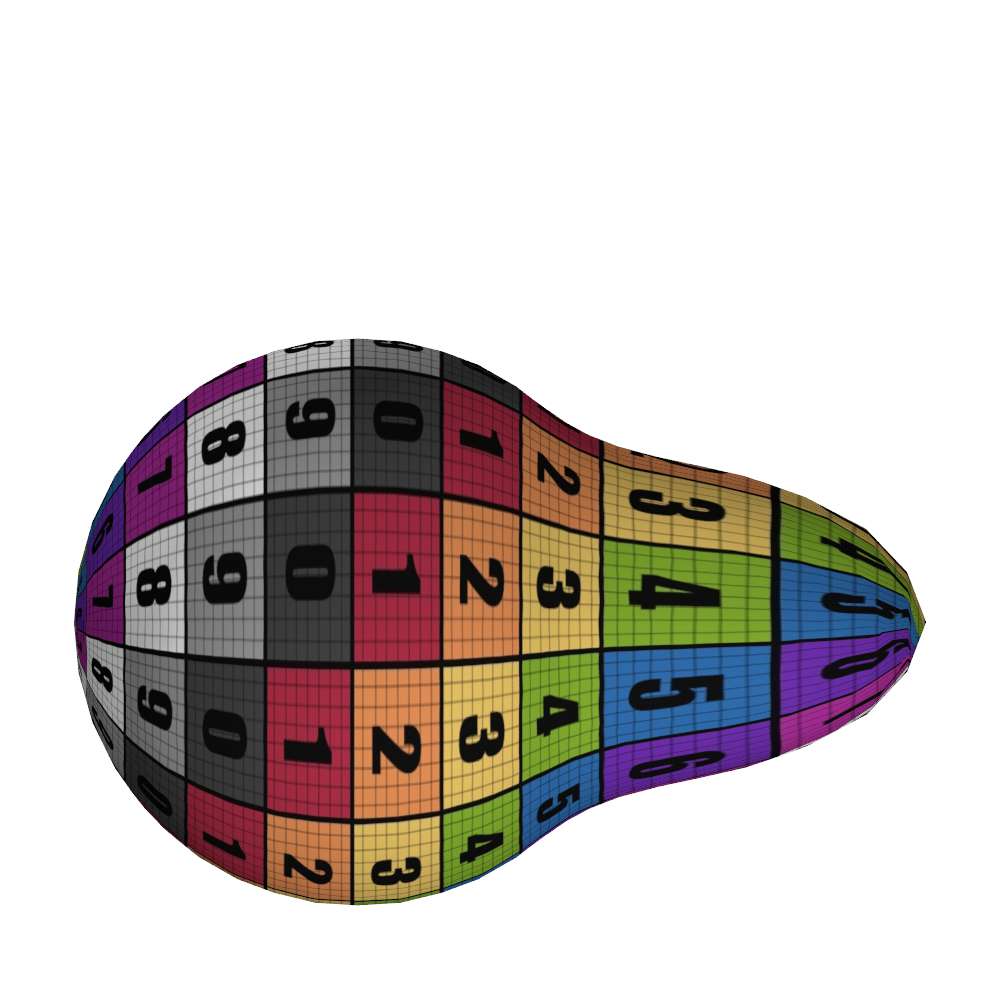}&
		\includegraphics[width=.16\textwidth]{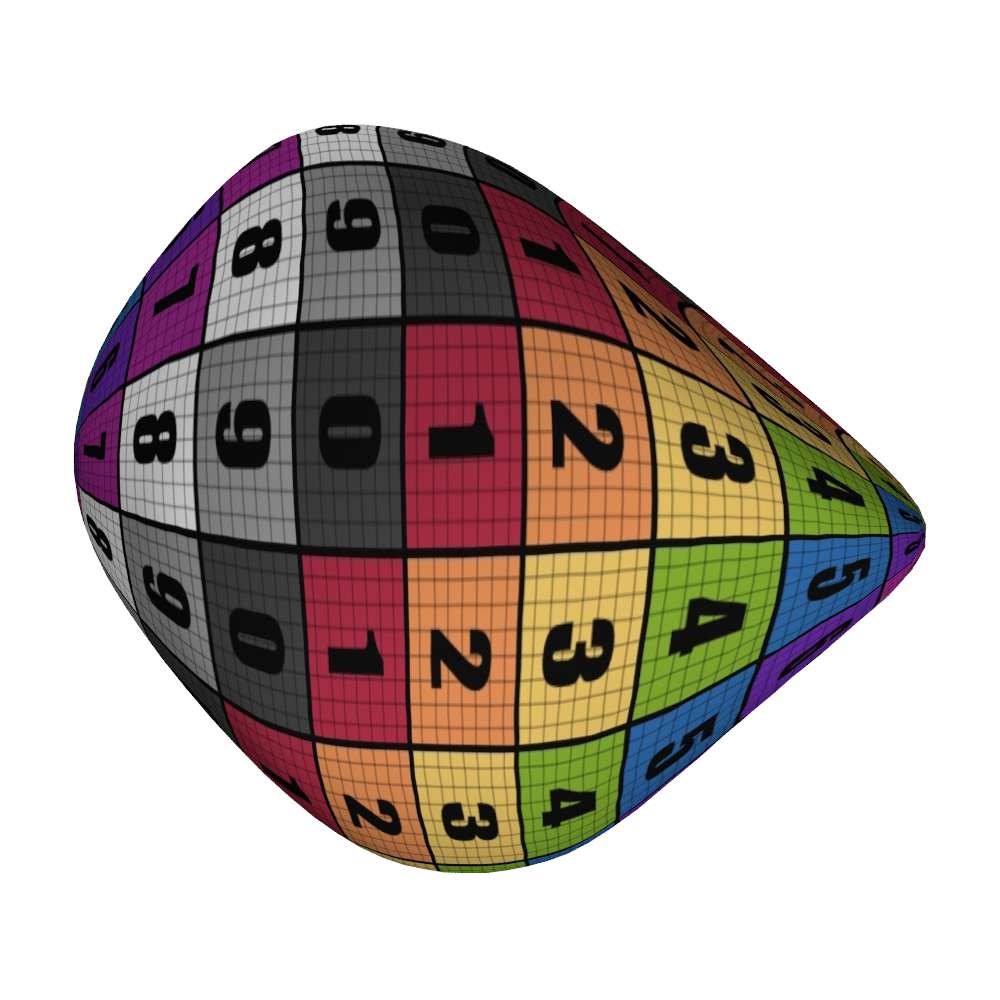}&
		\includegraphics[width=.16\textwidth]{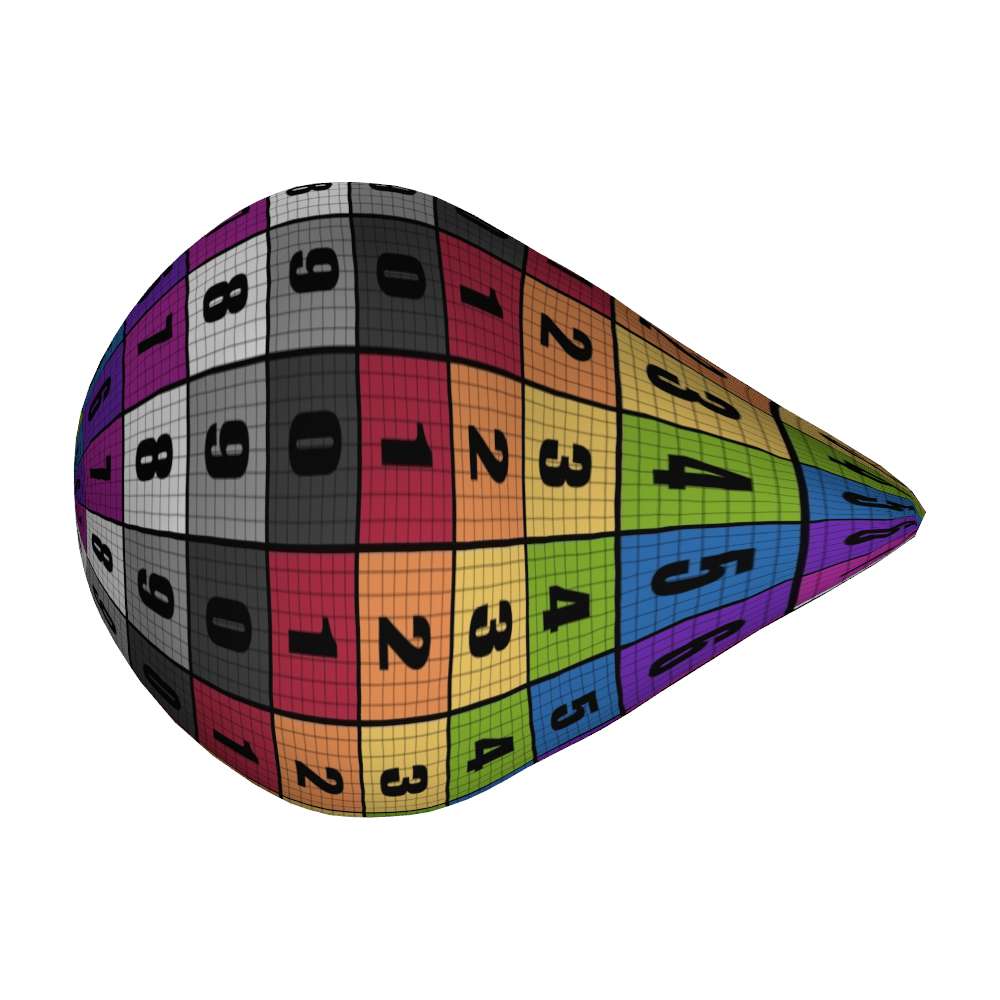}&
		\includegraphics[width=.16\textwidth]{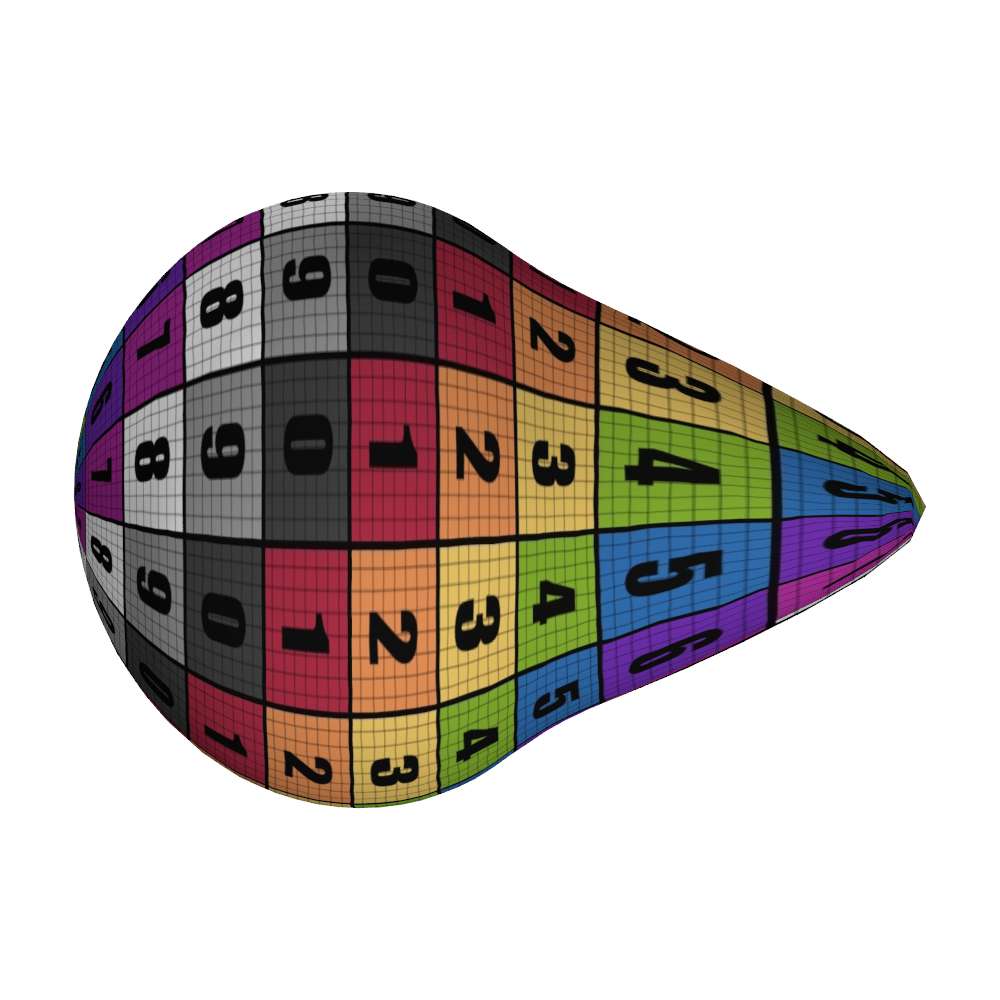}&
		\includegraphics[width=.16\textwidth]{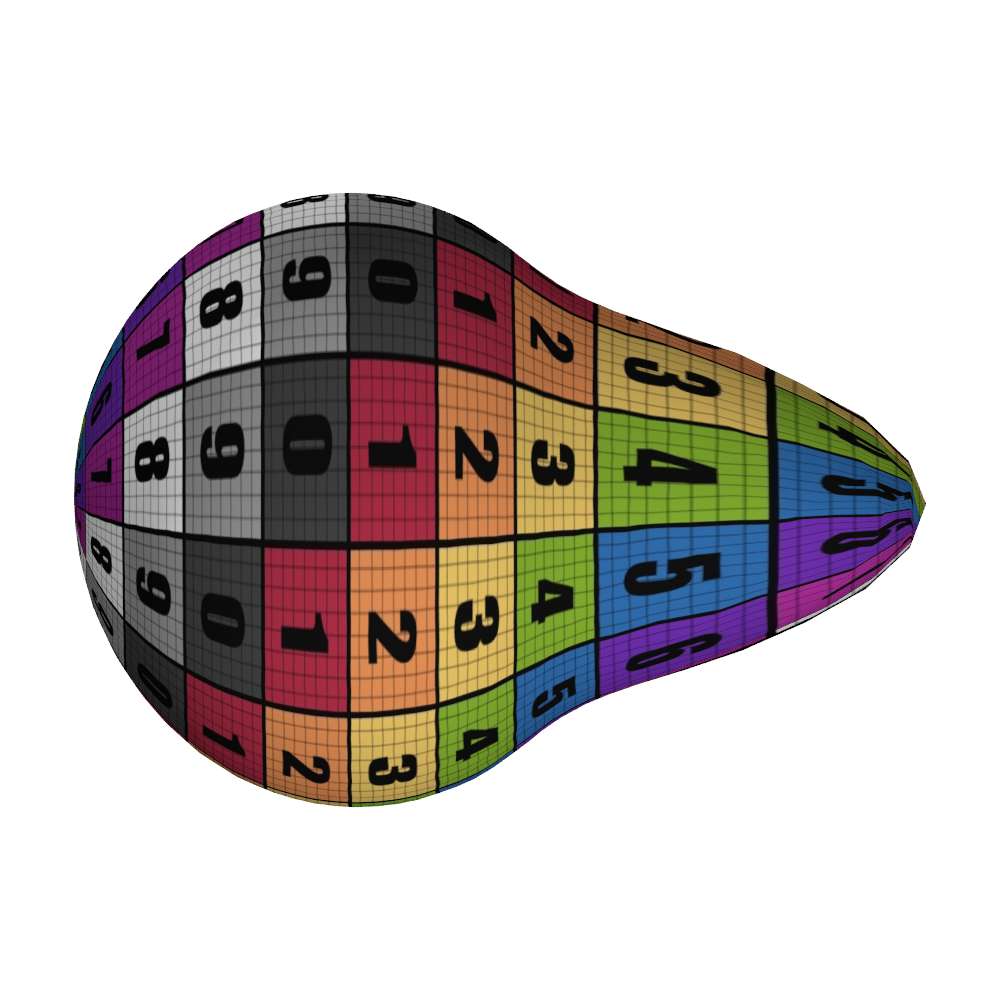}\\
		Source & Target & $10$ & $30$ & $60$ & $100$ \\
	\end{tabular}
        \vspace{-1mm}
	\caption{Two deformation fields (left) resulting from the correspondence encoded
          via the texture map are represented as operators with increasing number of basis
          functions and then recovered by solving a least-squares problem. Increasing the
          size of the basis leads to more accurate representation of high-frequency
          deformations. Note, our representation can be used to successfully recover almost
          all deformation fields (up to \emph{infinitesimal} rigid deformations) including global
          rotation (bottom row) regardless of the scale of the deformation.\vspace{-3mm}}
	\label{fig:basisSize}
\end{figure*}

{Figures~\ref{fig:jointVF} and \ref{fig:teaser} present an example of joint deformation
  design. Namely, we impose a set of directional constraints $U(p_j) = u_j$ and $V(q_j) = v_j$ on
  two different shapes $M$ and $N$ and we solve for two deformation fields, one on each shape, that
  are ``informed'' by the deformation of the other shape. On a single shape, our objective is
  designed to promote smoothness of the resulting deformation field and sparsity in the coefficients
  of the deformation field basis:
\begin{align*}
	\mathbb{E}_M(\alpha) := \sum_{i,j} \alpha_i \langle U_i, \Delta_M U_j \rangle \alpha_j + \tau \|\alpha\|_1 .
\end{align*}
Therefore, on a single shape, the optimization becomes:
\begin{align*}
	& \min_{\alpha} \, \mathbb{E}_M(\alpha) , %= \sum_{i,j} \alpha_i \langle U_i, \Delta_M U_j \rangle \alpha_j + \tau \|\alpha\|_1, \\
	& \text{s.t.} \, \sum_i \alpha_i U_i(p_j) = u_j ,
\end{align*}
where the constraints enforce the given pointwise directions.}

To design the deformation fields jointly, we propose to find a field $U$ on shape $M$ and $V$ on shape $N$ such that for a given functional map $C$ we have $\diffU{U} C \approx C \diffU{V}$ while respecting the local constraints on the respective shape. The resulting optimization problem reads:
\begin{align*}
	& \min_{\alpha,\beta} \, \| \sum_i \alpha_i \diffU{U_i} C - C \sum_j \beta_j  \diffU{V_j} \|_F^2 + \mathbb{E}_M(\alpha) + \mathbb{E}_N(\beta) \\
	& \text{s.t.} \, \sum_i \alpha_i U_i(p_j) = u_j , \sum_i \beta_i V_i(q_j) = v_j .
\end{align*}

As a result, the constraints as well as the structure of one shape is transferred onto the other. Moreover the area that could lead to contradictory deformation remains still.

\subsection{Functional Deformation transfer} \label{exp:defTrans}

\begin{figure}[t!]
	\centering
	\begin{tabular}{cc|ccc}
		\rotatebox{90}{Source}&
		\includegraphics[width=.16\columnwidth]{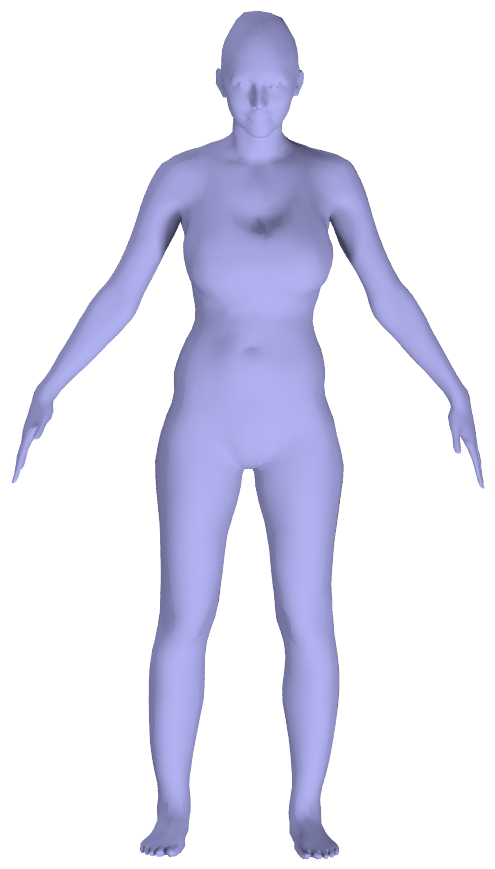}&
		\includegraphics[width=.16\columnwidth]{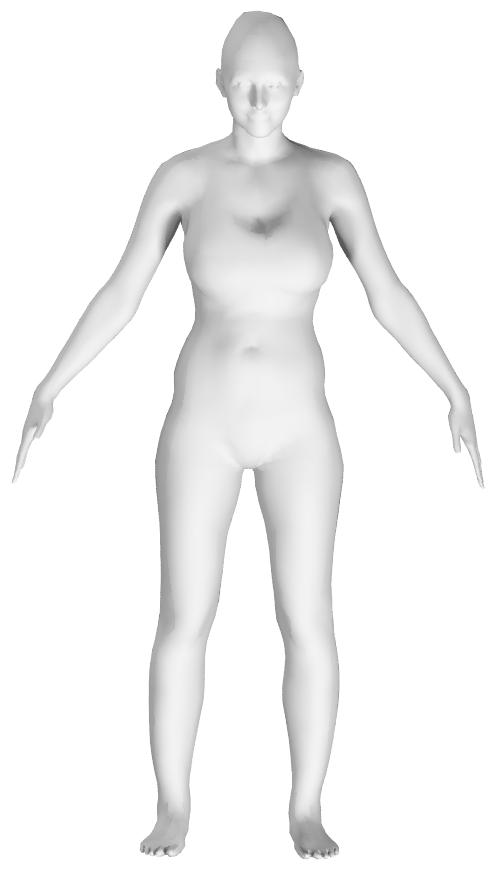}&
		\includegraphics[width=.16\columnwidth]{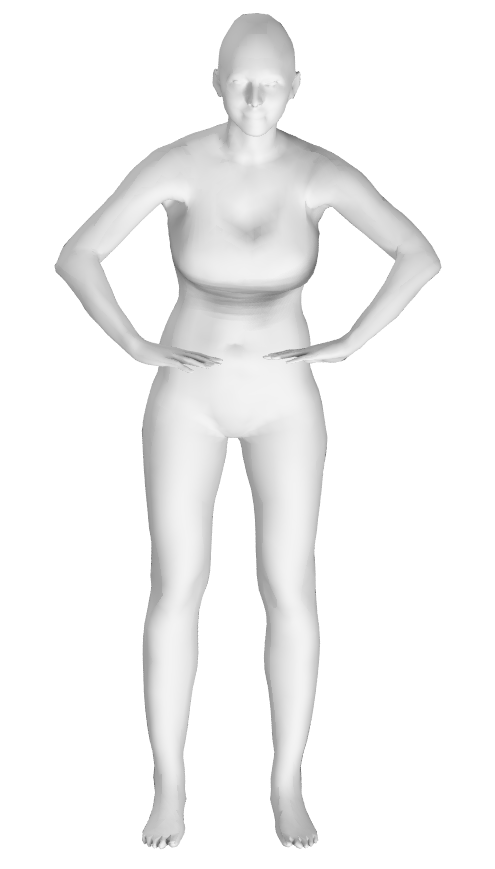}&
		\includegraphics[width=.16\columnwidth]{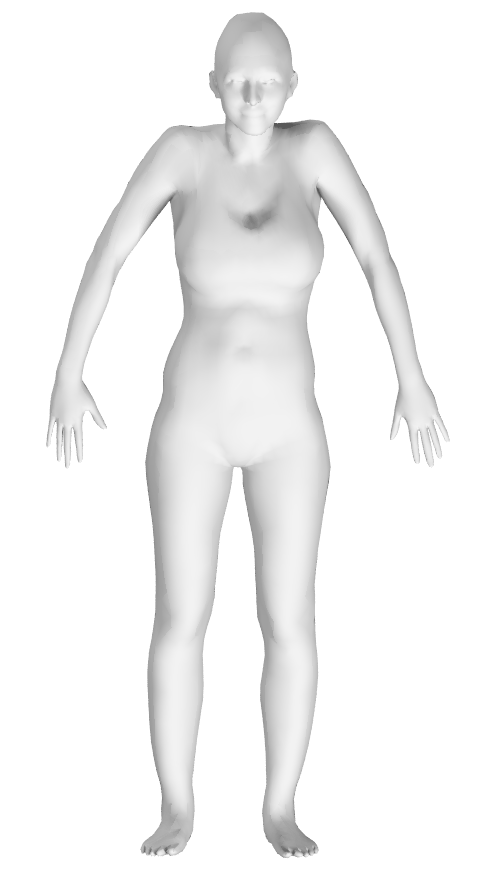}\\
		\rotatebox{90}{Target}&
		\includegraphics[width=.16\columnwidth]{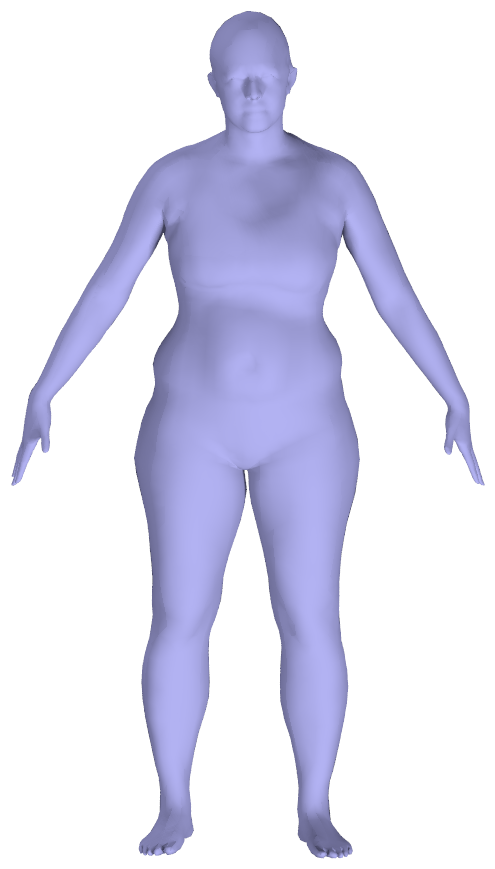}&
		\includegraphics[width=.16\columnwidth]{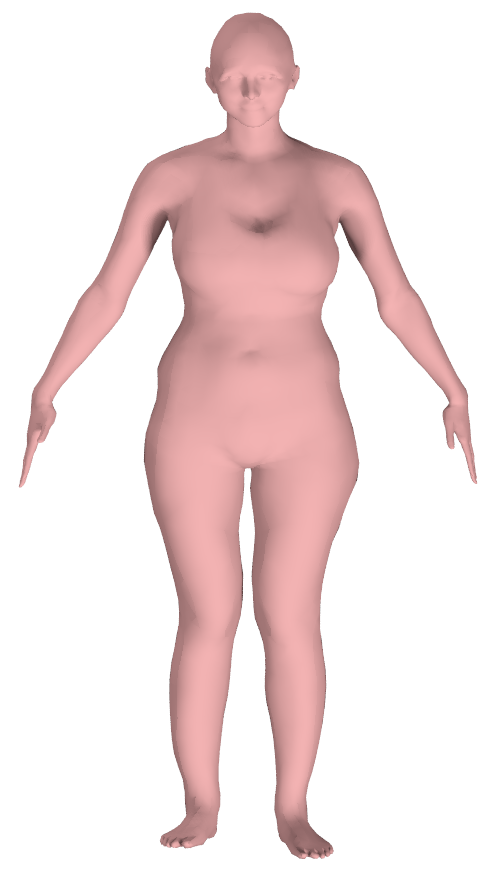}&
		\includegraphics[width=.16\columnwidth]{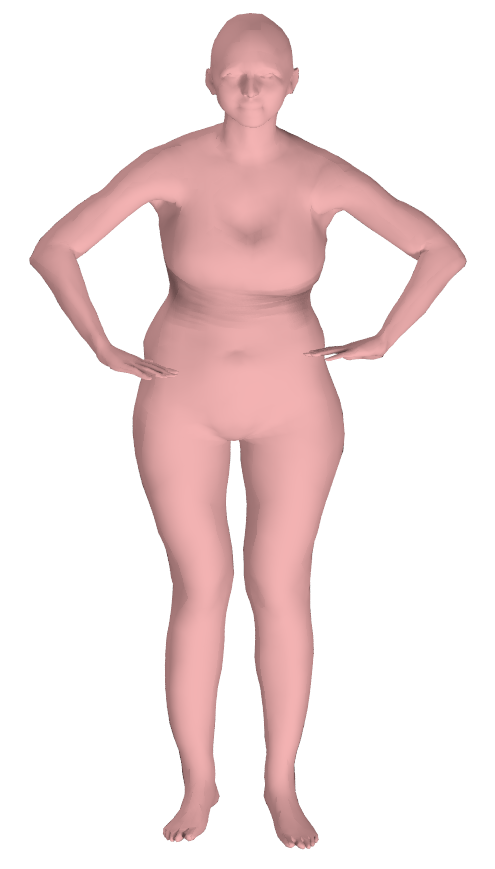}&
		\includegraphics[width=.16\columnwidth]{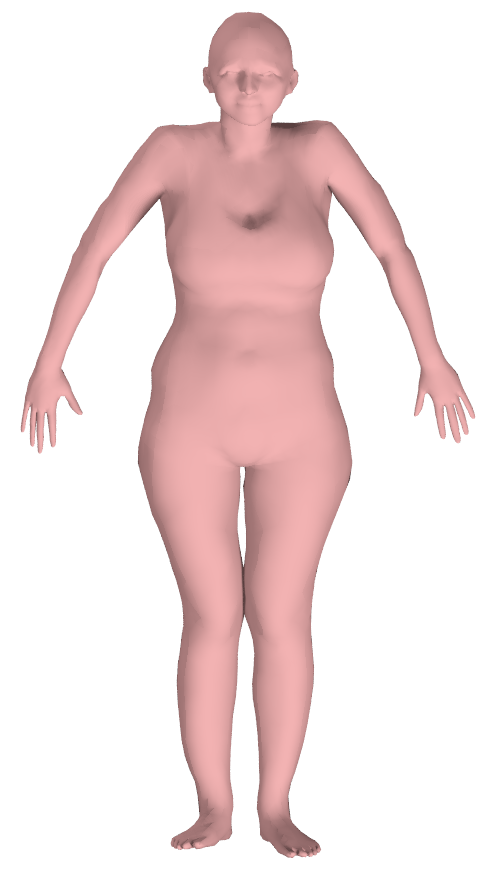}\\
		\rotatebox{90}{In collection}& &
		\includegraphics[width=.16\columnwidth]{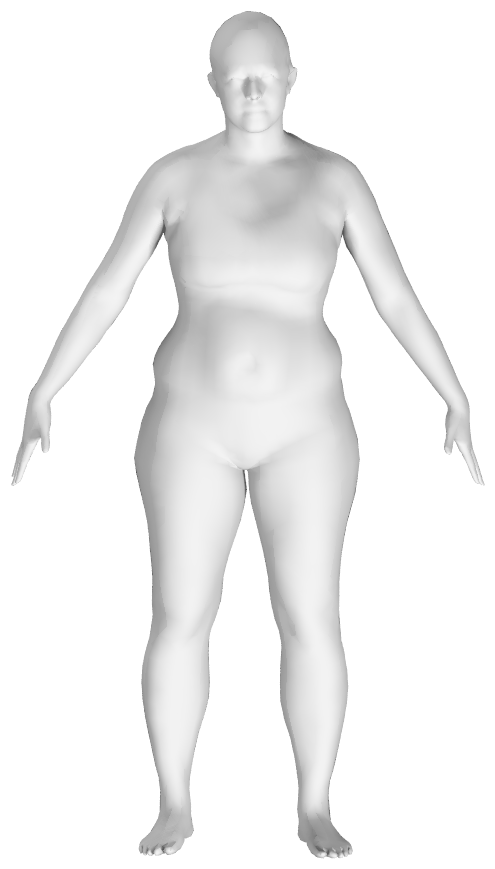}&
		\includegraphics[width=.16\columnwidth]{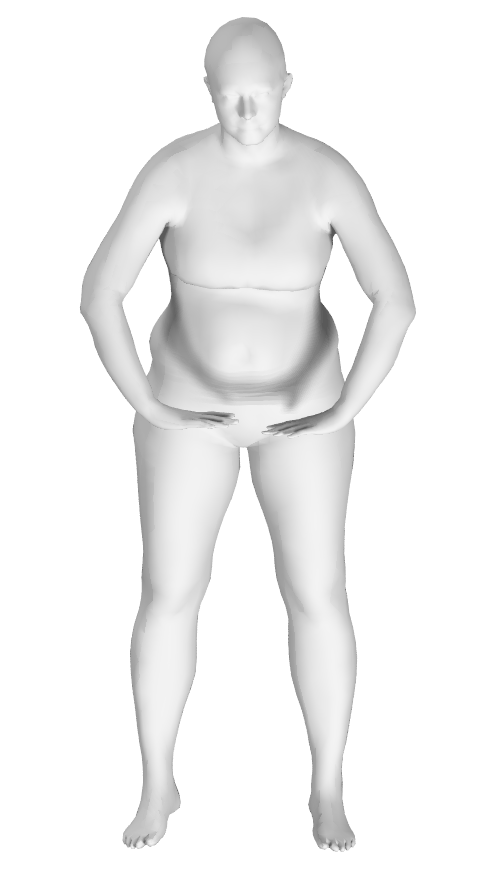}&
		\includegraphics[width=.16\columnwidth]{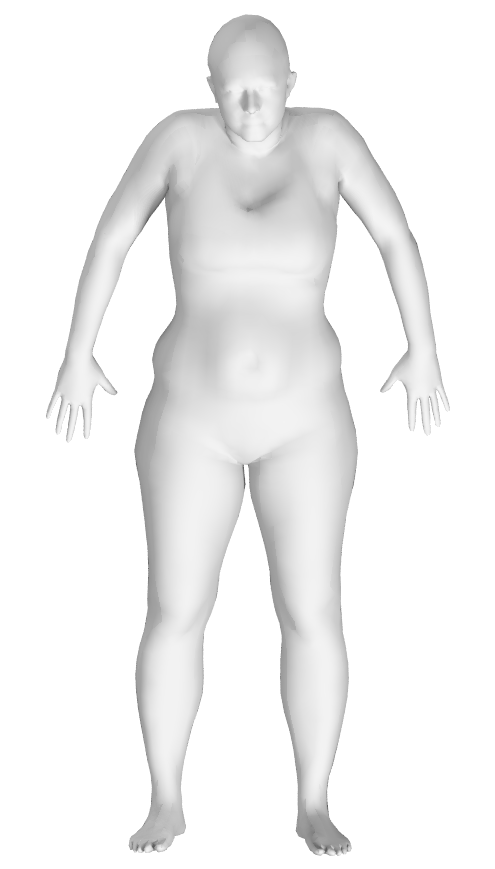}\\
	\end{tabular}
\vspace{-1mm}
	\caption{The deformation field defined by the blue shapes (first column) is transferred to the same shape in different poses (top white shapes). While the style is consistent across the poses (red shapes) some details of the deformation are lost due to the basis representation. The style transfer are compared to the corresponding shape in the collection (bottom white shapes).\vspace{-2mm}}
	\label{fig:faustTransfer}
\end{figure}

Given a deformation field $U$ on shape $M$ represented as an operator and a functional map $C$ from $N$ to shape $M$, we can use our method to transfer the deformation to an arbitrary mesh. The transferred deformation $V = \sum_i \alpha_i V_i$ on shape $N$ by solving: \begin{align}
	\min_\alpha \, \| \diffU{U} C - C \sum_i \alpha_i \diffU{V_i} \|^2 + \tau \| \alpha \|_1 .
	\label{eq:optiTransfer}
\end{align}
In all of our experiments below, we represent the linear operators $C$ and $\diffU{V}$ in a reduced functional basis, consisting of the first $200$ eigenfunctions of the Laplace-Beltrami operator. We parameterize the space of deformations by computing the $180$ extrinsic vector fields in each of the three categories described in Section \ref{sec:results} to build an over-complete dictionnary. This implies that the number of unknowns is relatively small: $\alpha \in \mathbb{R}^{540}$ and is independent of the resolution of the underlying mesh. We choose the parameter $\tau$, controlling the sparsity of the representation, to be $10^{-4}$ times the largest singular value of the linear map $V \mapsto C \diffU{V}$.

We solve the optimization problem described in Eq. \eqref{eq:optiTransfer} with CVX \cite{cvx}, using the default approach based on the interior point method.

\begin{figure}[t!]
	\centering
	\begin{tabular}{ccc}
		\includegraphics[width=.25\columnwidth]{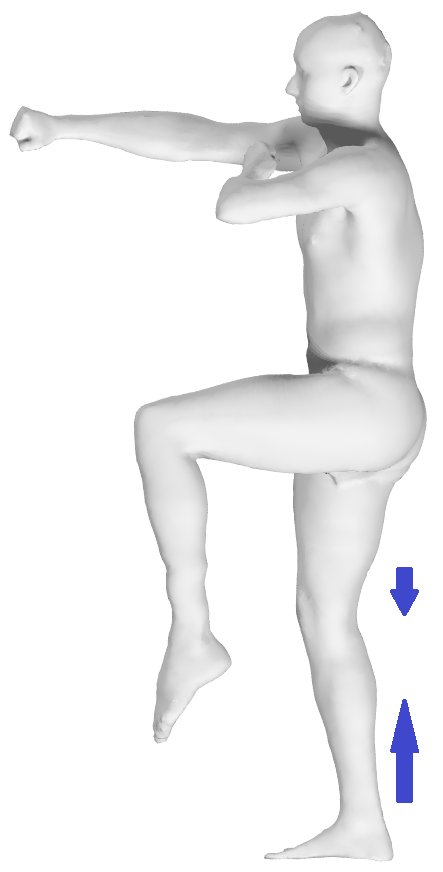}&
		\includegraphics[width=.25\columnwidth]{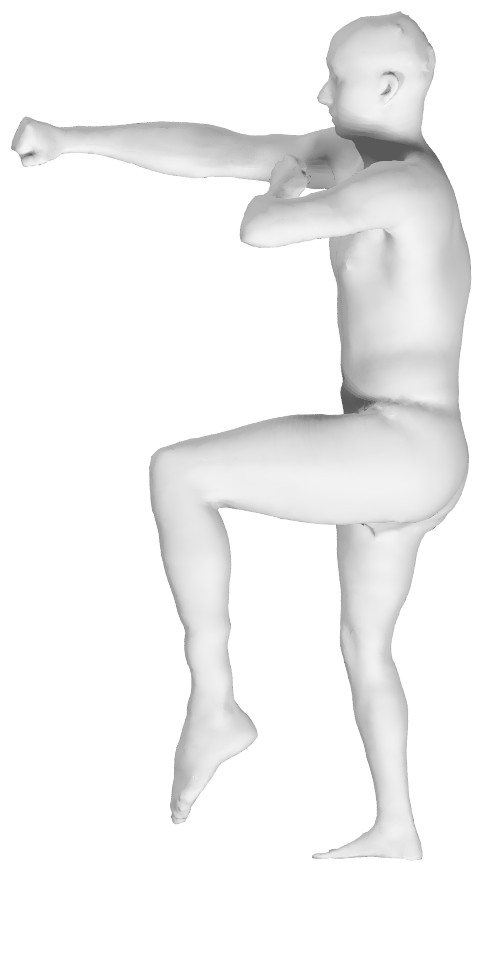}&
		\includegraphics[width=.25\columnwidth]{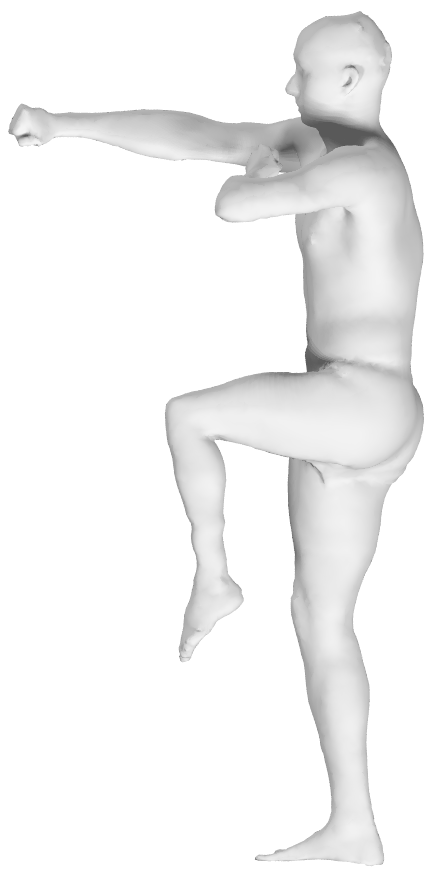}\\
		&%\includegraphics[width=.25\columnwidth]{Images/SymTransfer/mesh004_init}&
		\includegraphics[width=.25\columnwidth,trim={0 0 0 7cm},clip]{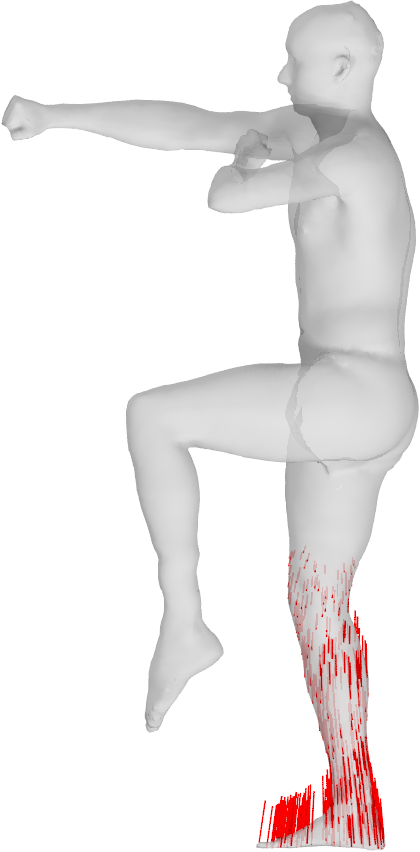}&
		\includegraphics[width=.25\columnwidth,trim={0 0 0 12cm},clip]{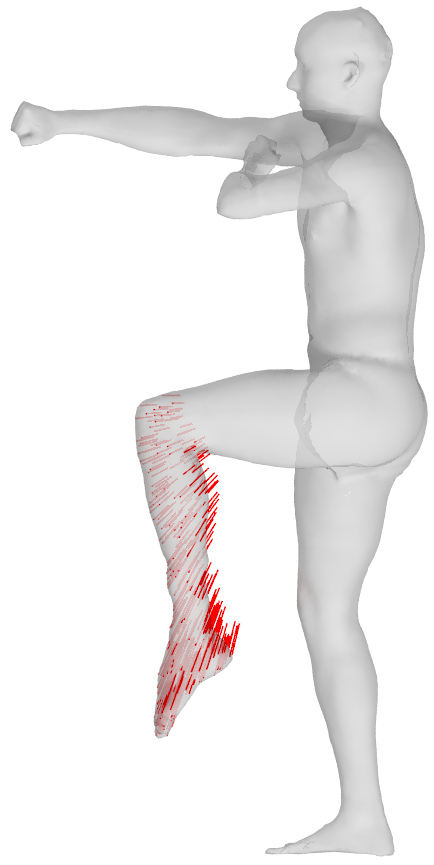}\\
		Initial shape & Deformation & Symmetric Def.
	\end{tabular}
\vspace{-1mm}
	\caption{An initial deformation (first two columns), corresponding to the shrinking of the right leg of a human model, is transferred to the left leg by imposing the commutativity between the infinitesimal shape difference and the symmetry map. Both legs are in different position so the transfer has to adapt to the geometry. \vspace{-1mm}}
	\label{fig:symTransfer}
\end{figure}

Using this setup we solve different instances of the deformation transfer problem, namely:
\begin{itemize}
%	\item Pose transfer: a reference shape is deformed to other poses and this deformation is transferred to a third mesh of different style (Figures~\ref{fig:armadilloTransfer} and \ref{fig:faceTransfer}).
	\item Style transfer: we transfer style across poses. Here, given two different shapes in a rest pose and a deformed version of one of them, we transfer the deformation to the other shape (Figure \ref{fig:faustTransfer}). This also shows that our vector field collection is not limited to a specific type of deformation.
	\item Symmetry transfer: we transfer a deformation from a shape onto itself using a symmetry map (Figure~\ref{fig:symTransfer}). Note that this task cannot be achieved with standard Jacobian-based methods such as \cite{sumner2004deformation}.
\end{itemize}

{We stress that although enabled by our representation, this is by no means the central application
  and therefore the results presented below simply serve as an illustration of the functionality
  that can be achieved using our functional deformation fields.}
%\paragraph{Pose transfer}
%Figure~\ref{fig:armadilloTransfer} shows an example of deformation transfer from an animation consisting of $20$ frames of a waving armadillo to armadillos with different connectivity and in a different starting position. The transferred animation induces a similar metric distortion across different shapes resulting in a consistent deformation. In Figure~\ref{fig:faceTransfer} we present a similar experiment for a collection of faces. The facial expressions of the first row are transferred to another face on the second row. For reference, we compare the results to the corresponding faces present in the collection.
%
%Unlike the approach of \cite{sumner2004deformation} our deformation transfer does not require a
%triangle-to-triangle map and is achieved using only an approximate functional map. Moreover the
%meshes do not need to be oriented or positioned consistently since only the metric changes are
%transferred as illustrated in the third row of Figure \ref{fig:armadilloTransfer}.

\paragraph{Style transfer}
We use our approach to transfer style across the poses of different shapes in the Faust dataset
\cite{Bogo:CVPR:2014}, shown in Figure \ref{fig:faustTransfer}. Here first consider the deformation
field $U$ given by the point displacements across two different shapes in approximately the same
reference pose. We then use our framework to transfer $U$ to another shape in a different pose and
with different mesh structure.  In Figure \ref{fig:faustTransfer} our method consistently preserves
the global structure, although some high frequency details of the deformation are lost due to
the projection onto a vector field  basis. %The deformation transfer for faces in Figure \ref{fig:faceTransferStyle} is successful because the deformation is very smooth and therefore well-represented in the basis.

\paragraph{Symmetry transfer}
One interesting feature of the functional representation of deformation fields is that it is ``shape aware.'' For example in Figure \ref{fig:symTransfer} we transfer the shrinking of the right leg to the left leg by looking for the operator which commutes with the operator representation of the symmetry map. Since both legs are in different positions this transfer is not easy to achieve by a simple point-to-point transfer of the vector field or even by transferring it using local coordinates. As shown in Figure \ref{fig:symTransfer} bottom row, our transferred deformation field adapts to the geometry. 

% Unlike the approach of \cite{sumner2004deformation} our deformation transfer does not require a triangle-to-triangle map and is achieved using only an approximate functional map. Moreover the meshes do not need to be oriented or positioned consistently since only the metric changes are transferred. Thus, our method naturally overcomes triangle flipping issues, unlike \cite{sumner2004deformation} which does not handle symmetries properly. 

\subsection{Relation to existing techniques}
{An important property of our deformation transfer algorithm is that it
  relies fully on the deformation of the \textit{metric}. This makes it
  fundamentally different from the spectral pose transfer described in
  \cite{levy06}. Those methods use the strong stability of the first
  eigenfunctions of the Laplace-Beltrami under deformation. Thus, a deformation
  field $V$ can be efficiently transferred by projecting its components into a
  reduced eigen-basis $\Phi$ of the initial shape and reconstructed using the
  basis $\Psi$ of the shape to be deformed. The new embedding $X'$ is computed
  from the old embedding $X$ simply by $X' := X + \Psi \Phi^\top V .$}

\begin{figure}
	\centering
\includegraphics[width=1.02\linewidth]{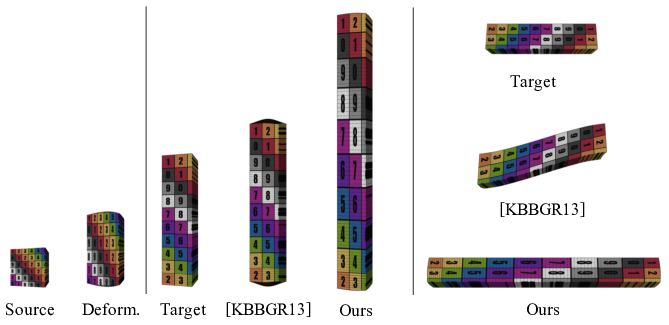}
	% \begin{tabular}{cc|ccc}
	% 	\includegraphics[width=.08\columnwidth]{Images/CoupledBasis/bar2}&
	% 	\includegraphics[width=.08\columnwidth]{Images/CoupledBasis/bar2_deform}&
	% 	\includegraphics[width=.08\columnwidth]{Images/CoupledBasis/bar1}&
	% 	\includegraphics[width=.08\columnwidth]{Images/CoupledBasis/bar1_coupled}&
	% 	\includegraphics[width=.08\columnwidth]{Images/CoupledBasis/bar1_transfer}\\
	% 	Source & Deform. & Target & [KBBGR13] & Ours\\
	% 	\\[-1.0em]%
	% 	 & &
	% 	\multicolumn{3}{c}{\begin{tabular}{c}
	% 	\includegraphics[height=.08\columnwidth]{Images/CoupledBasis/bar1_rot}\\
	% 	Target\\
	% 	\includegraphics[height=.16\columnwidth]{Images/CoupledBasis/bar1_rot_coupled}\\
	% 	~[KBBGR13]  \\
	% 	\includegraphics[height=.08\columnwidth]{Images/CoupledBasis/bar1_rot_transfer}\\
	% 	Ours
	% 	\end{tabular}}
	% \end{tabular}
	\caption{{An initial deformation (first two columns), corresponding to the expansion
            of a bar is transferred to another longer bar. Using the method described
            in~\protect\cite{kovnatsky2013coupled} ([KBBGR13]) the additional height is exactly the same as in the original deformation. Using our method the deformation is indexed on the metric thus the height of the model doubles. Furthermore, our method is rotation invariant (second row).} \vspace{-3mm}}
	\label{fig:coupledBasis}
\end{figure}

  {Recent improvements of this technique
    \cite{kovnatsky2013coupled,yin2015spectral} include pre-alignment of the
    spectral basis but the shortcoming are essentially the
    same. Figure~\ref{fig:coupledBasis} shows that this deformation transfer is
    by definition extrinsic, orientation dependent and furthermore completely
    agnostic to the intrinsic structure of the shape. Our method in contrast
    is rotation-invariant and directly linked to the induced changes in the geometry.}

{We also compare our method with the algorithm for deformation transfer described in
\cite{sumner2004deformation}. This method is based on reallocating Jacobian matrices defined on
triangles of the source mesh to those of the target mesh. This method, however, is not without
limitations. First, this transfer does not take into account changes in orientation from the source
to the target thus ruling out any possibility of symmetric transfer and requiring a pre-alignment of
the source and target meshes. This can be challenging to achieve in practice in case of non-rigid
deformations (e.g. Figure~\ref{fig:symTransfer}). Secondly, it assumes as input a
triangle-to-triangle map which can be cumbersome to obtain.}

These limitations do not apply to our representation as our approach is based on transferring metric
information, and is therefore immune to changes of orientation. Moreover, instead of a
triangle-to-triangle map, an approximate functional map is enough. Furthermore, note that although
in general reconstructing geometry from metric tensors is more difficult than a reconstruction from
Jacobians as local rotations are no longer available (see e.g. \cite{boscaini2015shape}) our method
relies on solving a moderately-sized convex optimization problem.

\begin{figure}[t!]
	\centering
	\begin{tabular}{ccc|cc}
		\includegraphics[width=.15\columnwidth]{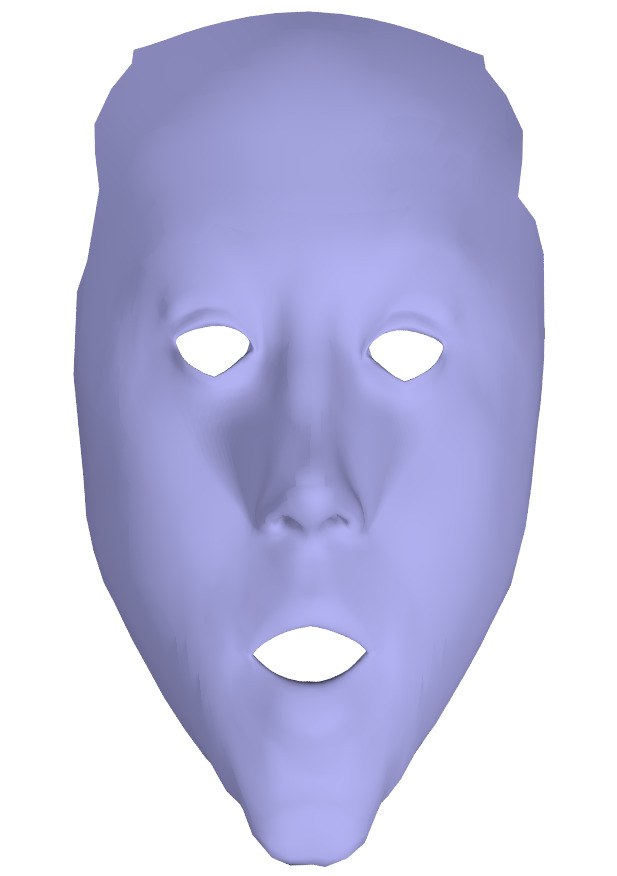}&
		\includegraphics[width=.15\columnwidth]{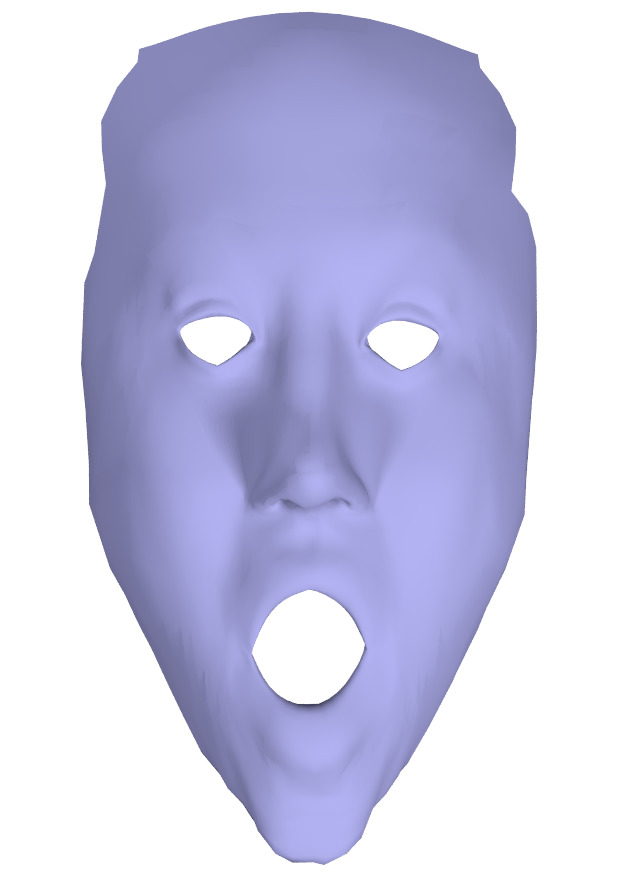}&
		\includegraphics[width=.15\columnwidth]{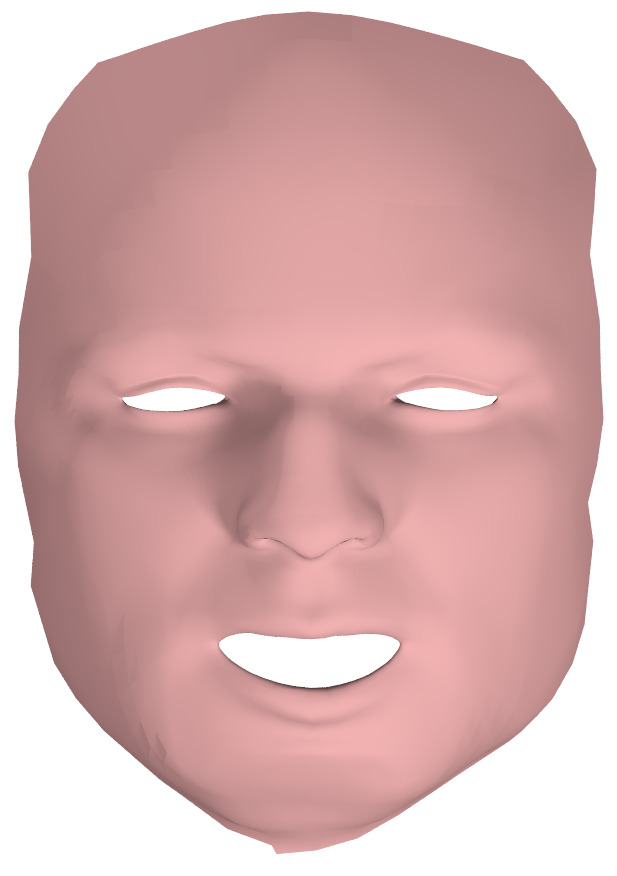}&
		\includegraphics[width=.15\columnwidth]{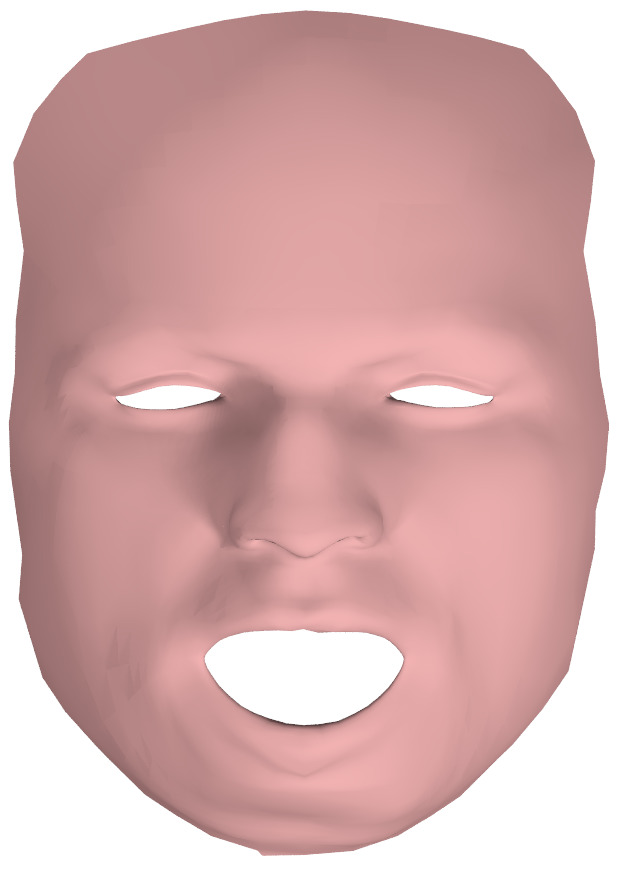}&
		\includegraphics[width=.15\columnwidth]{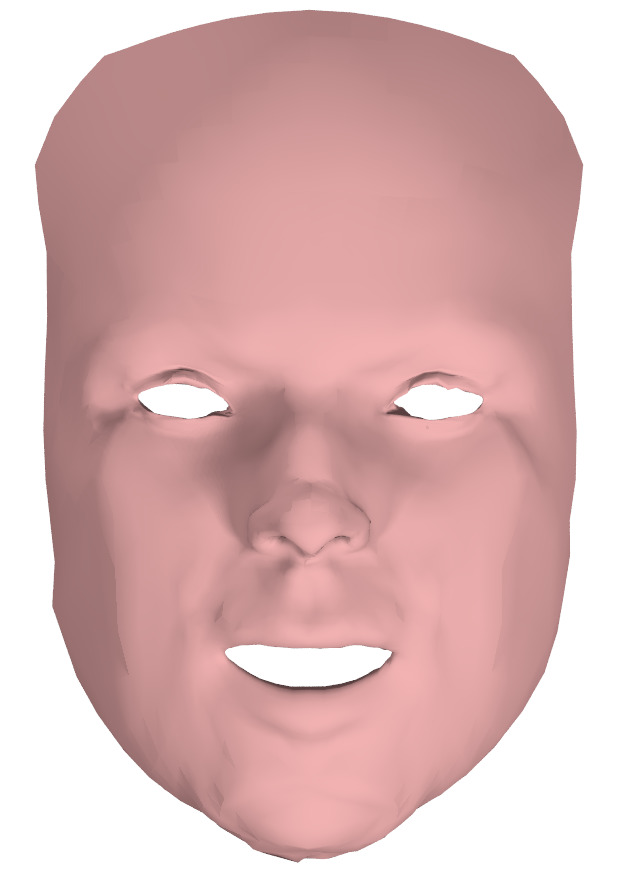}\\
		Source & Deformed & Target & Ours & Sumner et al. \\
	\end{tabular}
	\begin{tabular}{ccccc}
%		\includegraphics[width=.15\columnwidth]{Images/Sumner/Face_hld}&
%		\includegraphics[width=.15\columnwidth]{Images/Sumner/Face_hld_3}&
%		\includegraphics[width=.15\columnwidth]{Images/Sumner/Face_tar}& \\
%		Source & Deformed & Target & & \\
%		\hline
%		\includegraphics[width=.15\columnwidth]{Images/Sumner/Face3_transfer_100}&
%		\includegraphics[width=.15\columnwidth]{Images/Sumner/Face3_transfer_75}&
%		\includegraphics[width=.15\columnwidth]{Images/Sumner/Face3_transfer_50}&
%		\includegraphics[width=.15\columnwidth]{Images/Sumner/Face3_transfer_25}&
%		\includegraphics[width=.15\columnwidth]{Images/Sumner/Face3_transfer_0}\\
%		$0$ & $0.25$ & $0.5$ & $0.75$ & $1$ \\
%		\multicolumn{5}{c}{Ours} \\
%		\includegraphics[width=.15\columnwidth]{Images/Sumner/Face3_sumner_100}&
%		\includegraphics[width=.15\columnwidth]{Images/Sumner/Face3_sumner_75}&
%		\includegraphics[width=.15\columnwidth]{Images/Sumner/Face3_sumner_50}&
%		\includegraphics[width=.15\columnwidth]{Images/Sumner/Face3_sumner_25}&
%		\includegraphics[width=.15\columnwidth]{Images/Sumner/Face3_sumner_0}\\
%		$0$ & $0.25$ & $0.5$ & $0.75$ & $1$ \\
%		\multicolumn{5}{c}{\cite{sumner2004deformation}} \\
		\hline \\ \hline
		\includegraphics[width=.15\columnwidth]{Images/Sumner/Face_hld}&
		\includegraphics[width=.15\columnwidth]{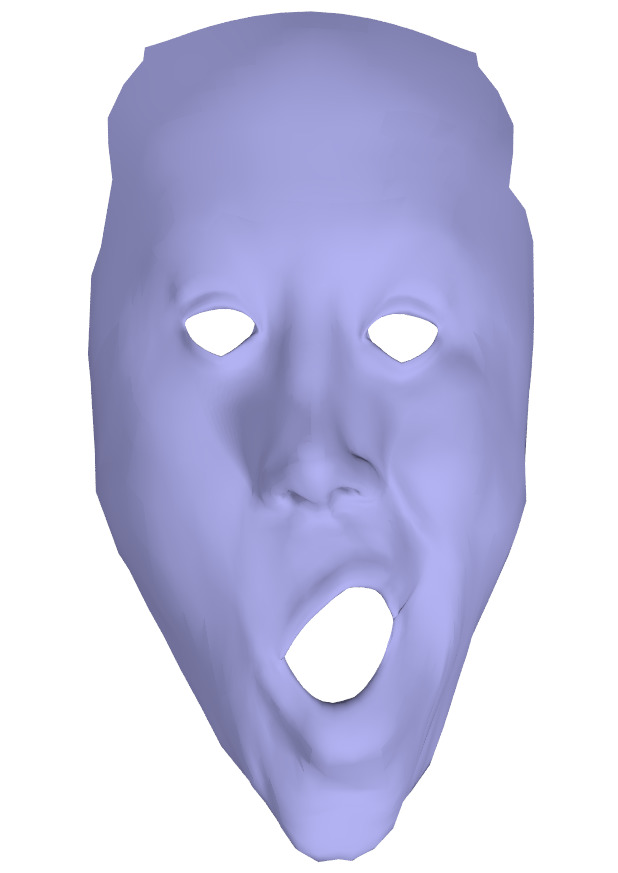}&
		\includegraphics[width=.15\columnwidth]{Images/Sumner/Face_tar}& \\
		Source & Deformed & Target & & \\
		\hline
		\includegraphics[width=.15\columnwidth]{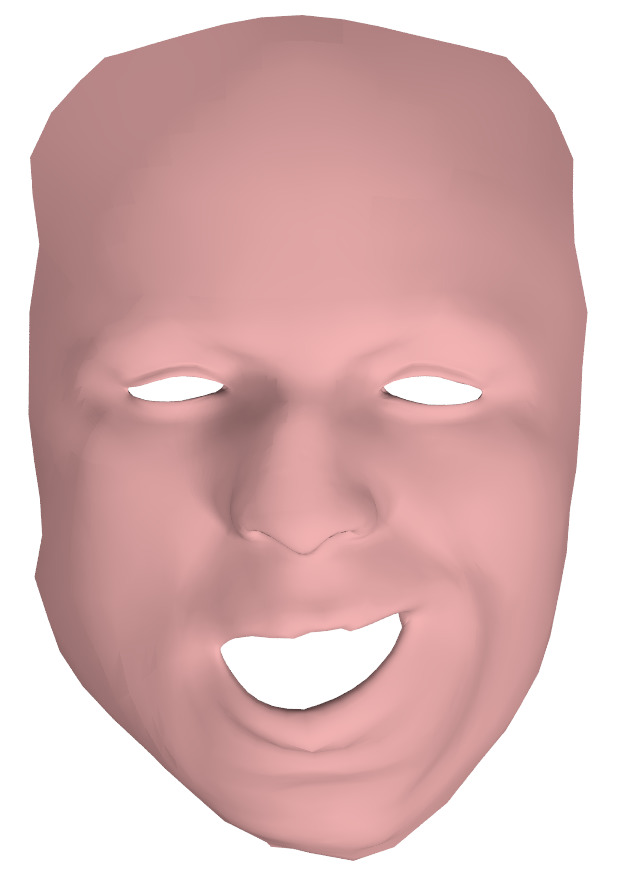}&
		\includegraphics[width=.15\columnwidth]{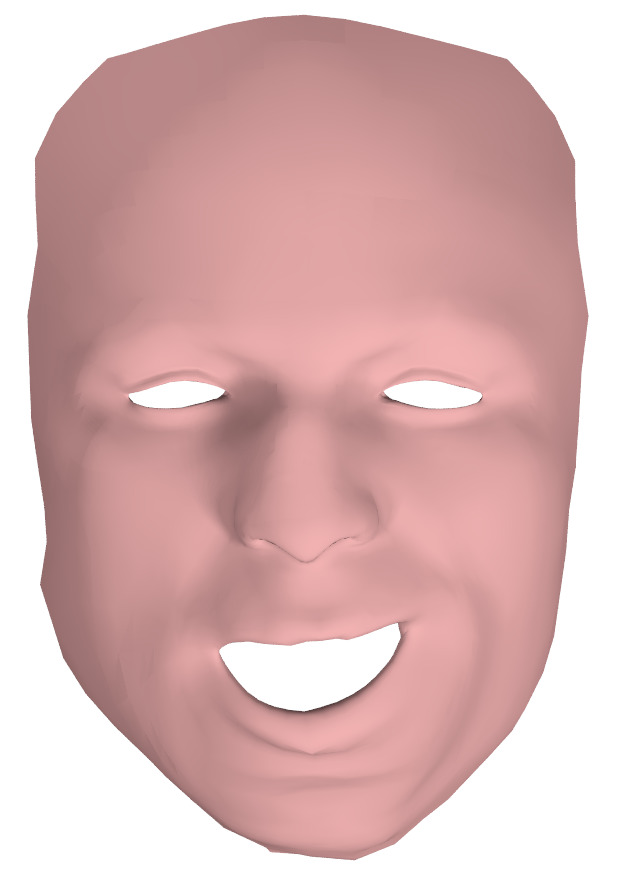}&
		\includegraphics[width=.15\columnwidth]{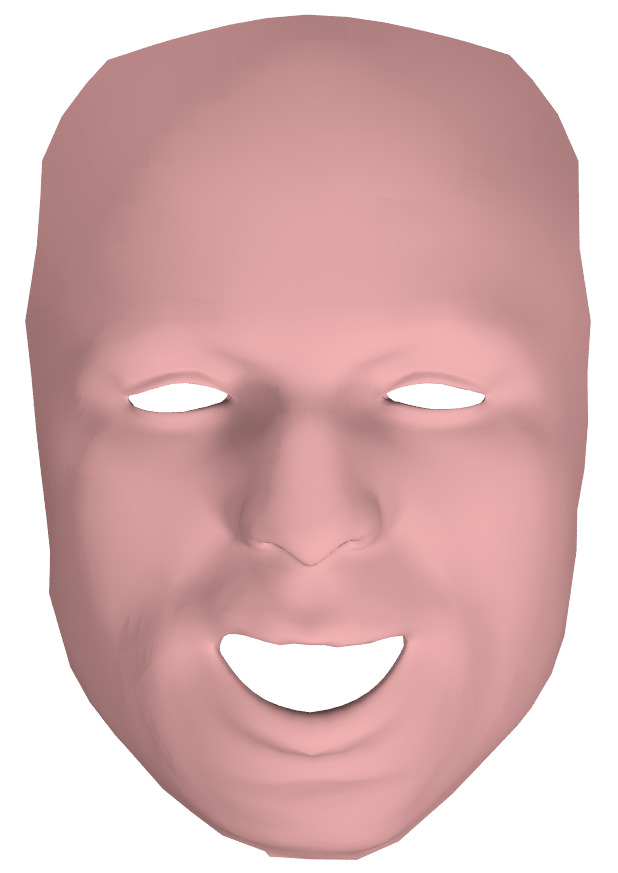}&
		\includegraphics[width=.15\columnwidth]{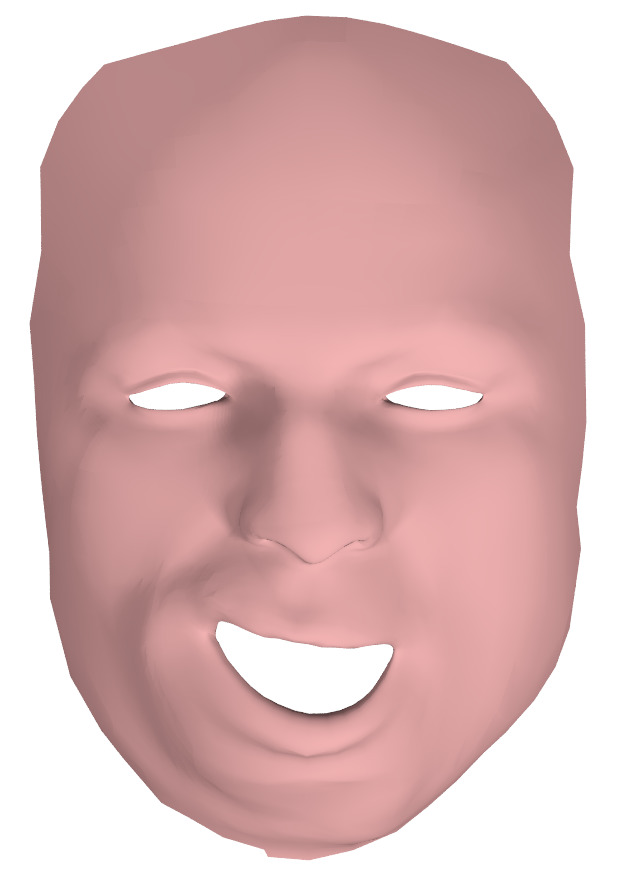}&
		\includegraphics[width=.15\columnwidth]{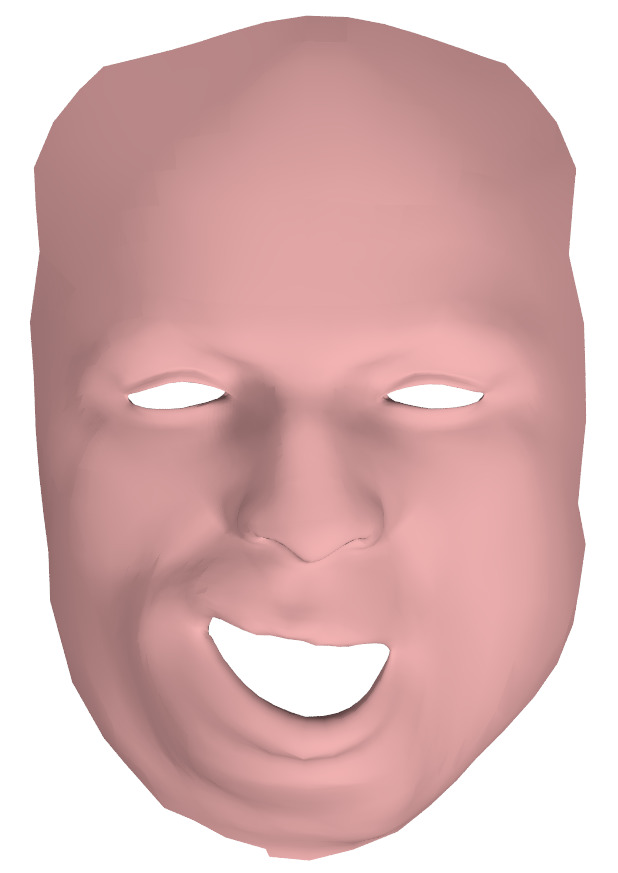}\\
		$0$ & $0.25$ & $0.5$ & $0.75$ & $1$ \\
		\multicolumn{5}{c}{Ours} \\
		\includegraphics[width=.15\columnwidth]{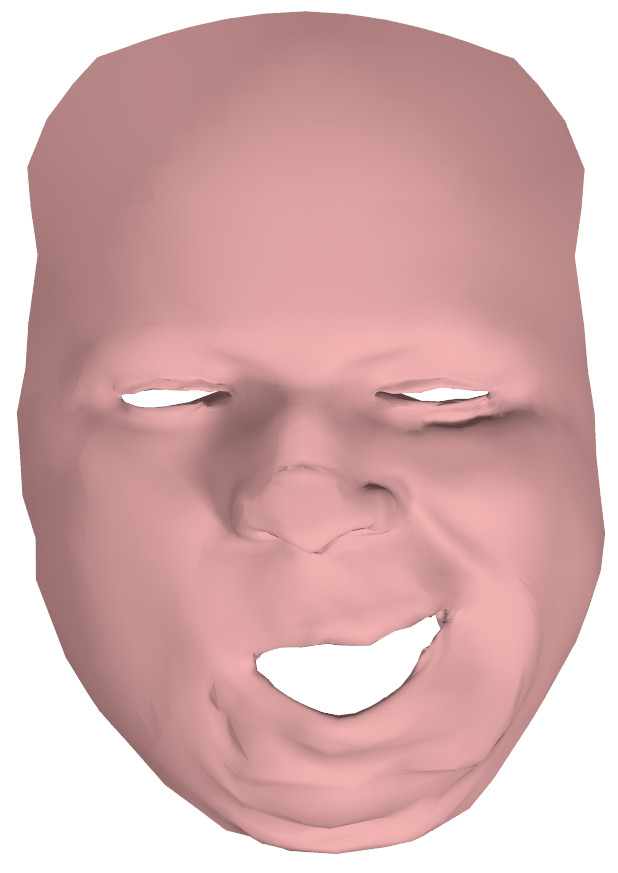}&
		\includegraphics[width=.15\columnwidth]{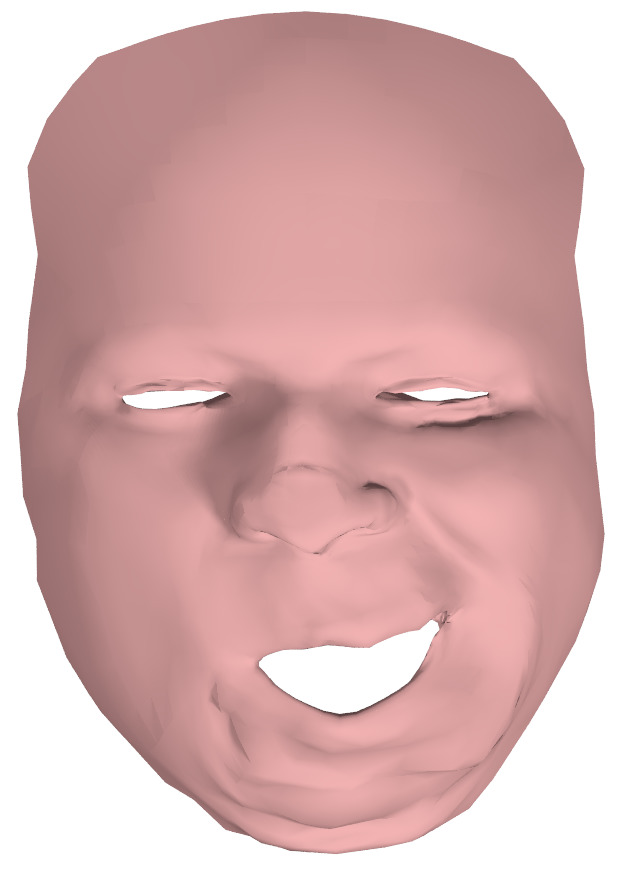}&
		\includegraphics[width=.15\columnwidth]{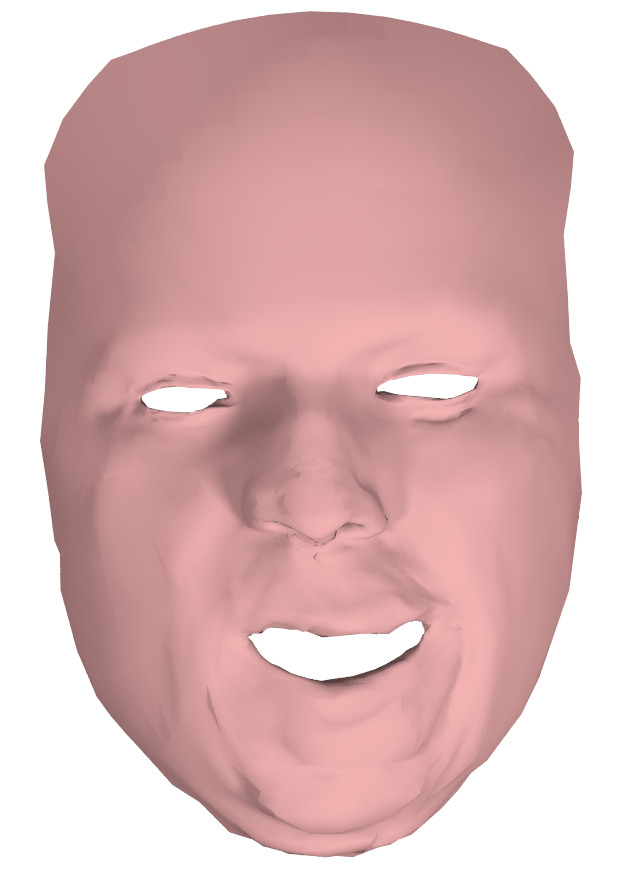}&
		\includegraphics[width=.15\columnwidth]{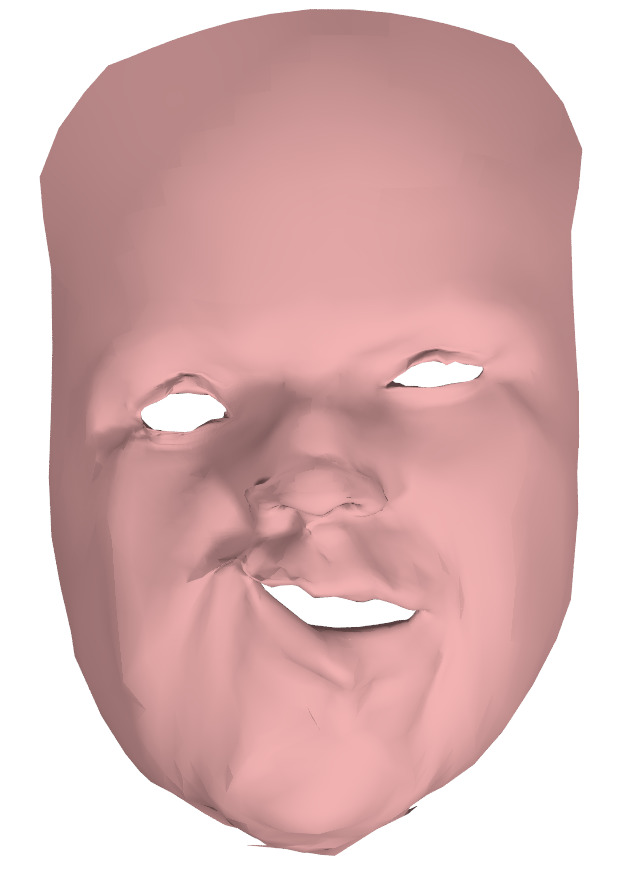}&
		\includegraphics[width=.15\columnwidth]{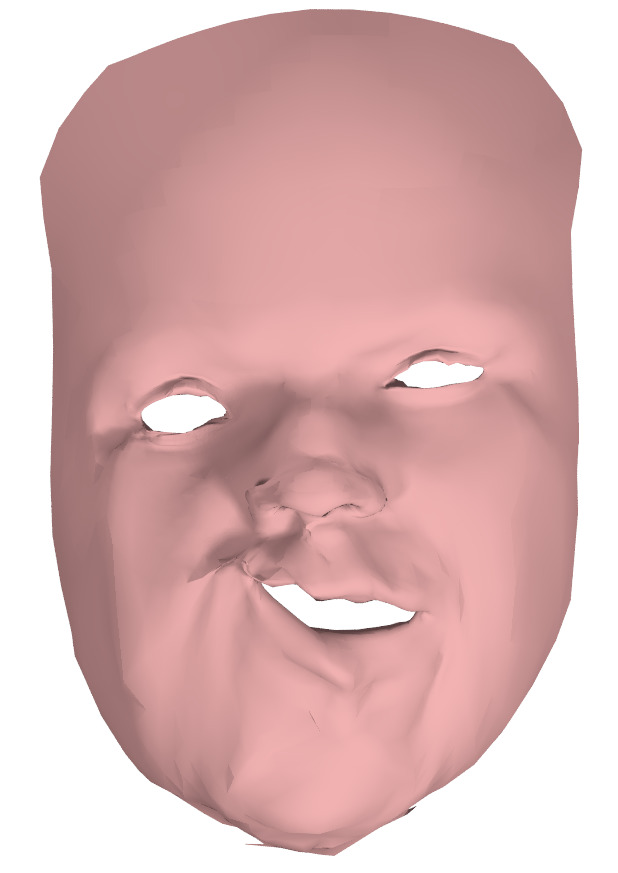}\\
		$0$ & $0.25$ & $0.5$ & $0.75$ & $1$ \\
		\multicolumn{5}{c}{\cite{sumner2004deformation}} \\
	\end{tabular}
	\caption{Deformation transfer using a noisy functional map consisting of a linear blending between the direct functional map ($\tau=0$) and the symmetric functional map ($\tau=1$). Our method is robust to this noise and outputs a non-linear interpolation of the deformation and its symmetric version. The method of in~\protect\cite{sumner2004deformation} fails when provided with a triangle-to-triangle map resulting of the conversion of the noisy functional map. \vspace{-1mm}}
	\label{fig:SumnerFmap}
\end{figure}

%Figure~\ref{fig:faceTransferStyle} shows that our method is able to recover deformations by using a functional map that are similar to the ones obtain with the algorithm proposed in \cite{sumner2004deformation} when given a precise triangle-to-triangle map. At the same time, 
Figure~\ref{fig:SumnerFmap} shows that working within the functional map framework makes our
algorithm more robust to noise usually encounter when using this representation. The computation of
functional maps, as described in \cite{ovsjanikov2012functional}, is done by solving a least squares
system incorporating intrinsic descriptors (HKS, WKS) therefore there often exists multiple
solutions in presence of an intrinsic symmetry $\pi$. We model a noisy functional map $C^\tau$ by a
linear blending between the direct map (mapping the left to the left and the right to the right) and
the anti-symmetric map (mapping the left to the right and the right to the left) represented as
operators: 
\begin{align*}
	C^\tau = \tau C_{\varphi \circ \pi} + (1-\tau) C_{\varphi} .
\end{align*}

Our method outputs a non-linear interpolation between the deformation and its symmetric version
while the method by Sumner et al. exhibits various artifacts.

\section{Conclusion and Future Work}
{In this paper we presented a method for representing extrinsic vector fields as linear
  operators acting on functions on the shapes, by considering the metric distortion induced by the
  deformation. In particular, we base our representation on the infinitesimal strain tensor and show
  how it leads to a linear functional operator that can naturally be combined with other such
  operators including functional maps and the Laplace-Beltrami. We showed how this representation
  can be used to analyze and design deformations and to introduce extrinsic information into the
  computation of functional correspondences. In the future, we are planning to use the newly
  introduced functional representation for shape animation. In this context, it would be interesting
  to establish a connection between the metric on the space of functional deformation fields and
  different inner products as suggested in \cite{eckstein2007generalized}. It would also very
  interesting to use our representation within the framework of shape spaces,
  e.g. \cite{kilian07}, for exploration and design.}

\bibliographystyle{acmtog}
\bibliography{fundeform}

%\appendix
%\input{sec-proofs}



\section*{Supplementary material}

# Online Appendix to:
# Functional Characterization of Infinitesimal Deformation

ETIENNE CORMAN and MAKS OVSJANIKOV
LIX, École Polytechnique, CNRS

## 1. EXTRINSIC VECTOR FIELDS AS OPERATOR

PROPOSITION 1. *For any extrinsic vector field $V$ there is a unique linear functional operator $E^V$ that satisfies:*

$$\int_M \langle \nabla g, \nabla E^V(f) \rangle \mathrm{d}\mu = \int_M \mathcal{L}_V \mathbf{g}(\nabla g, \nabla f) \mathrm{d}\mu. \qquad (1)$$

*Moreover, this operator is linear in both the vector field $V$ and function $f$.*

PROOF. Let $H_0^1(M)$ be the space of square integrable functions with $L^2(M)$ gradients and zero integrals:

$$H_0^1(M) = \left\{ f \in L^2(M) : \int_M \|\nabla f\|^2 \mathrm{d}\mu < +\infty, \int_M f \, \mathrm{d}\mu = 0 \right\}.$$

This space seems natural when studying this operator as $E^V$ maps any constant function to zero. When equipped of the scalar product $\langle .,. \rangle_{L^2} + \langle .,. \rangle_{H_0^1}$, $H_0^1$ is a Hilbert space.

The bilinear form $(f, g) \mapsto \langle f, g \rangle_{H_0^1(M)}$ is continuous and coercive thanks to the Wirtinger's inequality [Brezis 2010]. Moreover, for a given function $g$ in $H_0^1(M)$ the linear form $f \mapsto \int_M \mathcal{L}_V \mathbf{g}(\nabla g, \nabla f) \mathrm{d}\mu$ is continuous assuming that $V$ is smooth enough (at least $H_1$) since we have the inequality:

$$\int_M \mathcal{L}_V \mathbf{g}(\nabla g, \nabla f) \leq \|\nabla g\|_{L^2}^2 \|\mathcal{L}_V \mathbf{g}\|_{L^2}^2 \|\nabla f\|_{L^2}^2.$$

Thus all conditions of the Lax-Milgram theorem [Brezis 2010] are satisfied therefore for any function $g$ there exists a unique $E^V(g)$ satisfying Eq. (1). □

## 2. RELATION TO SHAPE DIFFERENCE OPERATORS

### 2.1 Unified Shape Difference

DEFINITION 2. *Assuming that $\varphi : N \to M$ is a diffeomorphism, the unified shape difference $D_I : \mathcal{C}^\infty(M) \to \mathcal{C}^\infty(M)$ is*

*defined implicitly by:*

$$\langle f, D_I(g) \rangle_{H_0^1(M)} := \int_M C_{\varphi^{-1}} \left( \langle \nabla C_\varphi(f), \nabla C_\varphi(g) \rangle \right) \mathrm{d}\mu^M.$$

Suppose that $\varphi : N \to M$ is a diffeomorphism. We denote $(\varphi_* X)_{\varphi(p)} = \mathrm{d}\varphi_p X_p$ the pullback of a vector field and $\mathrm{d}\varphi_p : T_p N \to T_{\varphi(p)} M$ the linear map between tangent spaces. Moreover the pullback with respect to $\varphi^{-1}$ of the metric field $\mathbf{g}^N : T_p N \times T_p N \to \mathbb{R}$ is given by $((\varphi^{-1})^\star \mathbf{g}^N)_p(X, Y) = \mathbf{g}^N_{\varphi^{-1}(p)}(\mathrm{d}\varphi^{-1} X, \mathrm{d}\varphi^{-1} Y)$. For the gradient of a function $f$ on $M$ at a point $q \in N$:

$$\nabla(f \circ \varphi)_q = (\mathrm{d}\varphi^{-1}\nabla f) \circ \varphi(q)$$

Let's note $\varphi^{-1}(p) = q \in N$, therefore the pullback metric reads

$$\langle \nabla(f \circ \varphi), \nabla(g \circ \varphi) \rangle_q = \mathbf{g}^N_{\varphi(q)}(\mathrm{d}\varphi^{-1}\nabla f, \mathrm{d}\varphi^{-1}\nabla g)$$
$$= ((\varphi^{-1})^\star \mathbf{g}^N)_{\varphi(q)}(\nabla f, \nabla g).$$

We can now rewrite Definition 2 with respect to the pullback metric:

$$\int_M \langle \nabla f, \nabla D_I(g) \rangle \mathrm{d}\mu = \int_M C_{\varphi}^{-1} \left( \langle \nabla C_\varphi(f), \nabla C_\varphi(g) \rangle \right) \mathrm{d}\mu$$
$$= \int_M C_{\varphi}^{-1} \left( ((\varphi^{-1})^\star \mathbf{g}^N)_{\varphi(p)} (\nabla f, \nabla g) \right) \mathrm{d}\mu$$
$$= \int_M ((\varphi^{-1})^\star \mathbf{g}^N)(\nabla f, \nabla g) \mathrm{d}\mu$$

This alternative definition leads to the characterization of the metric change:

PROPOSITION 2. *$D_I(f) = f$ for all $f \in \mathcal{C}^\infty(M)$ if and only if $\varphi$ is an isometry.*

PROOF. If $\varphi$ is an isometry then $(\varphi^{-1})^\star \mathbf{g}^N = \mathbf{g}^M$ so

$$\int_M \langle \nabla f, \nabla D_I(g) \rangle \mathrm{d}\mu^M = \int_M \langle \nabla f, \nabla g \rangle \mathrm{d}\mu$$

Using the fundamental lemma of calculus of variations: $\Delta D_I(g) = \Delta g$.

If $D_I(f) = f$ then

$$\int_M \langle \nabla f, \nabla g \rangle \mathrm{d}\mu = \int_M ((\varphi^{-1})^\star \mathbf{g}^N)(\nabla f, \nabla g) \mathrm{d}\mu$$

Using a result from [Schumacher 2013], it implies $\mathbf{g}^M = (\varphi^{-1})^\star \mathbf{g}^N$ so $\varphi$ is an isometry. □

### 2.2 Infinitesimal Shape Difference Operators

To define the infinitesimal shape differences we first need to introduce the correct framework and notation. Let's assume that the

family of oriented surfaces $M_t$ without boundary of intrinsic dimension 2 are isometrically immersed in $\mathbb{R}^3$ by the local mappings $F_t : U \subset \mathbb{R}^2 \to M_t \subset \mathbb{R}^3$. This family of manifolds is generated by the displacement of the points along the smooth vector field $V(p) \in T_pM \times T_pM \simeq \mathbb{R}^3$:

$$\frac{\partial F_t}{\partial t}(p) = V(p), \quad (p,t) \in M \times \mathbb{R}^+ \quad (2)$$

The metric of the embedded surface is by definition $\mathbf{g}_{ij}^t = \langle \partial_i F_t, \partial_j F_t \rangle$ and the area form is $\mu^t = \sqrt{\det \mathbf{g}^t}$. The Riemannian connection on the ambient space $\mathbb{R}^3$ is denoted $\tilde{\nabla}$. As mentioned in [do Carmo 2013], the projection of the ambient connection into the tangent space of $M$ coincides the unique Levi-Civita connection on $M$. Therefore the connection $\nabla$ on $M$ is naturally extended to extrinsic vector fields by $\nabla_i V_j = \langle \partial_i V, \partial_j F_t \rangle$. Once the connection is defined other differential operator can extended to extrinsic vector fields for example the divergence is defined as the trace of the connection $\text{div}(V) = \mathbf{g}^{ij} \nabla_i V_j$.

We consider the family of diffeomorphisms $\varphi_t : M_t \to M$ given by $\varphi_t(p) : F_t - t(p)V(p)$.

The derivative of local quantities links the Lie derivative with the Strain tensor.

LEMMA 1. *Given a one parameter family of surfaces described in Eq. (2), for a fixed point $p$, the first-order change in the metric tensor $\mathbf{g}$ and in the local area element $\mu = \sqrt{\det(\mathbf{g})}$ are given as:*

$$\frac{\partial \mathbf{g}(t)}{\partial t}\Big|_{t=0} = \mathcal{L}_V \mathbf{g} \quad (3)$$

$$\frac{\partial \mu(t)}{\partial t}\Big|_{t=0} = \text{div}(V)\mu. \quad (4)$$

PROOF. Those properties are easily proven when using local coordinates. Given a family of diffeomorphisms $\varphi_t$ the Lie derivative of the metric tensor with respect to the vector field $V$ denoted $\mathcal{L}_V \mathbf{g}$ is by definition:

$$\mathcal{L}_V \mathbf{g} := \frac{\partial}{\partial t} \left( (\varphi_t^{-1})^* \mathbf{g}(t) \right)\Big|_{t=0}.$$

Since the local immersion $F_t$ use a common chart system, the coordinates of the pullback metric $((\varphi_t^{-1})^* \mathbf{g}^t)_{ij}$ are equal to the metric on $M_t$ in local coordinates $\mathbf{g}_{ij}(t) = \langle \partial_i F_t, \partial_j F_t \rangle$. The computation of derivative is then straigthforward:

$$\frac{\partial}{\partial t} \left( ((\varphi_t^{-1})^* \mathbf{g}(t))_{ij} \right)\Big|_{t=0} = \frac{\partial}{\partial t} \langle \partial_i F_t, \partial_j F_t \rangle \Big|_{t=0}$$
$$= \nabla_i V_j + \nabla_j V_i$$

From there, Eq. (4) is easily obtained:

$$\frac{\partial \mu(t)}{\partial t}\Big|_{t=0} = \frac{\partial}{\partial t} \left( \sqrt{\det(\mathbf{g}(t))} \right)\Big|_{t=0}$$
$$= \frac{1}{2\mu} \det(\mathbf{g}) \mathbf{g}^{ij} (\nabla_i V_j + \nabla_j V_i) = \text{div}(V)\mu$$

□

We then obtain the derivative of the shape differences.

PROPOSITION 3. *Let $V$ be a smooth deformation field on $M$, the derivatives of $D_A$, $D_C$ and $D_I$ at time zero satisfy for all smooth*

*functions $f, g$:*

$$\langle f, E_A^V(g) \rangle_{L^2}^M = \int_M \text{div}(V) f g \, d\mu,$$

$$\langle f, E_C^V(g) \rangle_{H_0^1}^M = \int_M \text{div}(V) \langle \nabla f, \nabla g \rangle - \mathcal{L}_V \mathbf{g} \langle \nabla f, \nabla g \rangle \, d\mu,$$

$$\langle f, E_I^V(g) \rangle_{H_0^1}^M = -\int_M \mathcal{L}_V \mathbf{g} \langle \nabla f, \nabla g \rangle \, d\mu.$$

PROOF. The first statement is obtained by using (4):

$$\langle f, \partial_t D_A(g) \rangle_{L^2} = \frac{\partial}{\partial t} \left( \int_{M_t} C_t(f) C_t(g) d\mu^t \right)\Big|_{t=0}$$
$$= \frac{\partial}{\partial t} \left( \int_M f g \, d((\varphi_t)_* \mu^t) \right)\Big|_{t=0}$$
$$= \int_M \text{div}(V) f g \, d\mu$$

For the second statement let's start with the evolution of the point-wise scalar product between gradient:

$$\frac{\partial}{\partial t} \left( \langle \nabla f, \nabla g \rangle \right)\Big|_{t=0} = \frac{\partial}{\partial t} (\mathbf{g}^{ik} \partial_k f) \mathbf{g}_{ij} (\mathbf{g}^{jl} \partial_l g)\Big|_{t=0}$$
$$= \frac{\partial}{\partial t} \partial_i f \mathbf{g}^{ij} \partial_j g\Big|_{t=0}$$
$$= -(\mathbf{g}^{ik} \partial_k f) \frac{\partial \mathbf{g}_{ij}}{\partial t} (\mathbf{g}^{jl} \partial_l g)$$
$$= -\langle \nabla f, \nabla_{\nabla g} V \rangle - \langle \nabla_{\nabla f} V, \nabla g \rangle.$$

It follows that:

$$\langle f, \partial_t D_C(g) \rangle_{H_0^1} = \frac{\partial}{\partial t} \left( \int_{M_t} \langle \nabla C_t(f), \nabla C_t(g) \rangle d\mu^t \right)\Big|_{t=0}$$
$$= \frac{\partial}{\partial t} \left( \int_M ((\varphi_t^{-1})^* \mathbf{g}^t)(\nabla f, \nabla g) d((\varphi_t)_* \mu^t) \right)\Big|_0$$
$$= \int_M \text{div}(V) \langle \nabla f, \nabla g \rangle d\mu$$
$$- \int_M \left( \langle \nabla f, \nabla_{\nabla g} V \rangle + \langle \nabla_{\nabla f} V, \nabla g \rangle \right) d\mu$$

Starting from Definition 2:

$$\langle f, \partial_t D_I(g) \rangle_{H_0^1} = \frac{\partial}{\partial t} \left( \int_M C_t^{-1} ( \langle \nabla C_t(f), \nabla C_t(g) \rangle ) d\mu \right)\Big|_{t=0}$$
$$= \frac{\partial}{\partial t} \left( \int_M ((\varphi_t^{-1})^* \mathbf{g}^t)(\nabla f, \nabla g) d\mu \right)\Big|_{t=0}$$
$$= -\int_M \langle \nabla f, \nabla_{\nabla g} V \rangle + \langle \nabla_{\nabla f} V, \nabla g \rangle d\mu$$

□

## 3. DISCRETE CONNECTION

The connection of the ambient space $\tilde{\nabla}_u V$ where $u$ is a tangent vector and $V$ is an extrinsic vector field :

$$\tilde{\nabla} : \mathbb{R}^{3|\mathcal{F}|} \times \mathbb{R}^{3|\mathcal{V}|} \to \mathbb{R}^{3|\mathcal{F}|}$$
$$(u, V) \mapsto \tilde{\nabla}_u V$$

Recall that we build the connection $\bar{\nabla}$ using finite differences as follows. Since extrinsic vector fields are defined at vertices the differences are taken along the edges.

**DEFINITION 3.** *In a given triangle $T \in \mathcal{F}$ the ambient covariant derivative along the edge $e_{ij}$ is defined by*

$$\left(\bar{\nabla}_{\frac{e_{ij}}{\|e_{ij}\|}} V\right)_T = \frac{V_i - V_j}{\|e_{ij}\|}.$$

*Thus the ambient connection in the directions $E = (e_{ij}, e_{jk})$ can be stored in a matrix*

$$(\bar{\nabla}_E V)_T = \begin{pmatrix} V_i - V_j & V_j - V_k \end{pmatrix}.$$

*Then, given any tangent vector $x = E\alpha$, the covariant derivative in its direction can be computed as $\bar{\nabla}_x V = (\bar{\nabla}_E V)\alpha$.*

Given the expression above, the discrete Lie derivative of the metric at triangle $T$ follows immediately. Namely for any pair of tangent vectors $x = E\alpha$, $y = E\beta$ in the triangle $T$, we have:

$$\mathcal{L}_V \mathbf{g}(x,y)_T = \langle x, (\bar{\nabla}_E V)\beta \rangle + \langle (\bar{\nabla}_E V)\alpha, y \rangle. \tag{5}$$

After integration we obtain the discrete infinitesimal shape difference:

$$f^\top W_M E^V g = -\sum_{T \in \mathcal{F}} \mathcal{L}_V \mathbf{g}(\nabla f, \nabla g)_T \mu(T).$$

The expression of the matrix $W_M E^V$ is more easily found using the derivative of the unified shape difference operator (see Section 4) and is proven later in Thm. 1. The same goes for the proof of Prop. 4 which is also postponed until Section 4.4.

## 4. DISCRETE INFINITESIMAL SHAPE DIFFERENCES

### 4.1 Discrete Unified Shape Differences

The discretization of the unified shape difference is straightforward when $N$ and $M$ are triangle meshes and share the same connectivity. In Definition 2 given above, the gradients and the point-wise scalar products are taken on $N$ while the measure $\mathrm{d}\mu^M$ comes from $M$. Therefore the right hand side can be discretized by a modified cotangent weight formula:

$$W_M D_I = W_N^M, \text{ where}$$

$$(W_N^M)_{i,j} = \frac{1}{2}\left(\frac{\mu^M(T_\alpha)}{\mu^N(T_\alpha)}\cot\alpha_{ij}^N + \frac{\mu^M(T_\beta)}{\mu^N(T_\beta)}\cot\beta_{ij}^N\right). \tag{6}$$

In Section 4.1 we derived an infinitesimal shape difference from a discrete connection. This discrete can be done by a time derivative of the Eq. (6). To do so, however, we need to introduce an alternative formalism for the cotangent-weights Laplacian.

### 4.2 Cotangent weights alternative

The usual cotangent weight formula is not well-suited to carry out the computations. Therefore we use an alternative formulation which makes more apparent the link with continuous properties.

We denote the local basis $E = (e_{ij}, e_{jk})$ formed by edges of triangle of a triangle $T = \{x_i, x_j, x_k\} \in \mathcal{F}$ where $e_{ij}$ is an oriented edge. We denote $\|e_{ij}\| = \ell_{ij}$ the edge length. Using this notation, the Finite Element gradient is given by the formula [Botsch et al. 2010]:

$$\nabla f = \frac{1}{2\mu(T_{ijk})} \mathcal{R}^{90^\circ} E_{T_{ijk}} \begin{pmatrix} f_k - f_j \\ f_i - f_j \end{pmatrix}. \tag{7}$$

where $\mathcal{R}^{90^\circ}$ denotes the counter-clockwise rotation by $90^\circ$. One can remark that the gradient of a function can be expressed in an alternative way depending on the $2 \times 2$-symmetric matrix $\mathbf{g}_T$ per triangle:

$$\nabla f = E_T \mathbf{g}_T^{-1} \begin{pmatrix} f_j - f_i \\ f_k - f_j \end{pmatrix}. \tag{8}$$

This matrix will be referred to as *discrete metric tensor* in the local basis $E = (e_{ij}, e_{jk})$ of the triangle $T$:

$$\mathbf{g}_T := \frac{1}{2}\begin{pmatrix} 2\ell_{ij}^2 & \ell_{ki}^2 - \ell_{jk}^2 - \ell_{ij}^2 \\ \ell_{ki}^2 - \ell_{jk}^2 - \ell_{ij}^2 & 2\ell_{jk}^2 \end{pmatrix} = E^\top E. \tag{9}$$

Note that $\mathbf{g}_T$ is defined such that $\begin{pmatrix} 1 & 0 \end{pmatrix}^\top \mathbf{g}_T \begin{pmatrix} 1 & 0 \end{pmatrix} = \ell_{ij}^2$ so the bilinear form of two adjacent triangle agrees along the edges. Moreover using Heron's formula one can verify that $\det(\mathbf{g}_T) = 4\mu(T)^2$.

It follows an alternative expression for the standard cotangent formula:

$$f^\top W g = \sum_{T \in \mathcal{F}} \begin{pmatrix} f_j - f_i \\ f_k - f_j \end{pmatrix}^\top \mathbf{g}_T^{-1} \begin{pmatrix} g_j - g_i \\ g_k - g_j \end{pmatrix} \mu(T).$$

This formulation is equivalent to the one found in [Boscaini et al. 2015]:

$$f^\top W g = \sum_{T \in \mathcal{F}} \frac{1}{4\mu(T)} \begin{pmatrix} f_j - f_k \\ f_j - f_i \end{pmatrix}^\top \mathbf{g}_T \begin{pmatrix} g_j - g_k \\ g_j - g_i \end{pmatrix}, \tag{10}$$

by noting that any $2 \times 2$ invertible symmetric matrix is linked to its inverse by the formula:

$$\begin{pmatrix} 0 & -1 \\ 1 & 0 \end{pmatrix}^\top \mathbf{g}_T^{-1} \begin{pmatrix} 0 & -1 \\ 1 & 0 \end{pmatrix} = \frac{1}{4\mu(T)^2} \mathbf{g}_T. \tag{11}$$

The classical cotangent weight formula is recovered by noting that $\cot\alpha_{ij} = (-\ell_{ij}^2 + \ell_{jk}^2 + \ell_{ki}^2)/(4\mu(T))$.

### 4.3 Discrete Metric Derivative

First let's remark that the derivative of the discrete metric can be expressed with respect to discrete connection:

**LEMMA 2.** *Given a one parameter family of meshes, the first-order change in the metric tensor $\mathbf{g}_T = E^\top E$ and in the area at a triangle $T \in \mathcal{F}$, is given as:*

$$\left.\frac{\partial_t \mathbf{g}_T}{\partial t}\right|_{t=0} = E^\top (\nabla_E V)_T + (\nabla_E V)_T^\top E,$$

$$\left.\frac{\partial_t \mu(T)}{\partial t}\right|_{t=0} = \mathrm{div}(u)_T \mu(T),$$

*where the divergence at triangle $T$ is defined as $\mathrm{div}(u)_T := \mathrm{Tr}\left(\mathbf{g}_T^{-1} E^\top (\nabla_E V)\right)$.*

**PROOF.** First let's remark that the derivative of the discrete metric can be expressed with respect to discrete connection:

$$\left.\frac{\partial \mathbf{g}_T^t}{\partial t}\right|_{t=0} = E^\top (\nabla_E V)_T + (\nabla_E V)_T^\top E.$$

Since the metric is linked to the triangle area by $\mu_t(T) = \frac{1}{2}\sqrt{\det(\mathbf{g}_T^t)}$ the statement obtained by a direct computation of the derivative. □

## 4.4 Discrete infinitesimal shape difference

Taking the derivative of the discrete unified shape difference in Eq. (6) might be challenging. However, using the formulation of Eq. (10) leads to an equivalent formulation of Eq. (6) is:

$$f^\top W D_I^t g := \sum_{T \in \mathcal{F}} \langle \nabla^t f, \nabla^t g \rangle_T^t \mu(T). \tag{12}$$

By taking the derivative of his expression at time $t = 0$, we obtain an alternative discretization of the infinitesimal shape difference $E$.

THEOREM 1. *The discrete infinitesimal shape difference reads* $E^V(u) = W_M^{-1} H$, *where $H$ is a Laplacian matrix whose weights depend on the extrinsic vector field:*

$$(H)_{ij} = \frac{1}{2} \sum_{j \sim i} (c(T_{\alpha_{ij}}) + c(T_{\beta_{ij}})),$$

$$c(T) = (\langle e_{jk}, V_j - V_i \rangle + \langle e_{ij}, V_j - V_k \rangle) \frac{1}{4\mu(T)} - \text{div}(V)_T \frac{\langle e_{jk}, e_{ki} \rangle}{\mu(T)}$$

PROOF. Using the FEM gradient, e.g. Eq. (7), to discretize the unified shape difference written in Eq. (12) leads to:

$$f^\top W D_I^t g = \sum_{T \in \mathcal{F}} \frac{1}{4} \begin{pmatrix} f_j - f_k \\ f_j - f_i \end{pmatrix}^\top \frac{\mathbf{g}_T^\top}{\mu_t(T)^2} \begin{pmatrix} g_j - g_k \\ g_j - g_i \end{pmatrix} \mu(T). \tag{13}$$

We can now compute the derivative with respect to time by using Lemma 2:

$$f^\top W E^V g = \sum_{T \in \mathcal{F}} \frac{1}{4\mu(T)} \begin{pmatrix} f_j - f_k \\ f_j - f_i \end{pmatrix}^\top \mathbf{L}_T \begin{pmatrix} g_j - g_k \\ g_j - g_i \end{pmatrix},$$

$$\mathbf{L}_T = E^\top (\nabla_E V)_T + (\nabla_E V)_T^\top E - 2\text{Tr}\left(\mathbf{g}_T^{-1} E^\top (\nabla_E V)\right) \mathbf{g}_T.$$

The matrices $\mathbf{L}_T$ can be written in a form similar to the discrete metric (see Eq. (9)):

$$\mathbf{L}_T = \frac{1}{2} \begin{pmatrix} 2a_{ij} & a_{ki} - a_{jk} - a_{ij} \\ a_{ki} - a_{jk} - a_{ij} & 2a_{jk} \end{pmatrix},$$

$$\text{where } a_{ij} = \langle e_{ij}, V_i - V_j \rangle - 2\text{div}(V)_T \ell_{ij}^2, \tag{14}$$

where the divergence is define as in Lemma 2. This leads to the point-wise formulation:

$$(W_M E^V)_{ij} = \frac{1}{2} \sum_{j \sim i} (c(T_{\alpha_{ij}}) + c(T_{\beta_{ij}})),$$

$$c(T) = \frac{-a_{ki} + a_{jk} + a_{ij}}{4\mu(T)}.$$

□

## 5. EQUIVALENCE OF THE TWO DISCRETIZATIONS

PROPOSITION 5. *The discretization of $E$ based on the discrete Levi-Civita connection is equivalent to the one obtained by differentiating the unified shape difference operator.*

PROOF. In Eq. (5) the tangent vectors in a given triangle have be to expressed in the basis form by two edges of the triangle.

Following the discussion in Section 4.2, the FEM gradient at a face $T$ can be written in two equivalent ways:

$$\nabla f_T = \frac{1}{2\mu(T)} \mathcal{R}^{90°} E \begin{pmatrix} f_j - f_k \\ f_j - f_i \end{pmatrix} = E\mathbf{g}_T^\top \begin{pmatrix} f_j - f_i \\ f_k - f_j \end{pmatrix}.$$

Therefore the discrete strain tensor at triangle $T$ in Eq. (5) follows immediately:

$$\mathcal{L}_V \mathbf{g}(\nabla f, \nabla g) = \begin{pmatrix} f_i - f_j \\ f_j - f_k \end{pmatrix}^\top \mathbf{g}_T^{-1} E^\top (\bar\nabla_E V) \mathbf{g}_T^{-1} \begin{pmatrix} g_i - g_j \\ g_j - g_k \end{pmatrix}$$

$$+ \begin{pmatrix} f_i - f_j \\ f_j - f_k \end{pmatrix}^\top \mathbf{g}_T^{-1} (\bar\nabla_E V)^\top E \mathbf{g}_T^{-1} \begin{pmatrix} g_i - g_j \\ g_j - g_k \end{pmatrix}$$

$$= \begin{pmatrix} f_i - f_j \\ f_j - f_k \end{pmatrix}^\top \frac{\partial}{\partial t} \left(\mathbf{g}_T^{-1}\right)\Big|_{t=0} \begin{pmatrix} g_i - g_j \\ g_j - g_k \end{pmatrix}.$$

From Lemma 2, we recognize the term the derivative of the inverse metric. Using Eq. (11), one can further modified the expression to:

$$\mathcal{L}_V \mathbf{g}(\bar\nabla f, \nabla g)_T = -\frac{1}{4} \begin{pmatrix} f_j - f_k \\ f_j - f_i \end{pmatrix}^\top \frac{\partial}{\partial t} \left(\frac{\mathbf{g}_T^\top}{\mu_t(T)^2}\right)\Big|_{t=0} \begin{pmatrix} g_j - g_k \\ g_j - g_i \end{pmatrix}. \tag{15}$$

The right had side term appears in the discrete isometric shape difference as written in Eq. (13). This leads to the equality between the different discretization:

$$\frac{\partial}{\partial t} \left(f^\top W D_I^t g\right)\Big|_{t=0} = -\sum \mathcal{L}_V \mathbf{g}(\nabla f, \nabla g)_T \mu(T).$$

## 6. VECTOR FIELDS REPRESENTATION

PROPOSITION 4. *For almost all triangle meshes $M$ without boundary, the operator $E^V$ uniquely defines the extrinsic vector field $V$ up to rigid motion.*

PROOF. The proof is organized as follow: we show that we can recover the matrices $\mathbf{L}_T$ from the infinitesimal shape difference in Eq. (14) then we use a standard results in combinatorix to prove that $\mathbf{L}_T = 0$ if and only if the extrinsic vector field is a rigid motion.

*Kernel of $\mathbf{L}_T \mapsto E^V$.* The information about the extrinsic vector field is solely contained by the matrices $\mathbf{L}_T$. Like the discrete metric those matrices agrees across edges so they can be reduced to the vector $a \in \mathbb{R}^{|\mathcal{E}|}$ as defined in Eq. (14). The application $a \mapsto E^V$ is linear and we will prove it is almost always invertible.

Extracting elements of $W E^V$ corresponding to edges on $M$ yields a linear operator $B : \mathbb{R}^{|\mathcal{E}|} \to \mathbb{R}^{|\mathcal{E}|}$ with matrix

$$B_{ij} = \frac{1}{8} \begin{cases} \mu(T_i)^{-1} + \mu(T_i')^{-1} & \text{if } i = j \\ -\mu(T_i)^{-1} & \text{if } i, j \text{ are edges of } T \\ 0 & \text{otherwise.} \end{cases}$$

Here, indices $i, j$ refer to edges on $M$; for a given edge $i$, we label its adjacent triangles $T_i$ and $T_i'$. Remark that $B$ can be written as a weighted sum: $B = \sum_k \frac{1}{8} \mu(T_k)^{-1} B^k$, where each $B^k$ is a matrix such that:

$$B_{ij}^k = \begin{cases} 1 & \text{when } i = j, \text{ and } i \text{ belongs to triangle } k. \\ -1 & \text{when } i, j \text{ are edges of triangle } k. \\ 0 & \text{otherwise.} \end{cases}$$

It is easy to see that the intersection of the kernels of all $B^k$ is empty. Moreover, by considering the determinant of $B$ as a multivariate

ACM Transactions on Graphics, Vol. VV, No. N, Article XXX, Publication date: Month YYYY.

ACM Transactions on Graphics, Vol. VV, No. N, Article XXX, Publication date: Month YYYY.

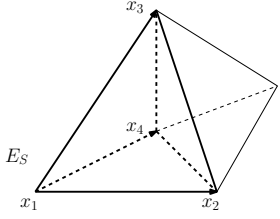

Fig. 1. Local basis in a tetrahedron.

polynomial with real coefficients, we conclude that $B$ is either singular for any choice of values of $\mu(T_k)^{-1}$, or for a finite set of coefficients, which thus have measure zero. The proof of the claim follows by noting that for a closed manifold mesh there must exist a non-singular weighted sum, which can be obtained by iteratively adding triangles, while maintaining that the corresponding matrix sum is non-singular on the incident edges.

*Rigidity Theorem.* As shown previously the kernel of $a \mapsto E^V$ is almost always reduced to the zero element. Going back to the matrices $\mathbf{L}_{Tr}$, the extrinsic vector field in the kernel should satisfy:

$$\mathbf{g}_T^{-1} E^\top (\nabla_E V)_T + \mathbf{g}_T^{-1} (\nabla_E V)_T^\top E - 2 \operatorname{Tr} \left( \mathbf{g}_T^{-1} E^\top (\nabla_E V) \right) \operatorname{Id} = 0.$$

Taking the trace in both sides implies that $\operatorname{div}(V)_T = \operatorname{Tr}(\mathbf{g}_T^{-1} \partial_t \mathbf{g}_T^\top|_0)$ should vanish so it is equivalent to have all matrices $E^\top (\nabla_E V)_T + (\nabla_E V)_T^\top E$ equal zero. It follows that the extrinsic vector field satisfies $\langle e_{ij}, V_i - V_j \rangle = 0$ at all edges. It has been proved in [Gluck 1975] that almost all simply connected closed surfaces only admit rigid deformation as solution of this equation. □

## 7. CONSTRUCTION FOR TETRAHEDRAL MESHES

Let's consider the case of a mesh whose constitutive elements are only tetrahedra. The set of tetrahedra is denoted $\mathcal{S}$. Figure 1 illustrates a tetrahedron $S$ defined by 4 vertices $(x_1, \ldots, x_4)$. The first 3 edges form the local basis: $E_S = (x_2 - x_1, \ldots, x_4 - x_1)$. The discrete metric tensor, now denoted $\mathbf{g}_S = E_S^\top E_S$, is expressed locally using the edge lengths:

$$\mathbf{g}_{ij} = \begin{cases} \ell_{1i}^2 & i = j \\ \frac{1}{2}(\ell_{1j}^2 - \ell_{1i}^2 - \ell_{1j}^2), & i \neq j. \end{cases}, \qquad 1 \leq i, j \leq k.$$

The volume of the simplex is accessible through the determinant of the metric by $\mu(S) = \sqrt{\det(\mathbf{g}_S)}/6$. The gradient of a piecewise linear function inside the simplex $S$ is now computed, by analogy with Eq. (8), with the formula:

$$\nabla f = E_S \mathbf{g}_S^{-1} \begin{pmatrix} f_2 - f_1 \\ \ldots \\ f_4 - f_1 \end{pmatrix}. \qquad (16)$$

This formula is strictly equivalent to other more classical FEM formulation as it relies only on computing the derivative of a piecewise linear function.

### 7.1 Connection for Tetrahedral Meshes

Now, the connection of the ambient space $\bar{\nabla}_u V$ where $u$ is a vector inside a tetrahedron and $V$ is an extrinsic vector field assigning a

vector per vertex:

$$\bar{\nabla} : \mathbb{R}^{3|S|} \times \mathbb{R}^{3|V|} \to \mathbb{R}^{3|S|}$$
$$(u, V) \mapsto \bar{\nabla}_u V$$

We build the connection $\bar{\nabla} V$ by analogy with triangle mesh case in Section 3 leading to add an extra vector in the local basis $E_S$.

DEFINITION 4. *In a given tetrahedron $S \in \mathcal{S}$ the ambient covariant derivative along the edge $e_{i1}$ is defined by*

$$\left( \bar{\nabla}_{\frac{e_{i1}}{\|e_{i1}\|}} V \right)_T = \frac{V_i - V_1}{\|e_{i1}\|}.$$

*Thus the ambient connection in the directions $E_S$ can be stored in a matrix*

$$(\bar{\nabla}_{E_S} V)_T = \begin{pmatrix} V_2 - V_1 & \ldots & V_4 - V_1 \end{pmatrix}.$$

*Then, given any tangent vector $x = E_S \alpha$, the covariant derivative in its direction can be computed as $\bar{\nabla}_x V = (\bar{\nabla}_{E_S} V) \alpha$.*

Given the expression above, the discrete Lie derivative of the metric at simplex $S$ follows immediately. Namely for any pair of tangent vectors $x = E\alpha, y = E\beta$ in the simplex $S$, we have:

$$\mathcal{L}_V \mathbf{g}(x, y)_S = \langle x, (\bar{\nabla}_x V)\beta \rangle + \langle (\bar{\nabla}_{E_S} V)\alpha, y \rangle.$$

Considering the local expression of the gradient in Eq. (16), we obtain:

$$\mathcal{L}_V \mathbf{g}(\nabla f, \nabla g)_S = \begin{pmatrix} f_2 - f_1 \\ \ldots \\ f_4 - f_1 \end{pmatrix}^\top \mathbf{g}_S^{-1} (\bar{\nabla}_{E_S} V) \mathbf{g}_S^{-1} \begin{pmatrix} g_2 - g_1 \\ \ldots \\ g_4 - g_1 \end{pmatrix}$$
$$+ \begin{pmatrix} f_2 - f_1 \\ \ldots \\ f_4 - f_1 \end{pmatrix}^\top \mathbf{g}_S^{-1} (\bar{\nabla}_{E_S} V)^\top \mathbf{g}_S^{-1} \begin{pmatrix} g_2 - g_1 \\ \ldots \\ g_4 - g_1 \end{pmatrix}$$

After integration we obtain the discrete infinitesimal shape difference on a tetrahedral mesh:

$$f^\top W_M E^V g = -\sum_{S \in \mathcal{S}} \mathcal{L}_V \mathbf{g}(\nabla f, \nabla g)_S \mu(S).$$

## 8. FUNCTIONAL MAP INFERENCE

PROPOSITION 6. *Given a pair of surfaces $M, N$ embedded in 3D, and a diffeomorphism $\varphi : N \to M$, let $C$ be the corresponding functional map $\mathcal{F}_M \to \mathcal{F}_M$. Then $M$ and $N$ are related by a rigid motion in space if and only if:*

$$\|C_\varphi \Delta_M - \Delta_N C_\varphi\| + \|C_\varphi E_M^n - E_N^n C_\varphi\| = 0,$$

*where $\Delta$ are the LB operators, while $E^n$ are functional deformation fields arising from the normal fields.*

PROOF. *Necessary condition.* The commutativity of the functional map with the Laplace-Beltrami operators immediately implies that $\varphi$ is an isometry.

The condition $C_\varphi E_M^n = E_N^n C_\varphi$ will provide an equality between the second fundamental form as $\mathcal{L}_n \mathbf{g} = -2\mathbf{h}$. Let $f, g$ be functions on $M$. Taking the inner product of the left hand side with the function $C_\varphi(f)$ and using the isometry property, allows to uncover the second fundamental form of $M$:

$$\langle C_\varphi(f), E_N^n C_\varphi(g) \rangle_{H_0^1(N)} = \langle C_\varphi(f), C_\varphi E_M^n(g) \rangle_{H_0^1(N)}$$
$$= \langle f, E_M^n(g) \rangle_{H_0^1(M)}$$
$$= -2 \int_M \mathbf{h}_M(\nabla f, \nabla g) \mathrm{d}\mu^M.$$

The right hand side leads to the pullback of the second fundamental form from $N$ to $M$:

$$\langle C_\varphi(f), E_N^\pi C_\varphi(g)\rangle_{H_0^1(N)} = -2\int_N \mathbf{h}_N(\nabla C_\varphi(f), \nabla C_\varphi(g))\mathrm{d}\mu^N$$
$$= -2\int_M ((\varphi^{-1})^\star \mathbf{h}_N)(\nabla f, \nabla g)\mathrm{d}(\varphi_\star \mu^N)$$
$$= -2\int_M ((\varphi^{-1})^\star \mathbf{h}_N)(\nabla f, \nabla g)\mathrm{d}\mu^M.$$

Thus for all functions $f, g$, we have:

$$\int_M \mathbf{h}_M(\nabla f, \nabla g)\mathrm{d}\mu^M = \int_M ((\varphi^{-1})^\star \mathbf{h}_N)(\nabla f, \nabla g)\mathrm{d}\mu^M.$$

Therefore, using a result from [Schumacher 2013], it implies $\mathbf{h}^M = (\varphi^{-1})^\star \mathbf{h}^N$.

The first and second fundamental forms of $N$ and $M$ agree, so as a consequence of the fundamental theorem of surface theory the two manifolds must relate by a rigid motion.

*Sufficient condition.* If $N, M$ are equal up to a rigid motion then the first and second fundamental forms are equal. It immediately implies that $C_\varphi \Delta_M = \Delta_N C_\varphi$. We show the second equality by reusing the computation done for the necessary condition:

$$\langle C_\varphi(f), C_\varphi E_M^\pi(g)\rangle_{H_0^1(M)} = \langle f, E_M^\pi(g)\rangle_{H_0^1(M)}$$
$$= -2\int_M \mathbf{h}_M(\nabla f, \nabla g)\mathrm{d}\mu^M$$
$$= -2\int_M ((\varphi^{-1})^\star \mathbf{h}_N)(\nabla f, \nabla g)\mathrm{d}\mu^M$$
$$= \langle C_\varphi(f), E_N^\pi C_\varphi(g)\rangle_{H_0^1(N)}.$$

Thus, we have $C_\varphi E_M^\pi = E_N^\pi C_\varphi$. □

## 9. INTRINSIC SYMMETRIZATION

The unified shape difference of composition of mapping can be computed from functional maps and shape differences of the independent maps as shown be the following Lemma.

LEMMA 5. *Assume that* $D_I^\varphi : H_0^1(M) \to H_0^1(M)$ *represents the distortion of the metric between the surfaces $M$ and $P$ induced by the diffeomorphism $\varphi : P \to M$ and $D_I^\phi : H_0^1(P) \to H_0^1(P)$ the distortion between the surfaces $P$ and $N$ linked through $\phi : N \to P$. The distortion $D_I^{\varphi \circ \phi} : H_0^1(M) \to H_0^1(M)$ associated to $\varphi \circ \phi : N \to M$ is given by*

$$D_I^{\varphi \circ \phi} = D_I^\varphi \circ C_\varphi^{-1} \circ D_I^\phi \circ C_\varphi.$$

PROOF. The proof relies only on Definition 2:

$$\int_P C_\varphi \left(\left\langle \nabla f, \nabla D_I^{\varphi \circ \phi}(g)\right\rangle\right)\mathrm{d}\mu$$
$$= \int_P C_\phi^{-1}\left(\langle \nabla(f \circ \varphi \circ \phi), \nabla(g \circ \varphi \circ \phi)\rangle\right)\mathrm{d}\mu$$
$$= \int_P \left\langle \nabla(f \circ \varphi), \nabla D_I^\phi(g \circ \varphi)\right\rangle \mathrm{d}\mu$$
$$= \int_P C_\varphi \left(\left\langle \nabla f, \nabla D_I^\varphi \left(D_I^\phi(g \circ \varphi) \circ \varphi^{-1}\right)\right\rangle\right)\mathrm{d}\mu.$$

This yields the equality $D_I^{\varphi \circ \phi}(g) = D_I^\varphi \left(D_I^\phi(g \circ \varphi) \circ \varphi^{-1}\right)$ for all $g \in H_0^1(M)$. □

Lemma 5 is used to compute the defining condition for intrinsic symmetrization. Namely, we are looking for the diffeomorphism $\varphi : M' \to M$ such that the self-map $\psi = \varphi^{-1} \circ \pi \circ \varphi : M' \to M'$ is an isometry or equivalently the unified shape difference $D_I^\psi$, computed with the map $\psi$, should be equal to identity. Using Prop. 5, $D_I^\psi$ becomes:

$$D_I^\psi = D_I^{\varphi^{-1}} C_\varphi D_I^{\pi \circ \varphi} C_\varphi^{-1}$$
$$= D_I^{\varphi^{-1}} C_\varphi D_I^\pi C_\pi^{-1} D_I^\varphi C_\pi C_\varphi^{-1}$$
$$= C_\varphi \left(D_I^\varphi\right)^{-1} D_I^\pi C_\pi^{-1} D_I^\varphi C_\pi C_\varphi^{-1}.$$

So the condition $D_I^\psi = I$ is equivalent to:

$$D_I^\pi C_\pi^{-1} D_I^\varphi C_\pi = D_I^\varphi.$$



\end{document}